\documentstyle[aps,prl,epsf,floats,multicol,amssymb,tighten]{revtex}

\begin{document}

\newcommand{\fig}[2]{\epsfxsize=#1\epsfbox{#2}} \reversemarginpar 
\newcommand{\mnote}[1]{$^*$\marginpar{$^*$ {\footnotesize #1}}}

\bibliographystyle{prsty}

\title{Nonequilibrium dynamics of random field Ising spin chains: exact results via real space RG \\
	{\it  \today}} 

\author{Daniel S. Fisher}
\address{Lyman Laboratory of Physics, Harvard University, Cambridge MA 02138, USA}
\author{Pierre Le Doussal}
\address{CNRS-Laboratoire de Physique Th\'eorique de l'Ecole\\
Normale Sup\'erieure, 24 rue Lhomond, F-75231
Paris}
\author{C\'ecile Monthus}
\address{Service de Physique Th\'eorique, CEA Saclay \\
91191 Gif-sur-Yvette, France}
\maketitle

\begin{abstract}
The non-equilibrium dynamics of classical 
random Ising spin chains with non-conserved magnetization are studied
using an asymptotically exact real space renormalization group (RSRG). We focus
on random field Ising spin chains (RFIM) with and without a uniform applied field,
as well as on Ising spin glass chains (SG) in an applied field.
For the RFIM we consider the universal regime where the random field 
and the temperature are both much smaller than the exchange 
coupling. In that regime, the Imry-Ma length that sets the scale of the equilibrium correlations is large and the
coarsening of domains from random initial conditions (e.g. 
a quench from high temperature) occurs over a wide range of length scales.
The two types of domain walls that occur
diffuse in opposite random potentials, of the form studied by Sinai, and domain walls annihilate when they  meet.
Using the RSRG we compute many universal asymptotic properties of
both the non-equilibrium dynamics and the equilibrium limit. We find that the configurations of the domain walls converge rapidly towards a set of system-specific time-dependent postitions that are {\it independent of the initial conditions}.  Thus the behavior of this non-equilibrium system is pseudodeterministic at long times because of the broad distributions of barriers that occur on the long length scales involved. Specifically, we obtain the time dependence of the energy, the magnetization and the
distribution of domain sizes (found to be statistically independent).
The equilibrium limits agree with known exact results.
We  obtain the exact scaling form of the two-point equal time correlation function
$\overline{\langle S_0(t) S_x(t) \rangle }$ and 
the two-time autocorrelations $\overline{ \langle S_0(t') S_0(t) \rangle}$.
We also compute the persistence properties of a single spin, of local magnetization,
and of domains. The analogous quantities for the $\pm J$ Ising spin glass
in an applied field are  obtained from the RFIM via a gauge transformation.
In addition to these we compute the two-point two-time correlation function
$\overline{  \langle S_0(t) S_x(t) \rangle  \langle S_0(t') S_x(t') \rangle}$
which can in principle be measured by experiments on spin-glass like systems.  The thermal fluctuations are  studied and found to be dominated by rare events; in particular all moments of truncated equal time correlations are computed. Physical properties which are typically measured in aging experiments are also studied, focussing on 
the response to a small magnetic field which is applied after waiting for the system to equilibrate for a time $t_w$.
The non-equilibrium fluctuation-dissipation ratio $X(t,t_w)$ is computed. We find that
for $(t-t_w) \sim t_w^{\hat{\alpha}}$ with $\hat{\alpha} <1$, it is equal to its equilibrium value $X=1$
although time translational invariance does not hold in this regime. It 
exhibits for $t-t_w \sim t_w$ an aging regime with a
non-trivial $X=X(t/t_w) \neq 1$, but the behaviour is markedly different from
mean field theory.
Finally the distribution of the total magnetization and of the number of domains
is computed for large finite size systems.   General issues about convergence towards equilibrium and the possibilities of weakly history-dependent evolution in other random systems are discussed.
\end{abstract}


\widetext




\newpage

\section{Introduction}

In many systems, the development of long range order is controlled by the dynamics of domain walls. 
The coarsening of domain structures evolving towards
equilibrium has been studied extensively in  pure systems
\cite{bray,majumdar} but little is known quantitatively about domain growth 
in the presence of quenched disorder. In random systems, the nonequilibrium dynamics plays an even more important role than in pure systems and is most
 relevant for understanding experiments, since many random 
systems become glassy at low temperatures with ultraslow 
dynamics which prevent full thermal equilibrium from being established 
within the accessible time scales. This slow dynamics is due, at least in cases in which it is qualitatively understood, to the large free energy barriers that must be overcome for order to be established on long length scales. Some of the best studied cases both experimentally and theoretically are a  
variety of random magnetic systems, particularly spin glasses
 and random field magnets \cite{Vihaoc,anderson,fisher_huse}.  Both of these systems have engendered a great deal of controversy about their equilibrium behavior and the resolution of these controversies has been greatly hampered by the inability of experiments to reach equilibrium.   One might well argue, however, that the most interesting properties of such random systems are, in fact, not their equilibrium properties, but the non-equilibrium dynamics involved in their Sisyphean struggle to reach equilibrium.
 
   Because of the dominance of the behavior of so many random systems by the interplay between equilibrium and non-equilibrium dynamic effects, it is important to find models with quenched randomness for which
 solid results about dynamics can be obtained.  In particular, one would like to be able to analyze the effects of activated dynamics caused by large barriers and compare results to predictions which come from either phenomenolgical scaling approaches to the dynamics --- often known somewhat misleadingly as  
``droplet models" \cite{fisher_huse}, or from mean field approaches \cite{Cuku}. Many of the interesting phenomena go under the general name of {\it aging} --- the dependence of measurable properties such as correlations and responses on the history of the system, in particular on how long it has equilibrated for: its ``age". For example, one would like to
compute quantities which probe the
violations of the fluctuation dissipation relations caused by non-equilibrium effects 
and compare them with results obtained in mean field models\cite{Cuku}; this has been done previously for coarsening of pure models \cite{purex1,purex2,purex3}.

One of the simplest non-trivial random models, and one in which  
coarsening occurs, is the random field Ising model (RFIM) in one dimension (1D).
Although this system does not have true long-range order, for weak randomness it exhibits a wide range of scales over which dynamics is qualitatively like those of other random systems, especially those, like two dimensional random field magnets and two-dimensional spin glasses, which do not have phase transitions but exhibit much dynamic behavior qualitatively similar to their three-dimensional counterparts which do have phase transitions. In particular, the weakly random 1D RFIM  has  a wide range of length scales over which the typical size of ordered domains grows
logarithmically  with time at low temperatures. In 1D, the RFIM is equivalent to
a spin glass in an applied magnetic field; this, or some analogous 1D systems should be conducive to 
experimental investigation. For a recent review on the RFIM see e.g. \cite{nattermann}.

The {\it equilibrium} properties of the 1D RFIM and of the 1D spin glass in a field have been
extensively studied \cite{nattermann,bruinsma,morgenstern,vilenkin,puma,brandt,%
gardner_derrida,bene,derrida_hilhorst,derrida_rfim,%
farhi_gutman,weigt_monasson,luck,luck_book,grinstein_mukamel}. Several thermodynamic
quantities such as the energy, entropy and magnetization 
have been computed exactly at low temperature for a binary distribution 
of the randomness in \cite{vilenkin,puma,brandt} and for continuous 
distributions in \cite{luck,luck_book,grinstein_mukamel}. Results are also available for
distributions of bonds with anomalous weight near the origin
\cite{gardner_derrida}. The free energy distribution has been studied in \cite{bene}. Equilibrium correlation functions
are harder to obtain and only a few {\it explicit} exact results
exist. For the binary distribution some results are 
presented in \cite{farhi_gutman}. Certain special limits have been solved,
such as the infinite field strength limit which is
related to percolation \cite{grinstein_mukamel}, but
for, e.g., higher cumulants of averages of truncated correlations,
only the general structure has been discussed \cite{weigt_monasson}.

In this paper we derive a host of exact results for the  
{\it nonequilibrium dynamics} of the RFIM and for the 1D Ising spin glass in a field, and, as a side benefit, also obtain many
equilibrium results, some new. Those of our results for equilibrium quantities
that can be compared with previously known ones are found to agree with these.

Here we focus on the universal regime of the RFIM in which both the random field
and the temperature are much smaller than the exchange 
coupling, the length and time scales are sufficiently long and the random field dominates the dynamics. With the random field being weak and the temperature low, the equilbrium correlation length is long and essentially the same as that at zero temperature --- the Imry-Ma length. The
coarsening of domains starting from initial conditions with only short distance correlations  --- such as
a quench from high temperature --- will, after a rapid initial transient, be dominated by the randomness over a wide range of time scales .

The basic tool that we use in this paper is a real space renormalization group (RSRG) method 
which we have developed recently to obtain exact results for the non equilibrium
dynamics of several 1D disordered systems \cite{us_prl}. Most of our previous results have been for the  {\it Sinai model} that describes the diffusion 
of a random walker in a 1D {\it random static force field} which is equivalent to a random potential that itself has the statistics 
of a 1D random walk \cite{sinai}. It can readily be seen that individual domain walls 
in the classical RFIM diffuse in a random potential of exactly this Sinai form, the complication being that they
annihilate upon meeting.  As shown in \cite{us_prl,us_rd} the RSRG can also be applied to many-domain-wall problems such as that which corresponds to the RFIM. A few  of the results of this paper have already been presented in a short paper \cite{us_prl};
the aim of the present paper is to show in detail
how the RSRG method applies to such disordered spin models and to explore more of its consequences. 
Although we will give here a detailed discussion of the
RSRG for the spin model, we will rely on Ref.
\cite{us_long} for many results about the
single particle diffusion aspects of the problem; these we will only sketch, referring the reader 
to \cite{us_long} for details.  

As for the single particle problem, the RSRG 
method allows us to compute a great variety of physical 
quantities,  remarkably, including even some which
are not known for the corresponding pure
model (e.g. the domain persistence exponents $\delta$ and $\psi$).
This provides another impetus for the study of the random models.
Using the RSRG we also obtain the equilibrium behavior which corresponds to a well defined scale
at which the decimation is stopped.

The RSRG method  is closely
related to that used to study disordered quantum spin chains 
\cite{ma_dasgupta,danfisher_rg1,danfisher_rg2,danfisher_rg3,hyman_yang,%
monthus_golinelli}. The crucial feature of the 
RG is coarse-graining the energy landscape 
in a way that  preserves the long time dynamics.
Despite its approximate character, the RSRG yields for many quantities
asymptotically exact results. As in \cite{danfisher_rg2} it works because
the width of the distribution of barrier heights grows without bound on long length scales,
consistent with rigorous results of \cite{sinai}. It is interesting to note that an exact RG has also been 
applied to the problem of coarsening of the {\it pure 1D} soft-spin Ising $\Phi^4$
model at zero temperature for which persistence exponents 
have been computed \cite{derrida_coarsening_phi4,majumdar_bray}.
Extensions to higher dimensions of the RSRG method
for equilbrium quantum models have recently been obtained
\cite{RQI}; these introduce hope that other dynamical models 
could be studied in higher dimensions as well.

The outline of this paper is as follows. In Section \ref{secmodel}
we define the RFIM and spin glass models and their dynamics and in Section \ref{summary} we summarize some of the main results and the physics involved. 
In Section  \ref{secrsrg} we explain how the RSRG method is applied to the RFIM,
give the corresponding equations and fixed point solutions and discuss
the properties of the asymptotic large time state. 
In Section 
\ref{secsingle} we derive exact large time results for single time 
quantities such as the energy, the magnetization and the domain size distribution.
 In Section \ref{secsinglecorr} we obtain the time dependent 
single time two spin correlation functions both with and without an 
applied field. In Section \ref{secaging} we compute the two time spin autocorrelation
as well as the two-point two-time correlations. In Section \ref{secrare} we study aging phenomena which necessitate considering 
 rare events which dominate all moments of the thermal (truncated) correlations, which we obtain, as well
as the response to a small uniform magnetic field applied after a waiting time $t_w$ as in a typical 
aging experiment. We also compute the fluctuation dissipation ratio $X(t,t_w)$. In Section \ref{secpersist}  
we compute the persistence properties --- the probabilities of no changes --- of a single spin, of the local magnetization
and of domains. Finally in Section \ref{secfinite}
we obtain the distribution of the total magnetization and of the number of domains
for large finite size systems. A brief discussion of the possibilities for experimental tests of the predictions and speculation on the applicability of the some of the general features to higher dimensional systems are presented at the end  of the paper.  

Various technical results are relegated to appendices. In Appendix \ref{cvfull} the convergence towards the
asymptotic state is shown. In Appendix \ref{autobias} the time dependent 
single-time two-spin correlation functions in an applied field 
is derived while in Appendix \ref{appssss}, the two-point two-time correlations 
are computed in detail and in Appendix \ref{appfinite} some of the finite size properties are 
analyzed.

\section{Models and notation}

\label{secmodel}

\subsection{Random Field Ising Model}

\subsubsection{Statics}

We consider the random field Ising chain consisting of N spins $\{ S_i=\pm 1\}_{i=1,N}$
with Hamiltonian :

\begin{eqnarray}
{\cal H } = - J \sum_{i=1}^{n=N-1}  S_i S_{i+1} - \sum_{i=1}^{i=N} h_i S_i
\label{rfimmodel}
\end{eqnarray}
with independent random fields $\{\{h_i\}\}$ with identical  distribution whose important moments are:
\begin{eqnarray}
&& H\equiv \overline{h_i}\\
&& g\equiv\overline{h_i^2} - H^2
\end{eqnarray}
so that $H$ can be considered as a uniform applied field and $g$ is the mean-square
disorder strength; here and henceforth we denote averages over the quenched randomness --- usually equivalent to averages over different parts of the system --- by overbars.

The statics and the dynamics of the random field Ising model (RFIM) can be studied
in terms of domain walls living on the dual lattice: on the bonds $\{(i,i+1)\}$ will be indexed by their left-hand site $i$. A configuration of spins $\{S_i\}$ can be represented as
a series of ``particles" $A$ corresponding to
domain walls of type $(+ \vert -)$ at positions $a_1,..a_{N_A}$,
and ``particles" $B$ corresponding to
domain walls of type $(- \vert +)$ at positions $b_1,..b_{N_B}$; these must of course occur in 
an alternating sequence $ABABAB...$.
[The relation between the numbers $N_A$ and $N_B$ of domain walls depends
on the boundary conditions : for instance
in a system with periodic boundary conditions, $N_A=N_B$,
while in a system with free boundary conditions, $|N_A-N_B| \leq 1$.]

It is very useful to introduce the {\it potential} felt by the domain walls:

\begin{eqnarray}
V(x) = - 2 \sum_{i=1}^{x} h_i
\end{eqnarray}
and rewrite the energy of a configuration $\{S_i\}$
in terms of the positions of the set of domain walls as:

\begin{eqnarray}
{\cal H } = {\cal H }_{ref} +2 J (N_A + N_B)  + \sum_{\alpha=1}^{N_A} V(a_\alpha) 
- \sum_{\alpha=1}^{N_B} V(b_\alpha) 
\end{eqnarray}
where ${\cal H }_{ref}$ is the energy of the reference configuration 
where all spins are (+) :
\begin{eqnarray}
{\cal H }_{ref} =  - J (N-1) -  \sum_{i=1}^{n=N} h_i
\end{eqnarray}
Each domain wall costs energy $2J$;
the domain walls A feel the random potential $+V(x)$; whereas the 
B walls feel the potential $-V(x)$, i.e. the two type of domain walls feel {\it opposite} 
random potentials.

Let us first recall some known features of the
statics. In the absence of an applied field ($H=0$)
the system is disordered at $T=0$ and contains domains with typical
size given by the Imry Ma length:
\begin{eqnarray}
L_{IM} \approx \frac{4 J^2}{g}
\end{eqnarray}
obtained from the following simple argument: the creation of a single pair of domain walls ($A$, $B$) a distance $L$ apart costs exchange energy $4J$ independent of $L$. But the random potential 
has typical variations
$|V(x)-V(y)|_{\text{typ}} \sim \sqrt{ 4 g \vert x-y \vert}$,
and thus the typical energy that the system can gain on a length $L$
by using a favorable configuration of the random potential is 
of order $ \sqrt{ 4 g L}$.
The two energies become comparable for $ L \sim L_{IM}$ and for
$L \ge L_{IM}$, it becomes favourable for the system
to create domain walls. Thus the ground state will contain domains of typical
size $L_{IM}$. One should note the difference between the case of e.g. bimodal distributions
(for which the ground state can be degenerate) and continuous distributions for which it is non-degenerate; we will generally consider continuous distributions for simplicity.

At positive temperature  without a random field,
the thermal correlation length $L_T \sim \exp(2 J/T)$ gives 
the typical size of domains in the system in equilibrium
at temperature $T$. But in the presence of the random field
the equilibrium state at 
finite temperature is dominated by the random fields and still given by the Imry Ma picture provided:
\begin{eqnarray}
1 \ll L_{IM} \ll L_T
\label{regime}
\end{eqnarray}
which is the regime studied in the present paper.

In presence of a uniform applied field $H>0$ the system at low temperature
will contain domains of both orientations but with different typical
sizes leading to a finite magnetization per spin $m(g,H)<1$.

\subsubsection{Non-equilibrium Dynamics}

In this paper we will study the magnetization non-conserving dynamics of the
RFIM model (\ref{rfimmodel}) starting from a random initial 
condition at time $t=0$ corresponding to a quench from a high 
temperature state.

Although our results are independent of the details of the
dynamics (provided it satisfies detailed balance and is
non-conserving) let us for definiteness consider Glauber dynamics
where the transition rate from a configuration ${\cal C}$ 
to the configuration ${\cal C}_j$ obtained from ${\cal C}$ by a flip of the spin $j$,
is:
\begin{eqnarray}
W({\cal C} \to {\cal C}_j) = \frac{ e^{-\beta \Delta E } }
{ e^{\beta \Delta E } + e^{-\beta \Delta E }}
=\frac{1}{2} \left[ 1- {\rm tanh} \left( \frac{\beta \Delta E}{2} \right)\right]
\end{eqnarray} 
which satisfies the detailed balance condition
and where $ \Delta E=2 J S_j (S_{j-1} + S_{j+1}) + 2 h_j S_j$
is the energy difference between the two configurations,
which takes the following possible values in terms of domain walls:
\begin{eqnarray}
&&  \Delta E \{ \hbox{ creation of two domain walls} \}= 4 J  \pm  2 h_j  \\
&&  \Delta E \{ \hbox{ diffusion of one domain wall} \}=  \pm  2 h_j \\
&&  \Delta E \{ \hbox{ annihilation of two domain walls} \}= -4 J \pm  2 h_j
\end{eqnarray} 

In this paper we focus on the regime:
\begin{eqnarray}
\{h_i\} \ll J \\
T\ll J
\end{eqnarray} 
in which (\ref{regime}) holds. The randomness will dominate on length scales that are sufficiently large so that the cumulative effect of the random field energies is greater than $T$; i.e. for scales
\begin{equation}
l\gg \overline{h^2}/T^2.
\end{equation}
So that there will be only one basic length and time scale, it is simplest to consider $\{h_i\} \sim T$. 

There will be a wide range of time scales during which the domains grow
from initial sizes of $O(1)$ to sizes of order $L_{IM}$ by which time the system
will be close to equilibrium. Since the Imry-Ma length $L_{IM}$ 
is very large, we expect universality of the long time dynamics,
as we indeed find.

\subsection{Ising spin glass in a magnetic field}
\label{isingsg}

In this paper we  also consider the 1D Ising spin glass in a uniform field $h$:

\begin{eqnarray}
{ \cal H }= -\sum_{i=1}^{i=N-1} J_i \sigma_i \sigma_{i+1} - \sum_{i=1}^N h \sigma_i
\end{eqnarray}

As is well known, in the case of a bimodal $(\pm J)$ distribution
with equal probabilities for either sign,
1D Ising spin glasses in a field are equivalent
via a gauge transformation to random field ferromagnets.
More precisely, setting $J_i = J \epsilon_i = \pm 1$ for $i=1,..N-1$,
and defining $\sigma_1=S_1$ and $\sigma_i = \epsilon_{1}..\epsilon_{i-1}  S_i$
for $i=2,..N$, the gauge transformation gives the new Hamiltonian
\begin{eqnarray}
{\cal H }= -\sum_{i=1}^{i=N-1} J S_i S_{i+1} - \sum_{i=1}^N h_i S_i
\end{eqnarray}
where $h_i = h \epsilon_{1}..\epsilon_{i-1}$. Since the $\epsilon_i$ are independent 
random variables taking the values $\pm 1$ with probability 1/2,
the fields $h_i$ are also independent random variables 
taking the values $\pm h$ with probability 1/2.
Hence the new Hamiltonian describes a $(\pm h)$ random field Ising model
with $H=0$ and $g=h^2$ in (\ref{rfimmodel}). 

The physical interpretation is as follows. The spin glass chain
in zero field ($h=0$) has two ground states, given by $\pm \sigma^{(0)}_i$,
where $\sigma_i^{(0)} = \epsilon_{1}..\epsilon_{i-1}$, which
correspond, via the gauge tranformation, to the pure Ising 
ground states $S^{(0)}_i=+1$ and $S^{(0)}_i=-1$. 
In the presence of a field $h>0$, the ground state of the spin glass is
made out of domains of either zero field ground states. Their 
typical size is thus given by the Imry-Ma length $L_{IM}=\frac{4 J^2}{h^2}$.
These domains correspond to the intervals between frustrated bonds
since at the position of a domain wall one has $J_i \sigma_i \sigma_{i+1}
=J S_i S_{i+1} <0$. Similarly each domain has magnetization:

\begin{eqnarray}
\vert \sum_{i \in \text{domain}} \sigma_i^{(0)} \vert
= \frac{1}{h} \vert \sum_{i \in \text{domain}} h_i \vert
= \frac{1}{2 h} |V(a_\alpha) - V(b_\alpha)|
\end{eqnarray}
which is thus proportional to the absolute value of the corresponding barrier of the Sinai
random potential, a property which will be used below.

Via the gauge transformation, the non equilibrium dynamics (e.g. Glauber dynamics)  
of the spin glass in a field starting either from (i) random initial conditions 
(corresponding to a quench from high temperature) 
or from (ii) the pure ferromagnetic state (obtained by applying a large
magnetic field that is quickly reduced to be $h<<J$)
corresponds to the non equilibrium dynamics (e.g. Glauber)
of the RFIM also starting from random initial conditions.
Thus in the regime $h \sim T \ll J$ 
many of the universal results obtained for the RFIM will hold directly for the 
spin glass in a field. Thus we will study the two models in parallel in the
present paper.


 An important complication in applying the results to 1D spin glasses is that the strengths of the exchange couplings will in general be random. This provides an extra random potential for the domain walls which has, however, only short range correlations.  The conditions for there to be a wide regime of validity of the universal coarsening behavior found here is that the distribution of exchanges be {\it narrow}: 
 \begin{equation}
 |J|_{max}-|J|_{min} \ll |J|_{min}, 
 \end{equation}
 ideally, with  $|J|_{max}-|J|_{min} \sim h$ or smaller.  Note that the equilbrium correlation length will be set by $|J|_{min}$ for small $h$ as the domain walls can easily find positions with $|J|\cong |J|_{min}$ that are near to extrema of the potential caused by the random fields. 

Having defined the models, before turning to the details of the calculations, we now briefly decribe the important physics and summarize our main results.

\bigskip

\section{Qualitative behavior and summary of results}

\label{summary}

As for any Ising system, the static and dynamic properties of random spin chains can be fully described in terms of domain walls between  ``up" and ``down" domains.  The crucial simplification in 1D is that the domain walls, which come in two types $A\equiv(+|-)$  and $B\equiv(-|+)$, are point objects.  As discussed above, a random field induces a potential that the walls feel which has the statistics of  a random walk so that its variations on a length scale $L$ are of order $\sqrt{L}$. The $A$ walls tend to minima of the potential while the $B$ walls tend to maxima.  If they meet they annihilate but, on the time scales of interest here, the probability that a pair is spontaneously created, $e^{-4J/T}$ can be ignored.  

As shown by Sinai \cite{sinai}, the motion of a single domain wall in such a random potential is completely dominated by the barriers that have to be surmounted to find low energy extrema.  Since the time to surmount a barrier of height $b$ is of order $e^{b/T}$, and to find a minima a distance $L$ away a barrier of order $b\sim h\sqrt{L}$ will have to be overcome, the typical distance a wall moves in time $t$ is only of order $l(t)\sim \ln^2 t$.  Although this motion is controlled by rare thermal fluctuations that take the wall over a large barrier, the position of a wall that started at a known point can, at long times, be predicted with surprising accuracy.  This is because the width of the distribution of barriers to go distances of order $L$ is as broad as the magnitude of the lowest barrier which enabled the wall to move that distance. Thus at long times, when the barriers become very high, the probability of going first over other than the lowest surmountable barrier that delimits the region in which the wall is currently, is extremely small. 
The position of a single wall at long times is thus determined by its initial position and by the height of the maximum barrier which it {\it could} surmount up to that time; this quantity, which we denote \begin{equation}
\Gamma\equiv T\ln t
\end{equation}
yields a well defined region that the wall can explore up to time $t$. The boundaries of the appropriate region that encompasses the wall's initial position are determined by $\Gamma(t)$.  The wall will be in local equilbrium in this ``valley" on time scales of order $t$ and thus tend to spend most of its time near the bottom of the valley.  This behavior, as proved by Golosov \cite{golosov}, implies that the long time dynamics of a single wall is pseudodeterministic: rescaled by the typical distance it has gone, $\ln^2 t$, the wall's position is {\it asymptotically deterministic} at long times. 

In a random field Ising chain, the two types of domain walls move in random potentials which are identical except for their sign.  When walls meet, they annihilate.  Not surprisingly, since each wall can move a distance of order $\ln^2 t$ in time $t$ and they cannot pass through each other or occupy the same position, the density of domain walls remaining at time $t$ is simply of order $1/\ln^2 t$ so that the correlation length starting from random (or short-range correlated) initial conditions grows as 
\begin{equation}
\xi(t) \sim \Gamma^2/g \sim \ln^2 t .
\end{equation}
This is in sharp contrast to the much faster power-law growth of the correlation length in most non-random systems, for example, the $\xi(t) \sim \sqrt{t}$  for Ising systems with non-conserved order parameter.

The time-dependent correlation length can be more precisely defined from the average non-equilibrium correlation function. This, and most other non-equilibrium properties, are scaling functions of the ratio of lengths to  powers of log-times, as is characteristic of ``activated dynamic scaling" \cite{dsf-act-dyn} which occurs in many random systems.  We find that in zero applied field the average equal-time correlation function behaves as
\begin{eqnarray} 
 \overline{\langle S_0(t)S_{x}(t) \rangle} 
= \sum_{n=-\infty }^{\infty } 
\frac{48 + 64 (2n+1)^2\pi^2 g \frac{x}{\Gamma^2 }}{(2n+1)^4 \pi^4}
e^{-(2n+1)^2\pi^2 2 g \frac{x}{\Gamma^2}}
\end{eqnarray}
whose Fourier transform is at large time:
\begin{equation}
\sum_{x= -\infty}^{+\infty}  e^{i q x} \overline{\langle S_0(t)S_{x}(t) \rangle} 
\sim \frac{8}{\pi^2 q^2 \xi(t)}\Re\biggl[\tanh^2(\pi \sqrt{iq\xi(t)}/2)\biggr] 
\end{equation}
with $\Re$ denoting the real part. These results obtain until the non-equilibrium correlation length
\begin{equation}
\xi(t)= \frac{T^2\ln^2 t}{2g\pi^2}
\end{equation}
reaches the equilibrium correlation length
\begin{eqnarray}
\xi_{eq}=
\frac{8 J^2}{\pi^2 g} =\frac{2}{\pi^2} L_{IM}
\end{eqnarray} 
It is intriguing that the {\it form} of the non-equilibrium correlations in the universal scaling limit are {\it identical} to those of the equilibrium correlations, the only difference being the correlation length.

A remarkable property  of the coarsening in the RFIM chain was conjectured in \cite{us_prl}: the positions of the {\it set} of domain walls and hence all of the correlations in the non-equilibrium state are {\it asymptotically deterministic} at long times. This means that while the evolution of the domain wall structure is only logarithmic in time, the domain wall configurations of two runs following quenches using the same sample converge to each other much more rapidly: as a power of time!  More precisely, the probability that the spin configurations of the two runs measured at the same time after the quench differ substantially in a region of size of order $\xi(t)$ decays to zero as a non-universal power of time, becoming negligible well before the system is anywhere near equilibrium. [Strictly speaking, because in rare region, as discussed later, fluctuations of the position of an individual or neighboring pair of domain walls will exist, one will need to do a certain amount of time averaging for this pseudodeterminism to become most evident] 


The asymptotic determinism will occur even if the initial conditions were macroscopically distinct in the two runs: for example, if one was quenched from a high temperature state in zero applied field, and the other from a high temperature state with a small net magnetization caused by a uniform applied field.  This pseudodeterminism  in a system with many degrees of freedom is a dramatic effect; it implies that history dependence can, even under strongly non-equilibrium conditions, be weak; one might thereby be fooled into thinking that such history independence implies equilibrium.  

While the equal-time correlations during coarsening converge to an equilibrium-like form, two-time quantities show interesting history dependence, generally depending on {\it both} times rather than just the time difference as they would do in equilibrium.  There are typically three regimes: the later time, $t$,  much longer than the earlier, $t'$, the two times of the same order, and the {\it difference} between the two times, $t-t'$ much smaller than either time.  The scaling variables in these three regimes are, respectively, $\ln t'/\ln t$, $t'/t$, and $\ln(t-t')/\ln t$. In the first and third of these, the condition for asymptopia, that the scaling variable is small, is very  difficult to attain in practice, thus knowing the full form of the scaling functions is essential for analyzing experimental or numerical data.  Note that only in the last of these regimes should one expect to find equilibrium-like behavior characteristic of the local equilibrium that is being probed. 
 
 In this paper we compute a variety of correlations and response functions that illustrate some of the interesting non-equilibrium behavior.  The simplest of these is the temporal autocorrelations of a single spin.  With no applied field, the average autocorrelation function is found to decay as a power of the ratio of the correlation lengths at the two times:
\begin{equation}
\overline{\langle S_x(t)S_x(t')\rangle} \sim  \biggl(\frac{\xi(t')}{\xi(t)}\biggr)^\lambda
\end{equation}
for $t>t'$ with the exponent $\lambda=1/2$. In a three dimensional spin glass, the autocorrelation function determines the non-equilibrium decay of the magnetization after a large uniform field is turned off below the transition temperature\cite{fisher_huse}. There are complications here due to the fact that a small applied field needs to be left on to provide the randomness, but the basic physics is similar.

The single spin autocorrelation function provides some information on how much memory the system has of its earlier history.  More information is provided by the two-spin two-time correlation function $\overline{\langle S_x(t)S_y(t)\rangle\langle S_x(t')S_y(t')\rangle}$ which converges to a scaling function of $(x-y)/\xi(t)$ and $\xi(t')/\xi(t)$ whose Fourier transform in $(x-y)$, $D(q,t,t')$,  we compute exactly, to our knowledge the first such computation for a physical model pure or random. For an ideal spin glass in which there are no correlations between the positions of positive and negative exchanges and the distribution of the exchange strengths is symmetric in $J\to -J$, this correlation function, $D$, is the average over the randomness  --- or equivalently a positional average over the regions being probed by the   scattering --- of the product of the magnetic scattering intensity at time $t$ and wavevector $k$ and that at time $t'$ and wavevector $q-k$.  

Other properties associated with ``aging" can also be computed exactly; we study the thermal fluctuations around the configurations at two different times and the dynamic linear response to a uniform magnetic field that is turned on after waiting for some time for the system to equilibrate.  In equilibrium, these are related by the fluctuation dissipation theorem but we find, as expected, that this relation generally fails except in the limit that the time difference is much less than the waiting time.  The behavior we find is, however, rather different from that found in mean-field models \cite{Cuku2} and we discuss the contrasts between these results.  

In random systems controlled by zero temperature fixed points, such as is the case in the regime of scales studied here, thermal fluctuations and linear response functions are both dominated by rare spatially isolated regions of the system --- although which regions dominate depends on the time scales and properties of interest.  The study of these thus involves the {\it corrections} to the deterministic approximation to  the dynamics that led, as discussed above, to exact asymptotic results for many other quantities.  The dominant events are rare by a factor of, typically, $1/\ln t$, but nevertheless still lead to universal results. 

``Persistence" properties provide another probe of how much memory a system retains of its initial configuration, for example, what the probability is that a spin has never flipped.  Surprisingly, this and other related quantities --- including some which are not known in pure systems --- can also be computed exactly for the RFIM chains.

\bigskip

\section{Real-space renormalisation method}

\label{secrsrg}

The coarsening process taking place in non equilibrium dynamics
of the RFIM starting from random initial conditions can be thought of 
as a reaction-diffusion process in a 1D random environment
(Sinai landscape) for the domain walls. Indeed, in the regime
$T \ll J$ considered in this paper, the creation of pairs of 
domain walls is highly suppressed. The dynamics of the
domain walls is thus dominated by the random field as follows.
The domains walls A quickly fall to the local
minima of the random potential $V(i)\equiv V_i$, 
whereas the B walls quickly move to the local
maxima of $V_i$. Then they slowly diffuse by going over barriers in {\it opposite} Sinai potentials $\pm V_i$.
When an A and a B meet, they annihilate preserving 
the alternating ABAB... sequence.

We can thus use the real-space renormalisation procedure
introduced in \cite{us_long} to study the Sinai diffusion of a
single particle, and extend it to take into account 
the annihilation processes. In the single particle case this
procedure was shown to be asymptotically exact at long length and time scales. Here it is also expected to be 
asymptotically exact at large time, since as in the Sinai case,
we find that the the effective distribution of the random barrier heights
become infinitely broad in the limit of  large scales.  Thus the motion over the barriers becomes more and more deterministic at long times.

\subsection{Definition of the real space renormalization procedure }

We briefly outline the renormalization procedure, detailed
in \cite{us_prl,us_long}, for the diffusion of a particle in the Sinai
landscape $V_i$. Grouping segments with the same sign of the random field,
one can start with no loss of generality from a "zigzag" potential 
$V_i$ where each segment (``bond") is characterized by an energy barrier
$F_i=|V_i - V_{i+1}|$ and a length $l_i$. From the independence of the
random fields on each site, 
the pairs of bond variables $(F,l)$ are independent from bond to bond 
and are chosen from a distribution $P(F,l)$ normalized to unity.

\begin{figure}[thb]

\centerline{\fig{8cm}{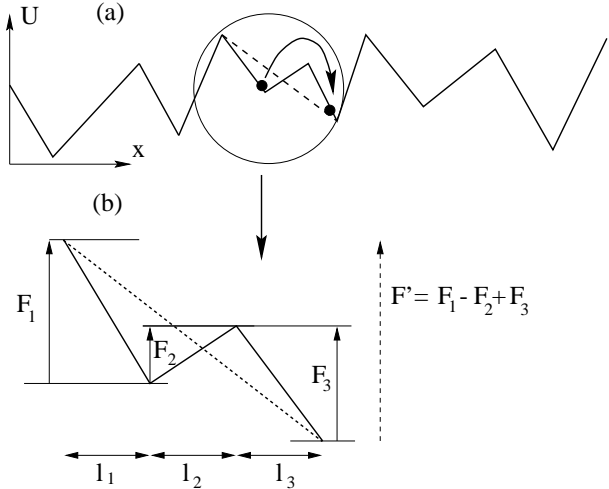}} 
\caption{ (a) Energy landscape in Sinai model (b) 
decimation method: the bond with the smallest barrier
$F_{min}=F_2$ is eliminated resulting in three bonds being grouped into
one. \label{fig1} } 

\end{figure}

The RG procedure which captures the long time behaviour in a given energy
landscape, is illustrated in
Fig. \ref{fig1}. It consists of the iterative decimation of 
the bond with the {\it smallest barrier} and hence the shortest time scale for domain walls to overcome it \cite{us_prl} \cite{us_long}. This smallest barrier, say,
$F_2$, together with its neighbors, the two bonds $1,3$, are replaced
by a single renormalized bond with barrier 
\begin{eqnarray}
F'=F_1 - F_2 + F_3
\end{eqnarray}
 and length 
\begin{eqnarray}
l'=l_1 + l_2 + l_3.
\end{eqnarray}
This defines a renormalized landscape at scale $\Gamma$ where all 
barriers smaller than  $\Gamma$ have been eliminated.    

Since the distribution of barriers is found to become broader and
broader \cite{danfisher_rg2}
an Arrhenius argument implies
that the diffusion of a particle
becomes better and better approximated at large time by the following
``effective dynamics'' \cite{us_prl,us_long}.
The position $x(t)$ of a particle that started at 
$x_0$ at $t=0$ coincides with -- or is at least very close to -- the bottom of the renormalized bond at scale
\begin{eqnarray}
\Gamma = T \ln t
\end{eqnarray}
which contains $x_0$. Note that we choose time units so as to set the
microscopic (non-universal) inverse attempt frequency to unity \cite{us_prl,us_long}.  This RG procedure is thus essentially decimation in {\it time}.  Processes that are faster than a given time scale are decimated away and assumed to be in local equilibrium.

\begin{figure}[thb]

\centerline{\fig{8cm}{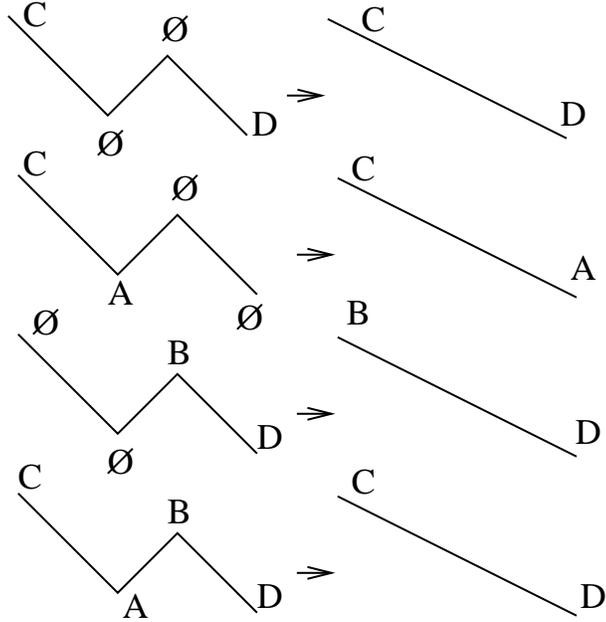}} 
\caption{ The four possible cases of the evolution under decimation of the middle bond of the domain walls, given the
constraint of alternating sequence of domains. $A$ and
$B$ denote respectively the domain walls $(+|-)$ and
$(-|+)$, $\emptyset$ denotes no domain wall present at
the top or bottom of the renormalized bond, while 
$C$ represents either a $B$ or $\emptyset$ and
a $D$ represents either a $A$ or $\emptyset$. \label{fig2} } 

\end{figure}

In the presence of domain walls of type A and B,
we must keep track of both the diffusion and the possible reactions
of the domain walls that occur during the decimation.
Upon the decimation of bond 2 (see figure \ref{fig1}),
there are four possible cases, illustrated on figure 
\ref{fig2}, according to whether or not there
is an A at the bottom of bond (2) and whether or not there is a B at the top of bond (2).
If (i) there are no A, no B, then one simply renormalise the bond
and nothing happens to the domain walls. If (ii) there is 
an A but no B then A goes to the bottom of the new renormalized bond. If
(iii) there is a B and no A then 
B goes to the top of the new renormalized bond.  And if (iv)
there is both an A and a B then the two domain walls meet and annihilate upon decimation.  This annihilation will occur at a time of order $e^{\Gamma/T}$ determined by the barrier of the decimated bond: by assumption, $\Gamma$ at this scale.

The only truly new process compared to the single particle diffusion
is the case (iv). The above RG rule is consistent with the large time
dynamics in this case for the following reason. Since 
A and B domains diffuse with independent thermal noises,
the time it takes for them to meet is again $T \ln t = F$
(accurate on a log scale). Indeed the probability that they meet at a point at potential
$V$ is $\sim \exp(-(F-V)/T) \exp(-V/T) 
= \exp(-F/T)$ (as can be seen from considering the equivalent $2d$ diffusion problem (of the pair of wall positions)
with an absorbing wall at $x+y=l$).

The second difference from the procedure in the case of 
the single particle, is that it must be stopped 
when the renormalisation scale $\Gamma$ reaches
the energy cost of creation of a pair of domain
walls:
\begin{eqnarray}
\Gamma = \Gamma_J  = 4J
\end{eqnarray}
Indeed beyond this scale $T \ln t > \Gamma_J$ one 
must take into account creation of pairs of domain
walls. As will be shown below, the state {\it at} 
$\Gamma = \Gamma_J$ gives the final 
equilibrium state.

\subsection{RG equations and statistical fixed point of the landscape}

Independently of whether the bonds contain domain walls (A or B) or not,
which is studied in the next section, one can study the evolution under
RG of the landscape. 
Since the RG rules preserve the statistical independence of
the variables $(F,l)$ from bonds to bonds, it is possible to write
closed RG equations for the landscape, i.e. for
$P_\Gamma^{+}(\zeta=F-\Gamma,l)$ 
and $P_\Gamma^{-}(\zeta=F-\Gamma,l)$ which denote the probabilities
that a $\pm$ renormalized bond at scale $\Gamma$ has a barrier
\begin{equation}
F=\Gamma + \zeta > \Gamma
\end{equation}
 and a length $l$, each normalized by
$\int_0^{\infty} d\zeta \int_0^{\infty} dl P_{\Gamma}^{\pm}(\zeta,l)=1$.

\begin{eqnarray}   \label{rgbonds}
&& (\partial_\Gamma - \partial_\zeta) P_\Gamma^{\pm} (\zeta,l)  =
 P_\Gamma^{\mp}(0,.)*_l P_\Gamma^{\pm}(.,.) *_{\zeta,l} P_\Gamma^{\pm}(.,.) \\ 
&& + P_\Gamma^{\pm} (\zeta,l) \int_0^{\infty} dl' \left(P_\Gamma^{\pm}(0,l') 
- P_\Gamma^{\mp}(0,l')\right)
\end{eqnarray}
where $*_l$ denotes a convolution with respect
to $l$ only and $*_{\zeta,l}$ with respect to both $\zeta$ and $l$
with the variables to be convoluted denoted by dots.
As discussed in \cite{danfisher_rg2} \cite{us_long},
 the solutions of these RG equations depend
on an assymetry parameter $\delta$ defined as the non vanishing root
of the equation 
\begin{eqnarray} \label{eqdelta}
\overline{ e^{- 4 \delta h} } = 1
\end{eqnarray}
which reduces in the limit of weak bias $H$ to

\begin{eqnarray}
\delta \simeq \frac{H}{2 g}
\end{eqnarray}
with $\delta=0$ in the absence of a uniform applied field.  Our results are valid for long times as long as
\begin{equation}
\delta T \ll 1
\end{equation} 

For large $\Gamma$, the Laplace transform of the distributions
$P^{\pm}_\Gamma$ take the following form, in the scaling  regime of small $\delta$ and
small $p$ with $\delta \Gamma$ fixed and $p \Gamma^2$ fixed \cite{danfisher_rg2}

\begin{eqnarray} 
\label{solu-biased}
\int_0^{\infty} dl P_{\Gamma}^{\pm}(\zeta,l) e^{-q l}  & = & U_{\Gamma}^{\pm}(
\frac{q}{ 2 g} ) e^{- \zeta u_{\Gamma}^{\pm}(\frac{q}{ 2 g})} \\
u_{\Gamma}^{\pm}(p) & = & \sqrt{p + \delta^2} \coth{[\Gamma \sqrt{p + \delta^2}]} 
\mp \delta  \nonumber \\
U_{\Gamma}^{\pm}(p)  & = & \frac{\sqrt{p 
+ \delta^2}}{\sinh{[\Gamma \sqrt{p + \delta^2}]}} e^{\mp \delta \Gamma}
\nonumber
\end{eqnarray}
This exact knowledge of the renormalized landscape will be used
below to extract physical quantities for the spin models.

\subsection{Convergence towards "full" states: asymptotic determinism}
\label{secfull}

Even armed with the statistical properties of the renormalized landscape, we still have to determine the long time distribution of the
occupation of the extrema of this landscape
by $A$ and $B$ domain walls. At first sight this seems a difficult problem.
Indeed the positions of A's and B's
can be correlated over many bonds
of the renormalized landscape since there are a priori empty maxima and minima
and the domain wall positions must respect the alternating constraint ABABAB.

\begin{figure}[thb]

\centerline{\fig{8cm}{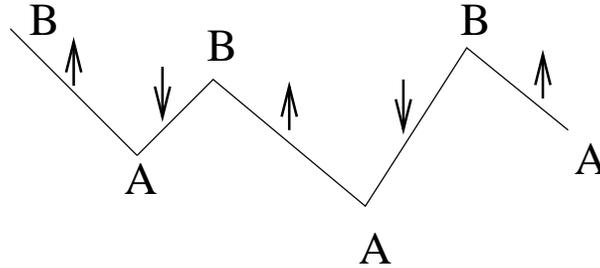}} 
\caption{Full state in the renormalized landscape; as 
we show, this corresponds to the state at large time (see text).
Each top and bottom of bonds are occupied by respectively a
$B$ and $A$ domain wall. The correponding spin orientations are 
also indicated for the RFIM. Decimation under increase of $\Gamma$ corresponds
to annihilation of the domain with the smallest barrier.
For the spin glass the position of the domain walls corresponds to the
frustrated bonds.\label{fig3} } 

\end{figure}

However, the RG analysis becomes simple if the system reaches at some stage 
a "full" state which has
one A wall at each minimum and one B wall at each maximum of the renormalized 
landscape as illustrated in figure \ref{fig3}.. It is easy to see that such a "full" state is preserved by the RG procedure.
Also note that this "full" state would obtain from the beginning
if, for instance, the initial condition were completly antiferromagnetic.

Generally, we consider random initial conditions and thus
the initial state is {\t not} a "full" state.
However, one can show that the system converges {\it exponentially} in $\Gamma$
towards a "full" state, as we now discuss.

The renormalisation procedure for the coarsening process of the RFIM
has the following important property :
the configuration for the spins at scale $\Gamma$ depends only
on the {\it renormalized} landscape at $\Gamma$ and on the {\it initial}
configuration of the spins (e. g. equilibrium at high temperature
before the quench at $t=0$).
In particular it does not depend on the initial landscape
or --- except occasionally at early times before the barrier distribution becomes very broad --- on the whole history of the reaction-diffusion processes
of domain walls. 

Let us first consider one ascending bond with extremities
$(x,y)$ of the renormalized landscape and assume that there were
initially $n$ domain walls in the interval $(x,y)$.
Neglecting for the time being the influence of
the two neighboring bonds, it is easy to see that 
the state at $\Gamma$ is determined only by $n$
and does not depend on the order in which the reactions between
domain walls have occured: indeed, it is determined by the {\it parity} of $n$ and
by the nature of the domain wall closest to the bottom end.
There are several cases 

i) If $n=0 $, then the final state of the occupation of the endpoints of the bond is $(\emptyset,\emptyset)$.

ii) If $n$ is odd, and if the domain wall closest to the bottom is of type A,
 then the final state is $(A,\emptyset)$

iii) If  $n$ is odd, and if the domain wall closest to the bottom is of type B,
 then the final state is $(\emptyset,B)$

iv) If $n$ is even with $n \geq 2$ and 
 the domain wall closest to the bottom is of type A,
 then the final state is $(A,B)$

v)  If $n$ is even with $n \geq 2$ and 
the domain wall closest to the bottom wall is of type B,
 then the final state is $(\emptyset,\emptyset)$.

\medskip
Of course, to obtain the real occupation of the top and the bottom
ends of one renormalized bond, one needs to consider what happens also on the two neighboring 
renormalized bonds and to compute the probabilities of the various states,
taking properly care of the alternating constraint ABAB.
This is done in Appendix \ref{cvfull}.  It is found that the crucial feature is that in order for a bond not to have both ends occupied by domain walls, either it or one of its neighboring bonds must have had {\it no} domain walls on it initially.  Since the bonds  tend to become progressively longer with time, the chance of this occuring drops rapidly with increasing $\Gamma$. 
This will be true even if the positions of the domain walls have some local correlations,  in particular for the case of an initial state  
that corresponds to a high temperature configuration in a small magnetic field so that there is an initial magnetization and the typical distances between a neighboring $A$ and $B$ wall will depend on which of the two is on the left.

We thus find that the system converges towards the "full" state
of the renormalized landscape exponentially fast in $\Gamma$,
with a non-universal coefficient that depends on the initial
concentration of domain walls, on the inital magnetization, and on the strength of the randomness.
This corresponds to a {\it power law} decay in time, as $t^{-\eta}$ with $\eta$ non-universal, of the probability that a maximum or minimum of the renormalized landscape at time $t$ is unoccupied.   The positions of the full set of domain walls at long times are thus asymptotically deterministic and independent of the initial conditions provided these have only short range correlations. Concretely, this asymptotic determinism can be characterized by the typical mean distance between ``missing" domain walls, i.e., deviations from the deterministic full state.  The distance between these ``errors" will grow exponentially in $\Gamma$ and hence as a power of time. The only exceptions to this are associated with fluctuations in the (nominally) {\it full} states: because of the rare times or configurations in which domain walls are either about to annihilate or are spending substantial fractions of their time in more than one valley; these fluctuation induced ``errors" we will study in detail later; they decay 
far more slowly with time than any persistent initial condition induced differences between runs.   

Since all the universal quantities that we will study have length scales that grow much more slowly -- typically logarithmically --- in time, or probabilites of occuring which decay much more slowly with time -- typically as powers of $1/\Gamma \sim 1/\ln^2 t$ -- we restrict consideration in what follows to
the analysis of  full states.  Note that our results for the biased case in which there is a small uniform field applied in addition to the random field do yield power law growth of the correlation length with time, but in the regime of validity of these results, the power law is very small and again the effects of  ``missing" domain walls are negligible at long times.

\subsection{Convergence towards equilibrium}

As mentioned earlier, the RG procedure must be stopped 
at 
\begin{equation}
\Gamma = \Gamma_J  = 4J
\end{equation}
 since at this scale one 
must start to take into account the energetic benfits of creation of pairs of domain walls.
At scale $\Gamma = \Gamma_J$ the typical domain size is the
Imry-Ma length $L_{IM}=4 J^2/g$, and the energy cannot 
be lowered further by {\it any} process in the full state.
Indeed, moving the domains walls without changing their number 
cannot lower the total energy since domain walls
already occupy all tops and bottoms of the renormalized potential. Decreasing the number
of domain walls by two can also not lower
the energy since the gain is $4 J$ while 
the loss due to the random field for walls separated by a bond of barrier $F$ is $F>\Gamma = 4 J$.
Similarly, adding two walls the cost is $4 J$ and the gain is 
$F<\Gamma =4 J$ since the only positions they can occupy
are by definition separated by a barrier $< \Gamma$ which has
already been decimated.
Thus if the renormalization is 
stopped at $\Gamma_J$, in 
the small field, low $T$ scaling limit the configuration of the walls 
corresponds precisely to the ground state and, up 
to negligible thermal fluctuations, to the thermal {\it equilibrium state}.
Thus we are able to compute equilibrium properties straightforwardly from the renormalization group analysis.

Thus the RG allows one to study the approach to equilibrium
starting from any initial condition characterized 
by typical domain sizes $L_0 e^{2 J/T_0} \ll L_{IM}$,
where $T_0$ represents the temperature before the quench. 
As explained above, under these conditions the relaxation towards
equilibrium always takes place by diffusion and annihilation of domain walls
before any domain creation can occur.

\subsection{Approach to equilibrium from more ordered initial conditions}

A qualitatively different type of evolution towards equilibrium
takes place in the other limit (not studied here):
\begin{eqnarray}
L_0 \sim e^{2 J/T_0} \gg L_{IM} = \frac{4 J^2}{g}
\end{eqnarray}
still imposing $h \ll T_0$. This is a regime of initial
temperatures low enough that the initial density of domains is
very small. In that case a very different relaxation
process towards the {\it same equilibrium state} discussed above
takes place: for $\Gamma < \Gamma_J= 4 J$ well separated
domain walls diffuse independently with very rare annihilations.
When $\Gamma$ reaches $\Gamma_J$ many domain walls are ``suddenly" -- within a factor of two or so in time -- created  
and the large initial domains break into many smaller ones of size $L_{IM}$.
Thus, the relaxation is more abrupt in this case than in the more interesting one studied in the present paper.

\section{Energy, magnetization and domain size distribution}

\label{secsingle}

From the knowledge of the fixed point for renormalized landscape (\ref{solu-biased})
and the fact that
the system reaches the full state (exponentially fast in $\Gamma$)
where each top is occupied by a B domain and each bottom by a A domain,
we can immediately compute several simple quantities. We will compare
these results with the existing exact results known in the statics.

\subsection{RFIM without applied field}

Specializing (\ref{solu-biased}) to the case $\delta=0$ in the absence
of an applied field the fixed-point of the RG equation is
\cite{danfisher_rg1,danfisher_rg2}
\begin{eqnarray}
\tilde{P}^*(\eta,\lambda) = LT_{s \to \lambda}^{-1}
\left(  \frac{\sqrt{s}}{\sinh \sqrt{s}  } \ 
e^{-\eta\sqrt{s}\coth \sqrt{s} }  \right)
\label{solu}
\end{eqnarray} 
where we have introduced the dimensionless rescaled variables 
for barriers 
\begin{equation}
\eta=(F-\Gamma)/\Gamma
\end{equation}
and for bond lengths
\begin{eqnarray}
\lambda= 2 g \frac{ l }{\Gamma^2}
\end{eqnarray}

\subsubsection{Number of domain walls per unit length}

Thus for large $\Gamma=T \ln t$, the average bond length $\overline{l}_{\Gamma}$,
equal to the average distance between two domain walls,
behaves as
\begin{eqnarray} \label{scaling}
\overline{l}_{\Gamma} =  \frac{1}{4 g }\Gamma^2 = \frac{1}{4 g } T^2 \ln^2 t
\end{eqnarray}
The number of domain walls per unit length decays as
\begin{eqnarray}
n(t) =  \frac{4 g}{T^2 \ln^2 t}
\label{density}
\end{eqnarray}
up to time $t_{eq} \sim \exp(4 J/T)$ at which equilibrium is reached 
and:
\begin{eqnarray}
n(t_{eq}) =  n_{eq}= \frac{1}{L_{IM}} 
\label{densityeq}
 \end{eqnarray}

\subsubsection{Energy density}

The energy per spin as a function of time (i.e of $\Gamma=T \ln t$) is
simply given by

\begin{eqnarray}
 E_\Gamma \simeq && - J + n_\Gamma ( 2 J - \frac{1}{2} <F>_{\Gamma}) \\
= && - J + \frac{4 g}{(T \ln t)^2} (2 J - T \ln t)
\label{eperspin}
\end{eqnarray}
where $<F>_{\Gamma}$ denotes the averaged barrier at scale $\Gamma$. This formula 
holds up to time $t_{eq} \sim \exp(4 J/T)$ where
the ground state energy at equilibrium is reached:

\begin{eqnarray}
E_{gs} \simeq - J - \frac{g}{2 J}
\end{eqnarray}
Since we consider the regime $g \ll J$, this result is 
expected to be exact to first order in $g/J$. It does indeed
agrees with the exact result of \cite{derrida_rfim}
concerning the bimodal distribution,
expanded to first order in $g$. To obtain higher orders
in the expansion in $g/J$ one would need to compute within the RG higher
orders in a $1/\Gamma$ expansion.

Note that the entropy per spin at $T=0$ computed in \cite{derrida_rfim}
for $\pm h$ distributions originates from degenerate configurations
occuring from short scales and is thus non universal.
If the distribution is continuous we expect $S \sim T$ from short scales, 
also non universal.

\subsubsection{Distribution of lengths of domains}

Since the bond lengths in the renormalized landscape are 
uncorrelated we obtain the  result that the 
lengths of the domains in the RFIM (both in the long time dynamics 
and at equilibrium) are {\it independent random variables}.
[Note that this is different from the exact result for the
dynamics of the {\it pure} Ising chain obtained in \cite{derrida_zeitak}.]

Moreover, during the coarsening process,
the probability distribution 
of the rescaled 
length $\lambda=\frac{2 g l}{\Gamma^2} = \frac{2 g l}{T^2 \ln^2 t}$
is obtained as:
\begin{eqnarray}  \label{pdel}
&& P^*(\lambda)  
= LT^{-1}_{p \to \lambda } \left( \frac{1}{ \cosh(\sqrt{p})}\right) \\
&&= \sum_{n = -\infty}^{\infty} \left(n+\frac{1}{2}\right)
\pi (-1)^n e^{- \pi^2 \lambda \left(n+\frac{1}{2}\right)^2}
= \frac{1}{\sqrt \pi \lambda^{3/2}}
\sum_{m = -\infty}^{\infty} (-1)^m (m+\frac{1}{2})
 e^{-  \frac{1}{\lambda} (m+\frac{1}{2})^2}
\end{eqnarray}

The distribution of the length of domains at equilibrium is also
given by $P(\lambda_{eq})$ where:
\begin{eqnarray}
\lambda_{eq}=\frac{2 g l}{\Gamma_J^2}= \frac{l}{2 L_{IM}}
\label{pdel2}
\end{eqnarray}

\subsection{Spin glass in a field}

Using the gauge transformation described in Section 
(\ref{isingsg}) the above results
readily apply also to the spin glass in a field. Let us recall
that a domain in the RFIM corresponds in the SG 
to an interval between two frustrated 
bonds. Using the above expressions
for the zero applied field RFIM and replacing $g \to h^2$,
we thus obtain the
averaged size of these domains from (\ref{scaling}),
their number per unit length from (\ref{density}) and (\ref{densityeq}),
and their distribution of lengths from (\ref{pdel}) and (\ref{pdel2}).

The distribution $P_\Gamma(M)$ of the magnetization $M=|\sum_{i \in domain} \sigma_i|$
of each domain is obtained from the distribution of barriers as:
\begin{eqnarray}
&& P_\Gamma(M) = \frac{1}{M_\Gamma} \exp( - \frac{M-M_\Gamma}{M_\Gamma} )
\theta(M-M_\Gamma) \\
&& M_\Gamma = \frac{\Gamma}{2 h}
\end{eqnarray}
with $\Gamma = T \ln t$. In equilibrium the same result holds with
$M_\Gamma \to 2J/h$. Note that since this variable is proportional
to the barrier, there are no domains of magnetization smaller 
than $M_\Gamma$. Similarly the joint distribution of magnetization and
length is given by (\ref{solu}) in Laplace transform and rescaled
variables $\eta=\frac{M-M_\Gamma}{M_\Gamma}$ and 
$\lambda=2 h^2 l/\Gamma^2$, with again the property of statistical independence 
of domains.

Finally, the energy per spin is given by the expression
(\ref{eperspin}) replacing $g \to h^2$.

Note that the results here are, strictly speaking, restricted to the case in which the mapping to a random field Ising model is exact:    the case in which all of the exchange interactions have the same magnitude and only differ in sign.  Nevertheless, in the more general case with a distribution of $|J|$'s, the universal aspects of the non-equilibrium behavior will be the same as long as there is a non-zero lower bound to this distribution, $|J|_{min}$.  However at times longer than $T\ln t=|J|_{min}$, domain walls will no longer necessarily be annihilated; whether they are or not will depend on the local $J$ as well as on the renormalized potential.  This will, of course, also affect the equilibrium positions of domain walls, but because there will always tend to be weak exchanges near to the extrema of the potential caused by the random fields, the changes in the positions of the walls will be negligible on the scale of the correlation length.

\subsection{RFIM in an applied field}

In a similar manner to the above analysis, we obtain results when a small uniform field $H$ is applied.
The solutions (\ref{solu-biased})
of the RG equations now depend on the parameter $\delta$ 
defined as the non vanishing root of (\ref{eqdelta}),
equal to $\delta=H/(2 g)$ in the small field limit for which our results will be asymptotically exact. 

\subsubsection{Number of domain walls and magnetisation}

The averaged sizes of domains $(\mp)$ (respectively with spins oriented
against and along the field) are found both to grow with time as:

\begin{eqnarray}
&&\overline{l}^{-}_{\Gamma} = \Gamma^2 \frac{1}{4 \gamma g} 
(1 - \frac{\sinh(\gamma)}{\gamma}e^{-\gamma}) \\
&&\overline{l}^{+}_{\Gamma} = \Gamma^2 \frac{1}{4 \gamma g} 
(e^\gamma \frac{\sinh(\gamma)}{\gamma}- 1)
\end{eqnarray}
where $\gamma = \Gamma \delta= \delta T \ln t$ and $\Gamma = T \ln t$.
In the long time limit it is:
\begin{eqnarray}
&&\overline{l}^{-}_{\Gamma} \approx \frac{T \ln t}{2 H} \\
&&\overline{l}^{+}_{\Gamma} \approx \frac{g}{2 H^2} t^{T H/g}
\end{eqnarray}
Note that the fact that the length of domains with spins in the opposite  direction from the applied
field still grow on average, is simply due to the fact that
the smallest ones (with barriers smaller than $T \ln t$) 
keep being eliminated.

The number of domain walls per unit length thus decays as
\begin{eqnarray}
n(t) \simeq \frac{2}{\overline{l}^{+}_{\Gamma}+\overline{l}^{-}_{\Gamma}} =
4 g \frac{\delta^2}{\sinh^2(\delta T \ln t)}
\label{densityb}
\end{eqnarray}
and the magnetisation per spin grows as
\begin{eqnarray}
&&  m(t) =  \frac{\overline{l}^+_{\Gamma}-\overline{l}^-_{\Gamma}}
{\overline{l}^+_{\Gamma} + \overline{l}^-_{\Gamma}} ={\cal M}[\frac{H}{2 g} T \ln t] \\
&& {\cal M}[\gamma] = \coth(\gamma) - \frac{\gamma}{\sinh(\gamma)^2}
\label{mt}
\end{eqnarray}
The function ${\cal M}[\gamma] $ starts as ${\cal M}[\gamma] \sim \frac{2}{3} \gamma$ 
for small $\gamma$ and goes exponentially to ${\cal M}=1$ for large $\gamma$.

These results hold up to time $t_{eq} \sim \exp(4 J/T)$ 
at which equilibrium is reached. The number of domain walls per unit length
in the equilibrium state is thus:
\begin{eqnarray}
n_{eq} \simeq 
4 g \frac{\delta^2}{\sinh^2( 4 \delta  J)}
\label{densityeqb}
\end{eqnarray}
with averaged sizes:
\begin{eqnarray}
&&\overline{l}^{-}_{eq} \approx \frac{2 J}{ H} \\
&&\overline{l}^{+}_{eq} \approx \frac{g}{2 H^2} e^{4 J H/g}
\end{eqnarray}
The equilibrium magnetisation per spin $m_{eq}$ is:
\begin{eqnarray}
&& m_{eq} = m(t_{eq}) = {\cal M}[\frac{2 J H}{g}]
\end{eqnarray}

\subsubsection{Energy per spin}

The energy per spin as a function of time (i.e of $\Gamma=T \ln t$) is
simply :

\begin{eqnarray}
&& E_\Gamma = - J - H + n_\Gamma ( 2 J - \frac{1}{2} <F>_-) \\
&& = - J - H + \frac{4 \delta^2 g}{\sinh^2(\delta \Gamma)}
\left[ 2 J - \frac{1}{2} \left( \Gamma + 
\frac{1}{2 \delta} (1 - e^{- 2 \Gamma \delta}) \right) \right] 
\end{eqnarray}
where $<F>_-$ denotes the averaged barrier of bonds against the field
at scale $\Gamma$. This obtains 
up to scale $\Gamma_{eq}=4 J$, at which equilibrium is reached
with ground state energy

\begin{eqnarray}
E_{gs} = - J - H - \frac{\delta g}{\sinh^2(4 J \delta )}
 (1 - e^{- 8 J \delta}) 
\end{eqnarray}

This result is compatible with the result Eq. (80) of
Derrida-Hilhorst (DH) \cite{derrida_hilhorst} 
obtained from studying products of random matrices,
as can be checked with the correspondance $\alpha_{DH} = 2 \delta$.
In addition we obtain here the explicit scaling form
in the small $\delta$ limit with $\delta J$ fixed. We have checked that 
this scaling form is also consistent with the exact result
(equation (8) in \cite{derrida_rfim})
for the bimodal distribution at leading order
in $\delta$.

\subsubsection{Distribution of domain lengths $P^{\pm}_{\Gamma}(l)$ }

As in the zero field case, the lengths of the domains are independent random variables.
Their probability distributions can be obtained from Laplace inversion of 
(\ref{solu-biased}).
They read respectively for $\pm$ domains:
\begin{eqnarray}
P^{\pm}(l) = \sum_{n=0}^{+\infty} c_n^\pm(\gamma) s_n^\pm(\gamma) 
e^{- l s_n^\pm(\gamma)}
\label{lengthdis}
\end{eqnarray}
where $\gamma = \delta T \ln t$ and the functions $c_n^\pm(\gamma)$, $s_n^\pm(\gamma)$
are given in equations (50,51,52,53) of
\cite{us_long}.

\section{Equal time two spin correlation function
$\overline{\langle S_0(t)S_{x}(t) \rangle}$ in the RFIM}

\label{secsinglecorr}

We can compute the disorder averaged two spin correlation function
by noting that, in a given environment, the equal time two spin thermal
correlation $\langle S_0(t)S_{x}(t) \rangle$ equals $+1$
if the points $0$ and $x$ are at scale $\Gamma=T \ln t$
on renormalized bonds of the same orientation (i.e. both ascending
or both descending), and equals $-1$ otherwise. 

\subsection{Zero applied field}

\begin{figure}[thb]

\centerline{\fig{8cm}{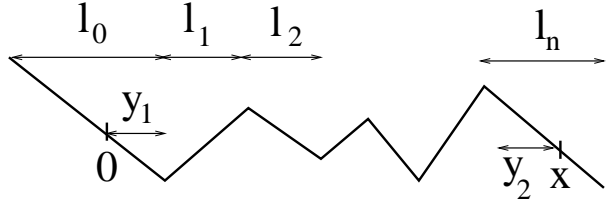}} 
\caption{Renormalized landscape at scale $\Gamma = T \ln t$ indicating the sites $0$ and $x$ of two spins and the bonds between them.\label{fig4} } 

\end{figure}

The average over the environments is, in zero field $H=0$:
\begin{eqnarray} \label{Ising}
 \overline{\langle S_0(t) S_{x}(t) \rangle} = 
\sum_{n = 0}^{\infty} (-1)^{n} Q_{\Gamma}^{(n)}(x)
\end{eqnarray}
where 
$Q_{\Gamma}^{(n)}(x)$ is defined as the probability that point $x$ 
belongs to the bond $n$ given that the point $0$
belongs to the bond $0$ of the renormalized landscape (see figure \ref{fig4}).
We have the normalization $\sum_{n = 0}^{\infty} Q^{(n)}_\Gamma(x)=1$.
For $n=0$, the probability that $x$ is on the same renormalized bond
as $0$ is
\begin{eqnarray}
Q_{\Gamma}^{(0)}(x)= && \frac{1}{\overline{l_{\Gamma}}}
\int_{0}^{\infty} dl_0 P_{\Gamma} (l_0) \int_0^{l_0} dy_1 
\int_{y_1}^{l_0} dy_2
\delta \left(x-(y_2-y_1) \right) 
\end{eqnarray}
and thus after rescaling the variables, we find
\begin{eqnarray}
Q_{\Gamma}^{(0)}(x) = q_0 \left(X=2 g \frac{x}{\Gamma^2} \right)
\end{eqnarray}
where
\begin{eqnarray}
q_0 (X)=  2
\int_{X}^{\infty} d\lambda_0 
(\lambda_0-X) P^* (\lambda_0)  
\end{eqnarray}
in terms of the fixed point solution $P^*(\lambda)$ given in (\ref{pdel})
for the distribution of rescaled lengths $\lambda=2 g l/\Gamma^2$.

For $n \geq 1$, we have
\begin{eqnarray}
Q_{\Gamma}^{(n)}(x) = q_n \left(X= 2 g \frac{x}{\Gamma^2} \right)
\end{eqnarray}
with
\begin{eqnarray}
&& q_n(X)  = 2
\int_{y_1,y_2,\lambda_i>0}  P^* (\lambda_0) 
 P^* (\lambda_1) ... 
P^* (\lambda_{n-1}) P^* (\lambda_n) \\
&&  \delta \left(X-(y_1+\lambda_1 + \lambda_2+...+\lambda_{n-1}+y_2) \right) 
\theta(\lambda_0-y_1) \theta(\lambda_n-y_2)
\end{eqnarray}
 The Laplace transforms read
\begin{eqnarray}
&& q_0(s)= \int_0^{\infty} dX e^{-sX} q_0(X)= 
\frac{2}{s^2} \left( P^*(s) - 1 + \frac{s}{2} \right) \\
&& q_n(s)= \int_0^{\infty} dX e^{-sX} q_n(X)=
2 (\frac{1-P^*(s)}{s})^2 \left[ P^*(s) \right]^{n-1}
 \qquad \hbox{for} \qquad n \ge 1 
\end{eqnarray}
and thus
\begin{eqnarray}
\sum_{n = 0}^{\infty} (-1)^{n} q_n(s)
=\frac{1}{s}-
\frac{4}{s^2} \left(\frac{1-P^*(s) }{1+P^*(s)}  \right)
\end{eqnarray}

Using now the explicit solution $P^*(s)=1/\cosh(\sqrt{s})$, we finally get

\begin{eqnarray} 
&& \overline{\langle S_0(t)S_{x}(t) \rangle}
= LT^{-1}_{s\to X= 2 g \frac{\vert x \vert }{\Gamma^2} }\left[ 
\frac{1}{s}-
\frac{4}{s^2} \rm{tanh}^2 \left( \frac{\sqrt s}{2} \right)\right]  \\
&& = \sum_{n=-\infty }^{\infty } 
\frac{48 + 64 (2n+1)^2\pi^2 g \frac{\vert x \vert}{\Gamma^2 }}{(2n+1)^4 \pi^4} 
e^{-(2n+1)^2\pi^2 2 g \frac{\vert x \vert}{\Gamma^2}} \label{Ising2}
\end{eqnarray}
with $\Gamma = T\ln t$.  
The leading long-distance behavior of the equal time correlation function is proportional to $x e^{-x/\xi(t)}$ rather than a simple exponential.  The Fourier transform of the spin-spin correlation function is simply
\begin{equation}
\sum_{x= -\infty}^{+\infty}  e^{i q x} \overline{\langle S_0(t)S_{x}(t) \rangle} 
\frac{8}{\pi^2 q^2 \xi(t)}\Re\biggl[\tanh^2(\pi \sqrt{iq\xi(t)}/2)\biggr] 
\end{equation}
with $\Re$ denoting the real part.
As explained previously, the renormalization
procedure has to be stopped at $\Gamma \simeq \Gamma_J  = 4J$, i.e
$t \sim t_{eq} \sim \exp(4 J/T)$ where the equilibrium state
has been reached.
Thus equation 
(\ref{Ising2}) with $\Gamma=\Gamma_J$
gives the mean equilibrium spin correlation 
function. The correlation length $\xi(t)$  is given by the decay
of the $(n=0)$ term which dominates at large distances 
\begin{eqnarray} 
\xi(t)=\frac{T^2 \ln^2 t} {  2 g \pi^2  } 
\label{xit}
\end{eqnarray}
up to scale $T \ln t_{eq} = \Gamma_J= 4 J$ and we obtain the
correlation length at equilibrium:
\begin{eqnarray}
\xi_{eq}=\xi(t_{eq})=
\frac{8 J^2}{\pi^2 g} =\frac{2}{\pi^2} L_{IM}
\label{xieq}
\end{eqnarray} 
This formula is in agreement with the limit $h \ll J$ of
the exact result for the equilibrium correlation of \cite{farhi_gutman}
in the case of a bimodal distribution $(\pm h)$.

Finally, note that the 
RFIM two-point correlation function $\overline{\langle S_i(t) S_{i+x}(t) \rangle}$
also corresponds for the spin-glass to the following correlation function
involving the zero-field $T=0$ ground-state $\sigma^{(0)}_i$

\begin{eqnarray} \label{Isingsg}
\overline{\langle S_i(t) S_{i+x}(t) \rangle}
 = \overline{ \sigma^{(0)}_i \sigma^{(0)}_{i+x}
\langle \sigma_i(t) \sigma_{i+x}(t) \rangle }
\end{eqnarray}

\subsection{Non-zero applied field}

In presence of an applied field $H>0$, one defines
$C^{\epsilon_1,\epsilon_2}_t(x)$ as the probability
that $sign[S_0(t)]=\epsilon_1$ and $sign[S_x(t)]=\epsilon_2$.
One has:
\begin{eqnarray}
 \overline{\langle S_0(t) S_{x}(t) \rangle} = 
C^{++}_t(x)+C^{--}_t(x)-2 C^{+-}_t(x)
\end{eqnarray}
We know already that
\begin{eqnarray}
&& C^{++}_t(x)+C^{+-}_t(x)= Prob\{ sign[S_0(t)]=+1 \}
=\frac{\overline{l}^+_{\Gamma}}{\overline{l}^+_{\Gamma} 
+ \overline{l}^-_{\Gamma}} \\
&& C^{--}_t(x)+C^{+-}_t(x)= Prob\{ sign[S_0(t)]=-1 \}
=\frac{\overline{l}^-_{\Gamma}}{\overline{l}^+_{\Gamma} 
+ \overline{l}^-_{\Gamma}}
\end{eqnarray}
and we can thus write
the correlation function as
\begin{eqnarray}
 \overline{\langle S_0(t) S_{x}(t) \rangle}
- \overline{\langle S_0(t) \rangle} ~~ \overline{\langle S_{x}(t) \rangle} = 
1- 4 C^{+-}_t(x) - 
\left( \frac{\overline{l}^+_{\Gamma}-\overline{l}^-_{\Gamma}}
{ \overline{l}^+_{\Gamma} 
+ \overline{l}^-_{\Gamma}}\right)^2
\end{eqnarray}
Performing an analysis similar to the case of zero field, we obtain
the Laplace transform
\begin{eqnarray}
\int_0^{\infty} dx e^{-q x}  C^{+-}_t(x)
=  \frac{(1-P_{\Gamma}^+(q))(1-P_{\Gamma}^-(q))}
{(\overline{l}^+_{\Gamma} 
+ \overline{l}^-_{\Gamma}) q^2 (1-P_{\Gamma}^+(q)P_{\Gamma}^-(q))}
\end{eqnarray}
The Laplace inversion can be performed as in \cite{us_long}.
The correlation decays as a sum of exponentials, and the term which
dominates the asymptotic decay gives a correlation length
\begin{eqnarray}
\xi(t) = \frac{\Gamma^2}{2 g s_0^+(\gamma)}
\end{eqnarray}
where $\Gamma= T \ln t$, $\gamma= \delta T \ln t$ and the function $s_0^+(\gamma)$ is defined by
equations (50) and (52) in \cite{us_long}. In particular the asymptotic behaviour for 
large $\gamma$ is:
\begin{eqnarray}
\xi(t) \sim \overline{l}^+_\Gamma \sim \frac{g}{2 H^2} t^{H T/g}
\end{eqnarray}
Note that even for the equilibrium
case ($\gamma = \delta \Gamma_J = 4 \delta J$), this correlation length:
\begin{eqnarray}
\xi_{eq}  \sim \frac{g}{2 H^2} e^{4 H J/g}
\end{eqnarray}
does not seem 
to have been obtained previously; it is very different from the
correlation length of the {\it truncated} correlations -- i.e. that of the thermal fluctuations -- computed for bimodal distribution 
in \cite{farhi_gutman} and discussed here
in Section (\ref{rareev}).

\section{Aging and two time correlations}

\label{secaging}

Some of the most interesting properties of random systems involve ``aging", the dependence of measured quantities on the history of the system, particularly on how long it has been equilibrated for. 
In this Section we study one of the fundamental properties which show the effects of aging: two time non-equilibrium correlations. As before, the system is quenched
from a random initial condition at time $t=0$ and we study the aging dynamics
at late times between time $t'=t_w$  -- the waiting time -- and $t$, both $t'$ and $t>t'$ being large.

We first consider the autocorrelation
function of a given spin, from which one can extract the autocorrelation 
exponent $\lambda$. Note that since this is a single site quantity 
it applies directly to both the random field and spin glass problems.

\subsection{Spin autocorrelations in zero applied field}

We consider the autocorrelations of the random field Ising model
in zero applied field, $H=0$:
\begin{eqnarray}
C(t,t')= \overline{\langle S_i(t) S_i(t') \rangle }
\end{eqnarray}

Except at short times the system is in the ``full" state and hence $C(t,t')$ is simply the
probability that the site $i$ -- which we take to be the origin -- belongs at both time $t'$ and time 
$t$ to renormalized bonds with the same orientation. Thus,
this quantity can in principle be obtained from the result in \cite{us_long}
for the probability $\overline{ P(x,t ; x't' \vert 0,0)}$ that
a particle diffusing in a Sinai landscape starting at $0$ at time 
$t=0$ is at $x'$ at $t'$ and $x$ at $t$. Indeed $C(t,t')$ (since it is
computed in the full state) is simply related to the probability that a particles postitions $X(t)$ 
and $X(t')$ have the same sign.
However, it can also be obtained through a much simpler direct computation
which we now present. 

We define $P^{++}_{\Gamma,\Gamma'}(\zeta)$ 
(resp. $P^{-+}_{\Gamma,\Gamma'}(\zeta)$ )
as the probability that the origin is on a descending  bond at $\Gamma'$,
 and is on a descending (resp. ascending) 
bond of strength $\zeta$ at a later stage, $\Gamma$.
The RG equations read
\begin{eqnarray}
\left(\partial_{\Gamma}-\partial_\zeta \right)
 P^{\pm +}_{\Gamma,\Gamma'} (\zeta)
=-2 P^{\mp }_{\Gamma}(0) P^{\pm +}_{\Gamma,\Gamma'} (\zeta) 
+ 2 P^{ \mp }_{\Gamma}(0) P^{\pm }_{\Gamma}(.) *_{\zeta} 
P^{\pm +}_{\Gamma,\Gamma'} (.)
+P^{\pm}_{\Gamma}(.) *_{\zeta} P^{\pm}_{\Gamma}(.) 
P^{\mp+}_{\Gamma,\Gamma'} (0)
\label{rg-auto}
\end{eqnarray}
together with the initial condition
\begin{eqnarray}
&&  P^{++}_{\Gamma',\Gamma'} (\zeta)
={ 1 \over 2}  \int_0^{\infty} dl { { l P_{\Gamma'}(\zeta,l) }
\over {\overline{l}_{\Gamma'} }} \\
&&  P^{-+}_{\Gamma',\Gamma'} (\zeta) =0
\end{eqnarray}

We introduce the scaling variable:
\begin{eqnarray} 
\alpha \equiv \frac{\Gamma}{\Gamma'}=\frac{\ln t}{\ln t'} .
\end{eqnarray}
Since for large $\Gamma'$, $P_{\Gamma'}$ has reached its fixed point 
value (\ref{solu}), one has in terms of the rescaled variable $\eta=\zeta/\Gamma$:
\begin{eqnarray}  
(\alpha \partial_\alpha - (1+\eta)
\partial_\eta  + 1)  P^{\pm+}_\alpha(\eta)  = 2 \int_0^{\eta} d \eta'
e^{-(\eta-\eta')} P^{\pm+}_\alpha(\eta')+\eta e^{-\eta} P^{\mp+}_\alpha(0)
\end{eqnarray}
together with initial conditions at $\alpha=1$
\begin{eqnarray}
&& P^{++}_{\alpha=1}(\eta)= { 1 \over 2} {{ \int_0^{\infty} d\lambda  \lambda P^*(\eta,\lambda) } \over { \overline{\lambda}}} 
=\frac{1}{6} \left(1 + 2 \eta \right) e^{-\eta} 
\end{eqnarray} 
and $P^{-+}_{\alpha=1}(\eta)=0$.
The solutions are
\begin{eqnarray}
 P^{\pm+}_{\alpha}(\eta) && = { 1 \over 2} \left(A^{\pm+}_{\alpha}+\eta
B^{\pm+}_{\alpha} \right) e^{-\eta} \\
 {\rm with}\ \ \ \  A^{\pm +}_{\alpha} && = {1 \over 6} \pm {1 \over {3 \alpha}} \mp {1 \over {6 \alpha^2}} \\
 B^{ \pm +}_{\alpha} && = {1 \over 3} \pm {1 \over {3 \alpha}}
\end{eqnarray} 
which obeys the normalisation condition
$\int_0^{\infty} d\eta \left( P^{++}_{\alpha}(\eta)+  P^{-+}_{\alpha}(\eta)
\right)= \frac{1}{2}$. Since with $H=0$ we have
$P^{\pm-}_{\alpha}(\eta)=P^{\mp+}_{\alpha}(\eta)$ we obtain
\begin{eqnarray}
&& C(t,t')=\int_0^{\infty} d\eta \left( P^{++}_{\alpha}(\eta)+   P^{--}_{\alpha}(\eta)- P^{+-}_{\alpha}(\eta)-P^{-+}_{\alpha}(\eta) \right) \\
&& = 
A^{++}_{\alpha} +B^{++}_{\alpha} -A^{-+}_{\alpha} -B^{-+}_{\alpha} 
={ 4 \over {3 \alpha}}-{1 \over {3 \alpha^2}}
\end{eqnarray}
and thus the autocorrelation function of the RFIM in zero applied field  
at large times $t \geq t'$ is
\begin{eqnarray}
\overline{ \langle S_i(t) S_i(t') \rangle }
={ 4 \over 3} \left({ {\ln t'} \over {\ln t}} \right)
-{ 1 \over 3} \left({ {\ln t'} \over {\ln t}} \right)^2
\label{resauto}
\end{eqnarray}

In particular the asymptotic behavior for fixed $t'$ is:
\begin{eqnarray}
\overline{ \langle S_i(t) S_i(t') \rangle }
\propto \left({ \overline{l}(t') \over \overline{l}(t)} \right)^{\lambda}
\end{eqnarray}
where $\overline{l}(t) \sim \ln^2 t$ is the characteristic length of the
coarsening in the RFIM and where the {\it autocorrelation exponent}
is:
\begin{eqnarray}
\lambda= \frac{1}{2}
\end{eqnarray}
The auto-correlation being invariant under gauge transformations,
we immediately obtain for the spin glass:

\begin{eqnarray}
\overline{\langle \sigma_i(t) \sigma_i(t') \rangle}
= \overline{\langle S_i(t) S_i(t') \rangle} 
\sim \left( \frac{ \overline{l}(t') 
}{\overline{l}(t)} \right)^{\lambda} \qquad \hbox{with}
\qquad \lambda = \frac{1}{2}
\end{eqnarray}

Note that this value of $\lambda$ saturates the lower bound of 
$d/2$ in contrast to the pure 1D Ising case which saturates the upper bound of 
$\lambda =d$ \cite{fisher_huse}.

\subsection{Autocorrelations for the RFIM in an applied field}

In presence of an applied field $H>0$ the calculation is similar to the above ; it is detailed
in  Appendix \ref{autobias}. Using the scaling variables:
\begin{eqnarray}
&& \gamma = \delta \Gamma \approx \frac{H}{2 g} T \ln t \\
&& \gamma' = \delta \Gamma' \approx \frac{H}{2 g} T \ln t' 
\end{eqnarray}
for small $H$ and the magnetization per spin,
\begin{eqnarray}
\overline{<S_i(t)>}= m(t) = {\cal M}(\gamma)=\coth \gamma - { {\gamma} \over {\sinh^2\gamma}},
\end{eqnarray}
the result for the autocorrelation function is:

\begin{eqnarray}
&&  \overline{<S_i(t') S_i(t)>} - \overline{<S_i(t)>} ~~ \overline{<S_i(t')>}
= { 1 \over {\sinh^2\gamma} }
\left((2 \gamma -\gamma' ) {\cal M}(\gamma')+ \gamma' \coth \gamma'-1 \right)
\end{eqnarray}

The long time asymptotic behaviour is:

\begin{eqnarray}
\overline{<S_i(t') S_i(t)>} - \overline{<S_i(t)>} ~~ \overline{<S_i(t')>}
= \frac{4 H T \ln t}{g t^{H T/g}} m(t').
\end{eqnarray}
This result is valid, strictly, to leading order in $\delta T$.  In general the exponent for the power law decay in time which occurs here and for other quantities in the presence of a uniform applied field will have $O(\delta^2 T^2)$ corrections.

\subsection{Two-point two-time correlation function for the spin glass}

In order to characterize the spatial aspects of aging dynamics in the
spin glass we have computed the following correlation function:

\begin{eqnarray}  \label{scalssss} \overline{ \langle S_0(t) S_x(t) \rangle
\langle S_0(t') S_x(t') \rangle  } = F \left( \frac{x}{\Gamma'^2} ;
\frac{\Gamma}{\Gamma'} \right) \end{eqnarray} 
with $t>t'$; this becomes, at long times, a scaling function
of  
\begin{equation}
X \equiv \frac{x}{T^2 \ln^2 t'}
\end{equation}
 and 
 \begin{equation}
 \alpha = \ln t/\ln t' \geq 1 .
 \end{equation}
  We have computed the scaling function $F[X,\alpha]$ in Appendix
\ref{appssss}. For simplicity we have set $2 g =1$. In Laplace transform variables we have 

\begin{eqnarray} 
\label{resssss}
&& \hat{F}[p,\alpha] = \int_0^{\infty} dX e^{-pX} 
F \left( X ; \alpha\right)
= \frac{1}{p} 
-\frac{4}{p^2 } {\rm tanh}^2 \left( \frac{\sqrt p }{2} \right) \\
&& + \frac{2}{\alpha^2 p^3 \sinh^2( {\sqrt p} ) } 
{\rm coth}^2 \left( \frac{\alpha \sqrt p }{2} \right)
 \left[ 8+3  p -16 \cosh( {\sqrt p})+
8  {\sqrt p} \sinh( {\sqrt p}) \right. \nonumber \\
&& \left. \qquad  \qquad  \qquad  \qquad  \qquad
+(8+5 p) \cosh(2  {\sqrt p})
-12 {\sqrt p} \sinh ( 2  {\sqrt p}) \right] \nonumber \\
&& -\frac{16}{\alpha^2 p^2 } 
\frac{ {\rm coth} \left( \frac{\alpha \sqrt p }{2} \right)} { \sqrt p}
\left( 1 - 
{\sqrt p} {\rm coth} ( {\sqrt p})\right)
\left( \frac{2}{ \sqrt p}  {\rm tanh} \left( \frac{ {\sqrt p}}{2} \right)-1\right) \nonumber \\
&& -\frac{16}{\alpha^2 p^3 \sinh^2\left( \frac{\alpha \sqrt p}{2} \right) }
\left( 1 - 
{\sqrt p} {\rm coth} ( {\sqrt p})\right)^2 \nonumber
\end{eqnarray}


Because the change  of correlations between two points is caused 
only by passing of domain walls through one of the end points,
the two-time correlation function decays at large $x$ to the square of the
autocorrelation function (\ref{resauto}), i.e 
$\lim_{p ->0} p \hat{F}[p,\alpha] = (4 \alpha-1)^2/(9 \alpha^4)$. 
The spatial decay to this constant value is determined by the closest 
poles to the imaginary axis in the complex $p$-plane.  This yields exponential decay with a characteristic length 
which is the maximum of $\xi(t') = (\Gamma'/\pi)^2$ and $\xi(t)/4 = (\Gamma/2\pi)^2$.
In the regime $\alpha \sim 1$, corresponding to $\ln t \approx \ln t'$, we have the following expansion
to order $O(\alpha-1)$:
\begin{eqnarray} 
 \int_0^{\infty} dX e^{-pX} 
F \left( X ; \alpha\right)
= \frac{1}{p} - (\alpha-1) \frac{8}{p^2} (1 -\frac{\sqrt{p}}{\sinh(\sqrt{p})})
\end{eqnarray}

\begin{figure}[thb]
\centerline{\fig{7cm}{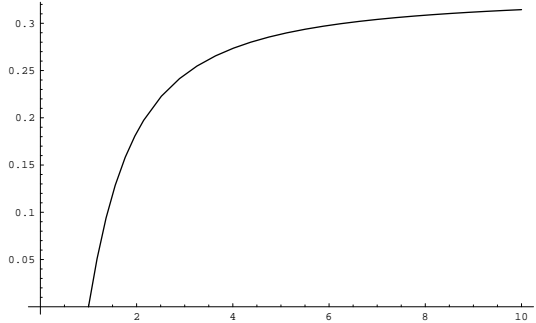}} 
\caption{Value at $q=0$ of the Fourier transform $\hat{H}(q=0, \alpha)$ ($y$ axis) as a function of $\alpha=\ln t/\ln t'$
($x$ axis). 
\label{tf2} } 
\end{figure}

\begin{figure}[thb]
\centerline{\fig{7cm}{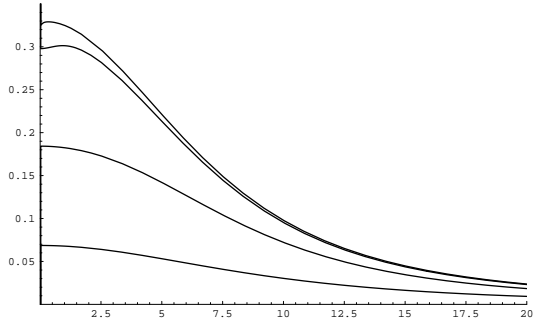}} 
\caption{Fourier transform $\hat{H}(q, \alpha)$ ($y$ axis) of the scaling function for the
two point two time spin glass correlation as a function of $q$ ($x$ axis) for four different values of 
$\alpha=1.25,2.,5.,20.$
\label{tf} } 
\end{figure}

Note that in experiments one could in principle measure the Fourier transform 
which is related to the cross-correlations of the scattering speckle patterns at two different times.
It reads:
\begin{eqnarray}  
&& \sum_{x=-\infty}^{+\infty} \overline{ \langle S_0(t) S_x(t) \rangle
\langle S_0(t') S_x(t') \rangle  } e^{i Q x}
= (T \ln t')^2 \hat{H}(q = Q (T \ln t')^2, \alpha =\frac{\ln t}{\ln t'}) \\
&&  \hat{H}(q, \alpha) = 2 {\cal R} (\hat{F}[p,\alpha] - (4 \alpha-1)^2/(9 \alpha^4 p) )|_{p= i q}) \quad q \neq  0
\end{eqnarray}
where ${\cal R}$ denotes the real part and the $1/p$ part has been substracted to get rid of the $\delta(q)$ part. 
The value of the Fourier transform as $q \to 0$ is:
\begin{eqnarray}  
\hat{H}(q=0 , \alpha) =  {\frac{-6 + 40\,\alpha - 59\,{\alpha^2} - 20\,{\alpha^3} + 45\,{\alpha^4}}{135\,{\alpha^4}}}
\end{eqnarray}
which is plotted in Fig \ref{tf2}. The Fourier transform is plotted in Fig \ref{tf} for 
several values of $\alpha$. Note the maximum at $q>0$ which develops 
for large $\alpha$ and which is related to the nonmonotonic behaviour 
of the correlation as a function of $x$. The fact that the above
correlation indeed reaches its limit by below can be seen as follows for large
$\alpha$. For $x \ll \Gamma^2$, $0$ and $x$ belong to the same domain 
and thus the above two points two time correlation is 
approximately equal to $ \overline{<S_0(t') S_x(t')>}$. This however decays exponentially
with $x/\Gamma'^2$ (and thus exponentially in $\alpha^2$ if one
chooses $x/\Gamma^2$ fixed but very small) while the asymptotic value 
(\ref{resauto}) decays only algebraically.

\newpage 

\section{Rare events, truncated correlations and response to a field}

\label{secrare}

In this Section we compute time dependent thermal (truncated) correlations 
as well as the response to a uniform field applied at time $t_w$.
For this one needs to go beyond the effective dynamics which just places each wall at a specific location at each time. Indeed,
in the effective dynamics, the local magnetization (i.e the thermal 
average) $\langle S_x(t) \rangle$ at a given point
$x$ is given by the orientation of the renormalized
bond containing the point $x$ at scale $\Gamma$,
and is thus either $+1$ or $-1$. Thus truncated correlations are
zero to leading order and to estimate them one needs to 
consider the rare events in which a domain wall can be found 
with substantial probabilities at two different positions. Such events occur with probability
$1/\Gamma$ and the two positions of the domain wall when they do occur are typically
separated by distance of order $\Gamma^2$. For example, 
for the single point Edwards-Anderson order parameter these lead to corrections to the zero temperature value of unity of order
\begin{eqnarray} 
1 - \overline{(\langle S_x(t) \rangle)^2} \propto \frac{1}{\ln t} 
\label{edand}
\end{eqnarray}

In this Section to simplify the notation somewhat, we set $2 g =1$.

\subsection{Description of the important rare events}

\label{rareev}

The rare events that are important for the RFIM turn out to be the same (as far as
the energy landscape is concerned) as
the ones that we considered in our previous study of the aging properties 
of the Sinai model \cite{us_long}, with a slightly different physical
interpretation and different observables to be computed. There are 
two types \cite{rareb} of such rare events that occur with probability $1/\Gamma$,
denoted (a) and (c) in \cite{us_long}. We now describe them.

\begin{figure}[thb]

\centerline{\fig{6cm}{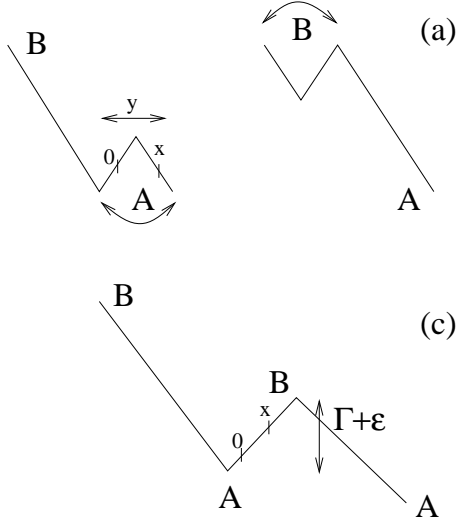}} 
\caption{Rare events (a) and (c) discussed in the text \label{fig5} } 
\end{figure}

{\it Events (a)} concern bonds which contain two almost degenerate extrema
(see figure \ref{fig5}).
In the following, we will need
the probability $D^{(a)}_{\Gamma,\Gamma'}(x)$ that two points $0$ 
and $x>0$ belong to a renormalized bond at $\Gamma$
which has two degenerate extrema separated by a barrier smaller than $\Gamma'$
such that the two points are both located between the two degenerate extrema.
This is, using the calculations of \cite{us_long}:
\begin{eqnarray}
D^{(a)}_{\Gamma,\Gamma'}(x) = && 
\frac{4}{\Gamma^2} \int_{l>y,y>x} P_\Gamma(l-y) (y-x) \int_0^{\Gamma'} d\Gamma_0
\frac{1}{\Gamma_0^4} \hat{r}(\frac{y}{\Gamma_0^2}) \\
&& = \frac{4}{\Gamma^2} \int_{y>x}
(y-x) \frac{1}{\Gamma'^3} G(\frac{y}{\Gamma'^2}) = \frac{16 \Gamma'}{\Gamma^2}
\sum_{k=1}^{+\infty} 
\frac{1}{k^2 \pi^2} e^{- \frac{x}{\Gamma'^2} k^2 \pi^2} 
\end{eqnarray}
where the functions $\hat{r}(Y)$ and $G(Y)$ have been defined in
(E5-8) and (174) of \cite{us_long}.

In such an event (a), the corresponding domain wall ($A$ or $B$)
will fluctuate thermally between the two extrema (see figure \ref{fig5}).
At equilibrium, the thermal probabilities of finding
the domain wall in the two positions are $p$ and $(1-p)$
with $p = 1/( 1 + e^{- u/T})$ and $u$
representing the energy difference between the two minima.
The random variable $u$ is distributed uniformly around $u=0$.
In the following, we will need 
\begin{eqnarray}
 c_n(T) = \overline{ (4 p (1-p))^n }
=  \frac{1}{2} T ~ 4^n \int_0^{+\infty} dz \frac{z^{n-1}}{(1 + z)^{2 n}} 
= T  \sqrt{\pi} \frac{\Gamma[n]}{\Gamma[n + \frac{1}{2}]} \label{cnT}
\end{eqnarray}
the factor $1/2$ arises as the integral is only \cite{misprint} over $u>0$.

{\it Events (c)} correspond to bonds about to be decimated,
with a barrier $\Gamma + \epsilon \approx \Gamma$.
The probability $D^{(c)}_{\Gamma}(x)$ that the segment $[0,x]$, $x>0$ belongs 
at scale $\Gamma$ to a bond of barrier $\Gamma$ (i.e. $\zeta=0$)
\begin{eqnarray}
D^{(c)}_{\Gamma}(x)=&&  \frac{2}{\Gamma^2} \int_{l>x}
P_\Gamma(\zeta=0,l) (l-x) = 
\frac{2}{\Gamma} \int_{0}^{+\infty} d \lambda \ \lambda
P^*(\eta=0,\lambda+\frac{x}{\Gamma^2} ) \\
&& = \frac{4}{\Gamma} \sum_{k=1}^{+\infty} 
(-1)^{k+1} \frac{1}{k^2 \pi^2} e^{- \frac{x}{\Gamma^2} k^2 \pi^2}
\label{cn}
\end{eqnarray}
For the events (c) associated with a barrier $\Gamma+\epsilon$,
the probability $p_c$ that the two corresponding domain walls 
have not yet annihilated at $\Gamma$ is given by
$p_c = \exp(-e^{-\epsilon/T})$. In the following, we will need
\begin{eqnarray}
 d_n(T) = \overline{ (4 p_c (1-p_c))^n }
= T ~ 4^n \int_0^{+\infty} \frac{dz}{z} e^{- n z} (1 - e^{-z})^{n}
 = T 2^{2 n} \sum_{k=0}^{n-1} C_{n-1}^k \ln(1 + \frac{1}{k + n})
\label{dn}
\end{eqnarray}
where we have used that the distribution in $\epsilon$ is uniform
around $\epsilon=0$.

\subsection{two point truncated equal time correlations}

\subsubsection{Non-equilibrium behavior}

We consider the various moments of the truncated equal time correlation function
defined as:
\begin{eqnarray}
C_n(x,t) = \overline{[ \langle S_0(t) S_x(t) \rangle - \langle S_0(t) \rangle
\langle S_x(t) \rangle ]^n };
\end{eqnarray}
$C_n(x,t)$ is the sum of two contributions. 
The first contribution $C^{(a)}_n(x,t)$ 
originates from events (a) described above, where 
we have $\langle S_0(t) S_x(t) \rangle = +1$ while
$\langle S_0(t) \rangle = \langle S_x(t) \rangle = \pm (1- 2p)$,
because the domain wall fluctuates between the two extrema
(see figure \ref{fig5}). We thus obtain the first contribution as

\begin{eqnarray}
&& C^{(a)}_n(x,t) =  c_n(T)  D^a_{\Gamma,\Gamma}(x)
\end{eqnarray}
with $\Gamma = T \ln t$.

The second contribution $C^{(c)}_n(x,t)$ originates from
events (c) described above, where we have 
$\langle S_0(t) S_x(t) \rangle = +1$ while
$\langle S_0(t) \rangle = \langle S_x(t) \rangle = \pm (1- 2p_c)$.
The second contribution is thus given by
\begin{eqnarray}
C^{(c)}_n(x,t) = d_n(T)  D^c_{\Gamma}(x)
\end{eqnarray}
The final result for the moments of the truncated equal time correlations is
thus:

\begin{eqnarray}
C_n(x,t) = \frac{4}{T \ln t} \sum_{k=1}^{+\infty}
\frac{(4 c_n(T) + (-1)^{k+1} d_n(T))}{k^2 \pi^2} ~
e^{- \frac{|x| k^2 \pi^2}{(T \ln t)^2}}
\label{ctruncated}
\end{eqnarray}
where $c_n(T)$ and $d_n(T)$ are given in (\ref{cnT},\ref{dn}).
Note that to this order in $1/\Gamma$ the $n$ dependence 
is only contained in the prefactor and in particular the
correlation length $\xi_{th}(t)= (T \ln t)^2/\pi^2$ extracted from (\ref{ctruncated})
does {\it not} depend on $n$ and is equal to the result 
(\ref{xit}).

For the spin glass, the above formula gives the {\it even} moments 
$C_{2 n}(x,t)$.

For the following sections, we will need:
\begin{eqnarray}
C_1(t)= \sum_{x=-\infty}^{x=+\infty}
C_1(x,t) = ( \frac{32}{45} + \frac{14}{45} \ln 2) T^2 \ln t
\label{c1t}
\end{eqnarray}

\subsubsection{Equilibrium truncated correlation}

To compute the equilibrium truncated correlations the method is
very similar. One must stop the RG at scale $\Gamma_J=4 J$ and
consider again events (a) (which give the same contribution as
above replacing $\Gamma$ by $\Gamma_J$) 
and events (c) but with a different interpretation and result
since there are now at equilibrium. Indeed in the renormalized landscape
at scale $\Gamma_J$ the only thermal fluctuations (apart from the
(a) events) come from barriers $\Gamma = \Gamma_J + \epsilon \approx \Gamma_J$.
The barriers much larger than $\Gamma_J$ are occupied by a pair
of domains with probability almost $1$ while the barriers well below 
$\Gamma_J$ (which have been decimated
at previous stages) are occupied with probability almost $0$.
The barriers with $\Gamma = \Gamma_J + \epsilon \approx \Gamma_J$
are occupied with probability $p  = 1/( 1 + e^{- \epsilon/T})$.
Thus we now have the equilibrium truncated correlations:

\begin{eqnarray}
C^{(eq)}_{n}(x) = c_n(T) ( D^a_{\Gamma_J,\Gamma_J}(x)
+D^c_{\Gamma_J}(x) ) = \frac{1}{J} c_n(T) \sum_{k=1}^{+\infty}
\frac{(4 + (-1)^{k+1})}{k^2 \pi^2} ~
e^{- \frac{|x| k^2 \pi^2}{(4 J)^2}}
\label{eqtruncated}
\end{eqnarray}

Here, as above, the correlation length $\xi^{eq}_{th} = 16 J^2/\pi^2$ 
extracted from (\ref{eqtruncated}) to this order in $\Gamma_J$
does {\it not} depend on $n$. Since it was argued in \cite{weigt_monasson} 
that the correlation lengths of the $C^{(eq)}_{n}(x)$ generically depend on 
$n$, our results suggest that here this dependence is subleading in $\Gamma_J$.
Our result for the correlation length of $C^{(eq)}_1(t)$ coincides with the 
result of \cite{farhi_gutman} and with the result (\ref{xieq}) with $2 g=1$
\cite{footnote_drift}
However the detailed form of the 
functions $C^{(eq)}_{n}(x)$ obtained here explicitly
depends on $n$.

\subsubsection{Approach to equilibrium for truncated correlations}

Since the equal time result for $\Gamma < \Gamma_J=4 J$ and
the equilibrium result differ only by substituting $d_n(T)$ 
by $c_n(T)$ in the (c) events, there should be a non trivial crossover
near $\Gamma=\Gamma_J$ towards equilibrium controlled by events
(c), which we now analyze.

Let us consider a bond with barrier $F=\Gamma_J + \epsilon$. 
When $\Gamma=T \ln t$ is close to $\Gamma_J$ this bond can
be either occupied (with probability $p(t)$) or empty
(with probability $1-p(t)$). One has:
\begin{eqnarray}
\frac{dp}{dt} = \frac{1}{\tau_1} (1-p(t)) - \frac{1}{\tau_2} p(t) 
\end{eqnarray}
where $\tau_1 = e^{\Gamma_J/T}$ is the inverse rate of creation of
a pair of domain walls (which immediately migrate to the
endpoints of the bond) and $\tau_2 = e^{(\Gamma_J + \epsilon)/T}$
is the inverse rate of annihilation of the pair of domain walls 
located at the endpoints of the bond. Thus, substituting
$t = e^{\Gamma/T}$ one finds:
\begin{eqnarray}
p(t) = \frac{1}{1 + e^{-\epsilon/T}} 
+ \frac{e^{-\epsilon/T}}{1 + e^{-\epsilon/T}}
\exp( - (1 + e^{-\epsilon/T}) \frac{t}{t_{eq}} )
\end{eqnarray}
where $\ln t_{eq} = \Gamma_J/T$ and 
one has $t/t_{eq} = e^{ - (\Gamma_J - \Gamma)/T }$.
Integrating over $\epsilon$,  one obtains that the
crossover to equilibrium for $t \sim t_{eq}$ is described by:
\begin{eqnarray}
&& C_n(x,t) = C^{(a)}_n(x) + C^{(c)}_n(x) \\
&& C^{(a)}_n(x) = c_n(T) D^{(a)}_{\Gamma_J,\Gamma_J}(x) \\ 
&& C^{(c)}_n(x) = e_n(T,t) D^{(c)}_{\Gamma_J}(x) 
\end{eqnarray}
with:
\begin{eqnarray}
&& e_n(T,t) = \overline{4 p(t)(1-p(t))} =
4^n T \int_{0}^{+\infty} dz \frac{z^{n-1}}{(1+z)^{2n}}
( 1 - e^{- (1+z) t/t_{eq}} )^n ( 1 + z e^{- (1+z) t/t_{eq}} )^n \\
&& = 4^n T \sum_{k=0}^n \sum_{p=0}^n (-1)^k C_n^k C_n^p 
\Gamma[n+p] U(n+p,1+p-n,(k+p) \frac{t}{t_{eq}})
e^{ - (k+p) \frac{t}{t_{eq}}}
\end{eqnarray}
where $\ln t_{eq} = 4 J/T$. This expression crosses over
from $e_n(T, t/t_{eq} \ll 1) \to d_n(T)$ and
$e_n(T, t/t_{eq} \gg 1) \to c_n(T)$.

\subsection{Two-point two-time truncated correlations}

We now consider the truncated two-point two-time correlations
\begin{eqnarray}
C_n(x,t,t_w) = \overline{( \langle S_0(t) S_x(t_w) \rangle - \langle S_0(t) \rangle
\langle S_x(t_w) \rangle )^n }
\end{eqnarray}

The calculation is very similar to the equal time truncated correlations.
We first consider the events of type (a), where we have now
to keep track of the barrier
$\Gamma_0$ between the two almost degenerate extrema.
There is a non-vanishing contribution if the barrier
$\Gamma_0$ is smaller than $\Gamma_w$ but bigger than
$\hat{\Gamma}=T \ln(t-t_w)$,
so that equilibration cannot take place between $t_w$ and $t$.
In that case we have $\langle S_0(t_w) S_x(t) \rangle = +1$ while
$\langle S_0(t_w) \rangle = \langle S_x(t) \rangle = \pm (1- 2p)$,
where $p = 1/( 1 + e^{- u/T})$ as introduced above in the description of
events of type $(a)$.
These events lead to the contribution
\begin{eqnarray}
&& C^{(a)}_n(x,t,t_w) = c_n(T) \frac{4}{\Gamma^2} \int_{l>y,y>x}
P_\Gamma(l-y) (y-x)
\int_{\hat{\Gamma}}^{\Gamma_w} d\Gamma_0
\frac{1}{\Gamma_0^4} \hat{r}(\frac{x}{\Gamma_0^2}) \\
&& = c_n(T) \left(  D^a_{\Gamma,\Gamma_w}(x) - D^a_{\Gamma,\hat \Gamma}(x) \right)
\end{eqnarray}
where $c_n(T)$ and $D^a_{\Gamma,\Gamma_w}(x)$ are given above.
Note that the contribution of these events vanishes when $\hat{\Gamma}$
becomes equal to $\Gamma_w$.

We now consider events of type (c), which give different contributions
and must be examined separately, in the scaling
regime $\hat{\Gamma} = T \ln (t-t_w) < 
T \ln t_w = \Gamma_w$ and the scaling regime $t-t_w \sim t_w$,
i.e $\hat{\Gamma} = \Gamma_w$.

We first consider $\hat{\Gamma} < \Gamma_w$. In that regime
the events of type (c) should be considered at scale $\Gamma_w$ 
where we have $\langle S_0(t_w) S_x(t) \rangle = +1$ while
$\langle S_0(t_w) \rangle = \langle S_x(t) \rangle = \pm (1- 2p)$
which, together with the (a) events, gives the total contribution
\begin{eqnarray}
C_n(x,t,t_w) =  C^{(a)}_n(x,t,t_w) + d_n(T) D^c_{\Gamma_w}(x)
\label{cxa}
\end{eqnarray}
this formula holds for $\hat{\Gamma} = T \ln (t-t_w) < T \ln t_w = \Gamma_w$.

In the regime $t-t_w \sim t_w$ i.e $\hat{\Gamma} = \Gamma_w$, the events
(c) also start equilibrating, which we now study.
Let $p_c(t_w)= \exp(e^{-\epsilon/T})$ the probability that
the domain walls separated by the barrier
$\Gamma_w + \epsilon$ have not yet annihilated at $t_w$.
Let $p_c(t) = p_c(t_w) \exp(-\frac{t-t_w}{t_w} e^{-\epsilon/T})
= \exp(-\frac{t}{t_w} e^{-\epsilon/T})$ the probability that
in addition they have also not yet annihilated at $t$.
One has (with $x$ and $0$ belonging to the bond being decimated):
\begin{eqnarray}
&& <S_x(t_w)> = 1 - 2 p_c(t_w) \\
&& <S_x(t)> = 1 - 2 p_c(t) \\
&& <S_0(t) S_{x}(t_w) > = 1 - 2 p_c(t_w) + 2 p_c(t)  \\
&& <S_0(t) S_{x}(t_w) > - <S_0(t)> <S_x(t_w)> = 4 p_c(t) (1-p_c(t_w))
\end{eqnarray}
Thus one gets, in the regime $t-t_w \sim t_w$ the total contribution 
\begin{eqnarray}
&& C_n(x,t,t_w) = d_n(T,t,t_w) D^c_{\Gamma_w}(x) \\
&& d_n(T,t,t_w)) = \overline{(4 p_c(t) (1-p_c(t_w))^n}
= T ~ 4^n \int_0^{+\infty} \frac{dz}{z} e^{- n \frac{t}{t_w} z} (1 - e^{-z})^{n}
\label{cxc}
\end{eqnarray}
In the limit $t \gg t_w$ these truncated correlations decay, for
fixed $x/(T \ln t_w)^2$, as $C_n(x,t,t_w) \sim (t_w/t)^n$.
They decay to zero as there are no other contributions for
later time $t$.

In the following we will need 

\begin{eqnarray}
C_1(t,t_w) &=& \sum_{x=-\infty}^{x=+\infty}
C_1(x,t,t_w) \\
&=&  \frac{32}{45} T^2 ( \ln t_w - \frac{(\ln(t-t_w))^3}{(\ln t_w)^2})
+ \frac{14}{45} T^2 \ln 2 \ln t_w 
 \qquad \text{for} ~~  0< \frac{\ln (t-t_w)}{\ln t_w} <1 
\label{c1ttw1} \\
&=&\frac{14}{45} T^2 \ln t_w \ln(1 + \frac{t_w}{t})
\qquad  \qquad \qquad \qquad \qquad \text{for} ~~ \frac{\ln (t-t_w)}{\ln t_w} \sim 1 \label{c1ttw2}
\\
&=&\frac{14}{45} T^2  \frac{t_w}{t} \ln t_w 
\qquad  \qquad \qquad \qquad \qquad \qquad \text{for} ~~ \frac{\ln (t-t_w)}{\ln t_w} > 1 \label{c1ttw3}
\end{eqnarray}

\subsection{Response to an applied field}

In order to compare with typical aging experiments, we will consider
the following two histories for the system and compare them:

(i) Apply $H>0$ starting from $t=0$ : the magnetization per spin $m(t)$
will then grow in time as computed in (\ref{mt}) up to time $t_{eq} \sim e^{4J/T}$,
where $m_{eq}$ is reached.

(ii) Keep $H=0$ between $t=0$ and $t_w$, and then apply $H >0$
for $t > t_w$ : in this case the magnetization per spin $m(t,t_w)$
remains $0$ up to time $t_w$, and then grows
to again reach $m_{eq}$ in the large time limit.

We now estimate $m(t,t_w)$ in the case (ii)
in the ``small applied field regime'',
where $H \sim 1/\Gamma_w^2$.
It is convenient to define 
\begin{equation}
\hat{t}\equiv t-t_w
\end{equation}
and introduce
the ratio between 
\begin{equation}
\hat{\Gamma} \equiv T \ln\hat{t} = T \ln (t-t_w) 
\end{equation}
 and 
$\Gamma_w = T \ln t_w $
\begin{eqnarray}
\hat{\alpha} = \frac{\hat{\Gamma}}{\Gamma_w} = \frac{\ln (t-t_w)}{\ln t_w} .
\end{eqnarray}
We discuss separately the three regimes $0< \hat{\alpha} < 1$,
$ \hat{\alpha} \sim 1$ and $\hat{\alpha} >1$.

\subsubsection{Response at early times $0< \hat{\alpha} < 1$ from degenerate
 wells}
 
We first study the scaling limit of small $H$ and
large $\Gamma_w$ with $H \Gamma_w^2$ fixed and
 $0< \hat{\alpha} = \frac{\ln \hat{t}}{\ln t_w} < 1$
fixed. In this regime the dominant contributions come
from bonds with near degenerate extrema, i.e the rare events
of type (a) described in (\ref{rareev}). 

Let us consider an ascending bond with a secondary minimum 
separated by a distance $y$ and a barrier $\Gamma_0$ at a potential
$u>0$ above the minimum. Using the results and notations of
\cite{us_long} we find that
the probability that an ascending bond 
of this type at $\Gamma_w$ (before the field is turned on)
is:
\begin{eqnarray}
r_{\Gamma_w}(y,\Gamma_0) d\Gamma_0 dy = 2 \theta(\Gamma_w - \Gamma_0) 
P_{\Gamma_0}(0, .)*_y P_{\Gamma_0}(0, .) d\Gamma_0 dy
\end{eqnarray}
Just before $\Gamma_w$, there is thermal equilibrium between
the two extrema and thus the probability that the primary
extremum is occupied by an A domain wall is $p_{eq}(u) = 1/(1 + e^{-u/T})$.
The thermally averaged total magnetization 
of the segment $y$ is then $<M> = y (1 - 2 p_{eq}(u))$.
One now turns on the field at $\Gamma_w$ and 
the thermally averaged magnetization of the segment remains
the same (up to negligible probability) 
until the time $\hat{\Gamma} = \Gamma_0$ when a new
equilibrium is attained (here we can neglect the equilibration 
time scale). Just after $\hat{\Gamma} = \Gamma_0$ the occupation 
probability of the left minimum
is now $p_{eq}(u') = 1/(1 + e^{-u'/T})$ where $u'=u - 2 H y$. 
The new magnetization is $<M>' = y (1 - 2 p_{eq}(u-2 H y))$. Similarly, the
contribution of a descending bond (with $y,\Gamma_0$) is
given by the same formula with $u <0$. Finally the contribution
of degeneracy of hills --- fluctuations of the $B$ domains --- 
yields an overall factor of two. The total contribution of all
these events to the magnetization per spin is:
\begin{eqnarray}
m(t,t_w) = m^{(a)}(t,t_w) = 2 \frac{1}{\Gamma_w^2} 
\int_0^{\hat{\Gamma}} d\Gamma_0 \int_0^{+\infty} dy
\int_0^{+\infty} du\  r_{\Gamma_w}(y,\Gamma_0) 
y ( 1 - 2 p_{eq}(u - 2 H y) + 1 - 2 p_{eq}(-u - 2 H y) )
\end{eqnarray}
This gives:
\begin{eqnarray}
&& m(t,t_w) = m^{(a)}(t,t_w) =  4 \frac{1}{\Gamma_w} 
\frac{\hat{\Gamma}}{\Gamma_w}
\int_0^{+\infty} dY Y G(Y) \int_0^{+\infty} du 
(\frac{1}{1 + e^{(u- 2 H \hat{\Gamma}^2 Y)/T} } - \frac{1}{1 + e^{(u + 
2 H \hat{\Gamma}^2 Y)/T}}) \\
&& 
= 8 (H \hat{\Gamma}^2) \frac{1}{\Gamma_w} 
\frac{\hat{\Gamma}}{\Gamma_w}
\int_0^{+\infty} dY Y^2 G(Y) \\
&& = 
\frac{32}{45} H \frac{\hat{\Gamma}^3}{\Gamma_w^2}
= \frac{32}{45} (H \Gamma_w^2) \frac{\hat{\alpha}^3}{\Gamma_w}
\label{maging}
\end{eqnarray}

In this regime, the magnetization as a scaling function of $H \Gamma_w^2$ 
is thus {\it exactly linear}. It can be shown that non linear response 
in $H \Gamma_w^2$ couples
to the curvature of the distribution of difference of potential
near zero and is of higher order in $1/\Gamma_w$.

\subsubsection{Response at times $\hat{\alpha} \simeq 1$}

When $t-t_w$ is of order $t_w$ a second effect adds
to the one computed above. It corresponds to the events of type
(c) described in Section \ref{rareev}
where the barrier of a bond at $\Gamma_w$ is equal to $\Gamma_w + \epsilon$
where $\epsilon = O(1)$ (of arbitrary sign). In the absence of the field
the pair of domains at the endpoints of the bond have not yet
annihilated at time $t_w$, with a probability 
$p_c(t_w) = \exp(- e^{- \epsilon/T})$. When adding the field 
at $t_w$ the barriers suddenly either increases (for descending bonds) 
or decreases (ascending bonds) by $2 H l$ where $l$ is the length of
the bond. For $t>t_w$ (and such that $t-t_w \sim O(t_w)$)
the probability $p_c(t)$ that the domain has not yet annihilated 
depends on $H$ and is:
\begin{eqnarray}
p_c(t) = p_c(t_w) \exp(- (\frac{t-t_w}{t_w})e^{(- \epsilon \mp 2 H l)/T})
\end{eqnarray}
for descending and ascending bonds respectively.
The events (c) thus result in a difference in magnetization compared to the zero
field case equal to:
\begin{eqnarray}
&& m^{(c)}(t, t_w) = \frac{2}{\Gamma_w^2} \int_{0}^{\infty} dl 
\int_{-\infty}^{+\infty} d \epsilon P^*(0,l) l
\exp(- e^{- \epsilon/T}) (
\exp(- \frac{t-t_w}{t_w} e^{\frac{1}{T}( - \epsilon - 2 H l)})
-
\exp(- \frac{t-t_w}{t_w} e^{\frac{1}{T} (- \epsilon + 2 H l)}) ) \\
&& = 
\frac{2 T}{\Gamma_w} 
\int_{0}^{\infty} d\lambda  P^*(\eta=0,\lambda) \lambda
\ln( \frac{t_w + (t-t_w) e^{2 H \lambda \Gamma_w^2/T}}{
t_w + (t-t_w) e^{- 2 H \lambda \Gamma_w^2/T} })
\label{complicated}
\end{eqnarray}
The total magnetization is now:
\begin{eqnarray}
m(t, t_w) = m^{(a)}(t, t_w) + m^{(c)}(t, t_w) =
 \frac{32}{45} H T \ln t_w
+ m^{(c)}(t, t_w) \end{eqnarray}
Note that in the present regime the magnetization as a
scaling function of $H \Gamma_w^2$ is complicated and
non linear. 

In the limit where $H \Gamma_w^2$ is small, one has
\begin{eqnarray}
&& m^{(c)}(t, t_w) = \frac{8}{\Gamma_w} H \Gamma_w^2
\frac{t-t_w}{t}
\int_{0}^{\infty} d\lambda  P^*(\eta=0,\lambda) \lambda^2
+ O((H \Gamma_w^2)^2)
= \frac{14}{45 } H \Gamma_w
\frac{t-t_w}{t} + O((H \Gamma_w^2)^2)
\label{alpha1}
\end{eqnarray}
where we have used $\int_{0}^{\infty} d\lambda  P^*(\eta=0,\lambda) \lambda^2
= 7/180$.

Although the above function (\ref{complicated}) is complicated,
at the special time such that $t -t_w = t_w$ it takes the 
simple value:
\begin{eqnarray}
m^{(c)}(2 t_w, t_w) = \frac{4}{\Gamma_w} H \Gamma_w^2
\int_{0}^{\infty} d\lambda  P^*(\eta=0,\lambda) \lambda^2
= \frac{7}{45} H \Gamma_w
\end{eqnarray}

\subsubsection{Response at times $\alpha=\hat{\alpha} >1$}

\label{responselarge}

For time differences $t-t_w \gg\gg t_w$, corresponding to $\hat{\alpha}=\alpha >1$, the response of the RFIM chain to an applied field will be dominated by the effective dynamics
described by the RSRG procedure. When the field $H>0$ is turned on, the descending bonds with $(F,l)$ become $(F+2 H l, l)$ and the ascending bonds become $(F- 2 H l, l)$
except if $\Gamma_w < F < \Gamma_w  + 2 H l$, since in this case they must be
immediately decimated.  Technically, from the point of view of the landscape,
it is more convenient to symmetrize the initial condition at ${\hat \Gamma}=\Gamma_w$
which amounts to artificially reintroduce the descending bonds $\Gamma_w -2 H l < F < \Gamma_w$
(these bonds being redecimated immediately do not introduce any errors for $\alpha >1$). 
This correspond to the following
initial distributions at ${\hat \Gamma}=\Gamma_w$ :

\begin{eqnarray}
&& P^{+}_{\Gamma_w}(F,l) = P_{\Gamma_w}(F,l) -
2 H l \partial_F P_{\Gamma_w}(F,l) + 2 H 
(P_{\Gamma_w}*_{F,l} P_{\Gamma_w} *_l (l P_{\Gamma_w}(0,l)) 
 - 4 H  P_{\Gamma_w}(F,l) 
  \int_{0}^{\infty} dl' l' P_{\Gamma_w}(0,l') ) \\
&& P^{-}_{\Gamma_w}(F,l) = P_{\Gamma_w}(F,l) +
2 H l \partial_F P_{\Gamma_w}(F,l) - 2 H 
(P_{\Gamma_w}*_{F,l} P_{\Gamma_w} *_l (l P_{\Gamma_w}(0,l)) 
 + 4 H  P_{\Gamma_w}(F,l) 
  \int_{0}^{\infty} dl' l' P_{\Gamma_w}(0,l') )
\end{eqnarray}
We now check that the magnetisation corresponding to this initial condition is the
one at the end of the $\hat{\alpha}=1$ regime (i.e the $t \to +\infty$ limit of \ref{alpha1}).
It  is, to first order in $H$\begin{eqnarray}
m_{eff}({\hat \Gamma}=\Gamma_w,\Gamma_w)  =
&& \frac{ \int_0^{+\infty} dl l \int_{0}^{\infty} d \zeta 
\left[ P^{+}_{\Gamma_w} (\zeta,l) - P^{-}_{\Gamma_w} (\zeta,l) \right] }
{\int_0^{+\infty} dl l \int_{0}^{\infty} d \zeta 
\left[ P^{+}_{\Gamma_w} (\zeta,l) + P^{-}_{\Gamma_w} (\zeta,l)\right] } \\
&& \simeq \frac{8 H }{\Gamma_w^2} \int_0^{\infty} dl l P_{\Gamma_w} (\Gamma_w,l)
=  \frac{14}{45} H \Gamma_w
\label{label}
\end{eqnarray}

We write to linear order $P^{\pm}_{\hat \Gamma}(F,l) = 
P_{\hat \Gamma}(F,l) + 2 H Q^{\pm}_{\hat \Gamma}(F,l)$ with the
initial condition in Laplace transform variables at $\hat \Gamma=\Gamma_w$:
\begin{eqnarray}
&& Q^{+}_{\Gamma_w}(\zeta,p) = - \partial_p( U_{\Gamma_w}(p) u_{\Gamma_w}(p) e^{- \zeta u_{\Gamma_w}(p)} )
- U_{\Gamma_w}'(0) U_{\Gamma_w}(p)^2 \zeta e^{- \zeta u_{\Gamma_w}(p)} 
+ 2 U_{\Gamma_w}'(0) U_{\Gamma_w}(p) e^{- \zeta u_{\Gamma_w}(p)} \\
&& Q^{-}_{\Gamma_w}(\zeta,p) = - Q^{+}_{\Gamma_w}(\zeta,p) 
\end{eqnarray}
To compute the magnetization we are interested only in
$P^{+}_{\hat \Gamma}(F,l) - P^{-}_{\hat \Gamma}(F,l)
= 2 H Q_{\hat \Gamma}(F,l)$ where $Q_{\hat \Gamma}(F,l) 
= Q^{+}_{\hat \Gamma}(F,l) - Q^{-}_{\hat \Gamma}(F,l)$ satisifes the
linearized RG equation:
\begin{eqnarray}
(\partial_{ \hat \Gamma} - \partial_\zeta) Q_{\hat \Gamma}(\zeta,p)
=  - Q_{\hat \Gamma}(0,p) P_{\hat \Gamma}(.,p) *_\zeta P_{\hat \Gamma}(.,p) 
+ 2 P_{\hat \Gamma}(0,p) Q_{\hat \Gamma}(.,p) *_\zeta P_{\hat \Gamma}(.,p)
+ 2 Q_{\hat \Gamma}(0,0) P_{\hat \Gamma}(\zeta,p)
\end{eqnarray}
The solution has therefore the form:
\begin{eqnarray}
Q_{\hat \Gamma}(\zeta,p) = 
 (A_{\hat \Gamma}(p) + \zeta B_{\hat \Gamma}(p) ) e^{- \zeta u_{\hat \Gamma}(p)}
\end{eqnarray}
where the coefficients $A_{\hat \Gamma}(p)$ and $B_{\hat \Gamma}(p)$
satisfy the RG equations 
\begin{eqnarray}
&& \partial_{\hat \Gamma} A_{\hat \Gamma}(p) = - u_{\hat \Gamma}(p) A_{\hat \Gamma}(p) +
B_{\hat \Gamma}(p) + 2 U_{\hat \Gamma}(p) A_{\hat \Gamma}(0) \\
&& \partial_{\hat \Gamma} B_{\hat \Gamma}(p) = - u_{\hat \Gamma}(p) B_{\hat \Gamma}(p)
\end{eqnarray}
with initial condition at ${\hat \Gamma}=\Gamma_w$
\begin{eqnarray}
&& A_{\Gamma_w}(p) = - 2 \partial_p(U_{\Gamma_w}(p) u_{\Gamma_w}(p))
+ 4 U_{\Gamma_w}(p) U_{\Gamma_w}'(0) \\
&& B_{\Gamma_w}(p) = 2 U_{\Gamma_w}(p) u_{\Gamma_w}(p) u_{\Gamma_w}'(p)
- 2 U_{\Gamma_w}(p)^2 U_{\Gamma_w}'(p)
\end{eqnarray}
The solutions are
\begin{eqnarray}
&& B_{\hat \Gamma}(p) = B_{\Gamma_w}(p) \frac{\sinh(\Gamma_w \sqrt{p})}{
\sinh({\hat \Gamma} \sqrt{p})} \\
&& A_{\hat \Gamma}(p) = 
( A_{\Gamma_w}(p) + ({\hat \Gamma} - \Gamma_w) B_{\Gamma_w}(p) )
\frac{\sinh(\Gamma_w \sqrt{p})}{\sinh({\hat \Gamma} \sqrt{p})} 
- 2 ({\hat \Gamma} - \Gamma_w) \frac{\sqrt{p}}{\sinh({\hat \Gamma} \sqrt{p})}
\end{eqnarray}
In the limit $\hat \Gamma \gg \Gamma_w$, we have
\begin{eqnarray}
&& B_{\hat \Gamma}(p) \simeq 
\frac{\sqrt{p} }{\sinh ({\hat \Gamma } \sqrt{p})}  \\
&& A_{\hat \Gamma}(p) \simeq 
- {\hat \Gamma }\frac{\sqrt{p} }{\sinh ({\hat \Gamma } \sqrt{p})}
\end{eqnarray}
and recover 
the first order linearisation in $\delta$ of the biased
fixed point solutions (\ref{solu-biased})
\begin{eqnarray}
P^{\pm}_{\Gamma}(\zeta,p) = U^{\pm}_{\Gamma}(p) e^{-\zeta u^{\pm}_{\Gamma}(p)} = P_{\Gamma}(\zeta,p) (1 \mp \delta \Gamma)  (1 \pm \delta \zeta)
= P_{\Gamma}(\zeta,p) + 2 (\delta \zeta- \delta \Gamma) 
\frac{\sqrt{p} }{\sinh ( \Gamma  \sqrt{p})} e^{- \zeta
\sqrt{p} \coth ( \Gamma  \sqrt{p}) }
\end{eqnarray}

The magnetisation at first order in $H$ is given by

\begin{eqnarray}
m_{eff}(\hat \Gamma) =
&& \frac{\int_0^{+\infty} dl l \int_{0}^{\infty} d \zeta 
\left[ P^{+}_{\hat \Gamma} (\zeta,l) - P^{-}_{\hat \Gamma} (\zeta,l)\right]}
{\int_0^{+\infty} dl l \int_{0}^{\infty} d \zeta 
\left[ P^{+}_{\hat \Gamma} (\zeta,l) + P^{-}_{\hat \Gamma} (\zeta,l)\right]} \\
&& \simeq \frac{2 H }{\hat \Gamma^2} \int_0^{\infty} dl l  
\int_{0}^{\infty} d \zeta Q_{\hat \Gamma}(\zeta,l)
= \frac{2 H }{\hat \Gamma^2} \left[ - \partial_p \left( 
\frac{A_{\hat \Gamma}(p)}{u_{\hat \Gamma}(p)} +
\frac{B_{\hat \Gamma}(p)}{u^2_{\hat \Gamma}(p)} \right)\right] \vert_{p=0} \\
&& = H {\hat \Gamma } \left[ \frac{2}{3} -\frac{16}{45}  
\left( \frac{\Gamma_w}{\hat \Gamma} \right)^3  \right]
\end{eqnarray}
It grows from
\begin{eqnarray}
m_{eff}(\hat \Gamma=\Gamma_w) = \frac{14}{45} H \Gamma_w
\end{eqnarray}
to the asymptotic regime for large $\hat \Gamma \gg \Gamma_w$
\begin{eqnarray}
m_{eff}(\hat \Gamma \gg \Gamma_w) = \frac{2}{3} H \hat \Gamma
\end{eqnarray}
that corresponds to the behavior of the magnetisation 
of the biased case at first order in $\delta$ (\ref{mt}).

We now summarize our results for the magnetisation in the regimes $\hat{\alpha} <1$, $\hat{\alpha} \simeq 1$ and $\hat{\alpha} >1$,
\begin{eqnarray}
&& m(\hat \Gamma, \Gamma_w) =
\frac{32}{45} H \Gamma_w \hat{\alpha}^3
\qquad \qquad \hbox{for} \qquad \hat{\alpha}=\frac{\hat{\Gamma}}{\Gamma_w} <1 
\label{mattw1} \\
&& m(\hat \Gamma, \Gamma_w) 
=\frac{32}{45} H \Gamma_w + \frac{14}{45 } H \Gamma_w
\frac{t-t_w}{t}
\qquad \qquad \hbox{for} \qquad 
\hat{\alpha} =\frac{\hat{\Gamma}}{\Gamma_w} \simeq 1 \label{mattw2} \\
&&  m(\hat \Gamma, \Gamma_w)= 
H {\hat \Gamma } \left[ \frac{2}{3} -\frac{16}{45 \hat{\alpha}^3 }  \right]+
\frac{32}{45} H \Gamma_w
\qquad \qquad \hbox{for} \qquad \hat{\alpha}=\frac{\hat{\Gamma}}{\Gamma_w} >1
\label{mattw3}
\end{eqnarray}

\subsection{Fluctuation-dissipation violation ratio}

Having computed truncated correlations and the response to an applied field,
we now discuss the fluctuation-dissipation violation ratio,
a measure of the nonequilibrium behaviour of the system. 
For two observables $A$ and $B$, the fluctuation-dissipation-theorem (FDT) violation ratio $X$ is defined as \cite{Cuku,Cuku2} 
\begin{eqnarray}
T \ R_{A,B}(t,t_w)= X_{A,B}(t,t_w) \ \partial_{t_w} C_{A,B}(t,t_w)
\end{eqnarray}
where $C_{A,B}(t,t_w)$ represents the truncated correlation
\begin{eqnarray}
C_{A,B}(t,t_w) = < A(t) B(t_w) > -< A(t)> <B(t_w)> 
\end{eqnarray}
and $R_{A,B}(t,t_w)$ represents the response in the observable $A$
at time $t$
to a field $H_B$ linearly coupled to the observable $B$ in the hamiltonian 
through a term of the form $-H_B B$
\begin{eqnarray}
R_{A,B}(t,t_w) = \frac{ \delta  < A(t) > } { \delta H_B (t_w)} \vert_{H_B=0} 
\end{eqnarray}

Here we have computed the magnetization resulting from an 
uniform magnetic field $H$, so that
the observables $A$ and $B$ are given by $A= (1/L) \sum_{i=1}^L S_i$
and $B= \sum_{i=1}^L S_i$ respectively. From the magnetization
\begin{eqnarray}
m(t,t_w)= \frac{1}{L} \sum_i <S_i(t)> = \overline{< S_0(t)>} = H \int_{t_w}^t du R(t,u)
\end{eqnarray}
and the truncated correlation
\begin{eqnarray}
C_1(t,t_w)=\frac{1}{L} \sum_{i=1}^L \sum_{j=1}^L
 ( < S_i(t) S_j(t_w) > -< S_i(t)> <S_j(t_w)> )
= \sum_x \overline{< S_0(t) S_x(t_w) > -< S_0(t)> <S_x(t_w)> } 
\end{eqnarray}
one obtains the fluctuation-dissipation ratio $X(t,t_w)$ as
\begin{eqnarray}
X(t,t_w) = - T \frac{\partial_{t_w} m(t,t_w)/H}{\partial_{t_w}  C_1(t,t_w)}
\end{eqnarray}
Note that we have used the infinite size limit to replace translational averages
by disorder averages. 

We have computed $C_1(t,t_w)$ (\ref{c1ttw1}-\ref{c1ttw3}) and $m(t,t_w)$
(\ref{mattw1}-\ref{mattw3}) in the three regimes $0< \hat{\alpha} < 1$, $ \hat{\alpha} \sim 1$ and $\hat{\alpha} >1$, 
and thus we get the following expression for the fluctuation dissipation violation ratio
at large times $t, t_w \gg 1$:

\begin{eqnarray}
X(t,t_w) &=& 1 
 \qquad  \qquad \qquad \qquad \qquad \qquad \text{for} ~~  0< 
\hat{\alpha}=\frac{\ln (t-t_w)}{\ln t_w} <1 
\label{xttw1} \\
X(t,t_w) &=&  \frac{t + t_w}{t} 
\qquad  \qquad \qquad \qquad \qquad \text{for} ~~ 
\frac{t-t_w}{t_w} ~~~ \text{a fixed number} ~~(\hat{\alpha}=1)  \label{xttw2} \\
X(t,t_w) &=& \frac{t}{t_w \ln t_w} ( 1 + \frac{24}{7} \frac{\ln^2 t_w}{\ln^2 t} )
\qquad  \qquad \text{for} ~~ 
\hat{\alpha}=\frac{\ln (t-t_w)}{\ln t_w} > 1 \label{xttw3}
\end{eqnarray}

The behaviour of $X$ upon increasing the time difference $t-t_w$ is as follows.
First of course one expects, when $t-t_w$ is a fixed number, a non universal equilibrium regime (not studied here)
where time translational invariance and FDT is obeyed. Increasing the time difference
as $t-t_w \sim t_w^{\hat{\alpha}}$, in the regime $0< \hat{\alpha} < 1$, we find that
{\it the FDT theorem is still obeyed} to leading order, which is quite interesting
since time translational invariance does {\it not} hold in that regime. 
Next, in the regime $\hat{\alpha}=1$, $t/t_w$ fixed, X becomes a 
non trivial scaling function of $(t-t_w)/t_w$,
which interpolates between 
$X=2$ for $(t-t_w)/t_w \to 0$ and 
$X=1$ for $(t-t_w)/t_w \to +\infty$. In order to match the value $X=1$
there is thus a non trivial crossover regime between the end of the quasi equilibrium regime $\hat{\alpha} <1$ 
and the beginning of the $\hat{\alpha}=1$ regime.
This crossover occurs for $(t-t_w )\sim t_w/\ln t_w$ 
and is given by
\begin{eqnarray}
&& X(t,t_w) = F_1( (t-t_w) \ln t_w / t_w ) \\
&& F_1(y) = \frac{14 y + 96}{7 y + 96}
\end{eqnarray}
Finally, for very separated times, in the regime $\hat{\alpha} >  1$, we find that $X$
grows towards $+\infty$. This occurs again after a crossover 
 between the end of the $\hat{\alpha}=1$ and the $\hat{\alpha}>1$ regime
which occurs on time scale $t \sim t_w \ln t_w$, where $X$ is given by
\begin{eqnarray}
&& X(t,t_w) = 1 + \frac{31}{7} \frac{t}{t_w \ln t_w}
\end{eqnarray}
which matches both the required limits.

We can now compare with the mean field models
\cite{footnotesg}. As in mean field
\cite{Cuku,Cuku2}, we find here an aging regime 
where $X$ is non trivial, and for $t/t_w$ fixed it is a function of this
scaling variable. On the other hand, contrarily to mean field models,
the ratio $X$ here is never a function of only $C_1(t,t_w)$, in the regime $\hat{\alpha}=1$
because of the extra power of $\ln t$ in $C_1(t,t_w)$,
and in general because of the presence of both scales $(t,t_w)$ and $(\ln t, \ln t_w)$.
In addition, $X$ is found here to be nonmonotonous and the
values taken by $X$ are not within the interval $[0,1]$. In
particular $X$ tends to $+\infty$  in the asymptotic regime
$t \sim t_w^\alpha$ with $\alpha >1$ since truncated correlations become
very small compared to the response in that regime. 
Since the ratio $X$ has an interpretation in some contexts as an inverse
effective temperature $X = 1/T_{eff}$ \cite{Cuku2}, one would find here
that $T_{eff} \to 0$ at large time separations, 
in contrast to the result that $T_{eff} \to +\infty$
in mean field models. Although this might appear surprising at first sight,
one should remember that in finite
dimensions many of the properties of the RFIM are controlled  --- in the the renormalization group sense --- by a zero temperature fixed point; 
this includes the intermediate regime of length scales studied 
here for a system  --- the 1D RFIM  --- with no phase transition.
 
It is also interesting to compare our results with
the fluctuation-dissipation ratio $X$ in pure systems
presenting domain growth, typically ferromagnets
below their critical temperature $T < T_c$ \cite{purex1,purex2,purex3}
or at $T=T_c$. The ratio $X$ is usually computed by considering the local observables $A=B=S_0$
but to compare with the present study we will translate these results for a
spatially uniform applied field. 

Let us briefly recall why $X=0$ in the large time scaling regime for $T<T_c$ ($X$ decays to $0$ for large $t_w$).
When initial conditions (high temperature $T > T_c$) before the quench are chosen to have
zero magnetization ($\sum_x <S_x(t)>=0$) the spin autocorrelation coincides with the truncated correlation $\sum_x <S_x(t)>=0$
and takes a simple scaling form \cite{bray}
\begin{eqnarray}
C_{1}^{pure}(t,t_w)= \sum_x < S_0(t) S_x(t_w) >  \simeq L(t_w)^d
f_1 \big( \frac{L(t_w)}{L(t)} \big) 
\label{cpure}
\end{eqnarray}
where $L(t)$ is the typical size of domains at time $t$
an $d$ the dimension of space. On the other hand, the magnetization
when a uniform field has been applied between  
 $t_w$ and $t$ behaves as \cite{purex2,purex3,braytrieste}
\begin{eqnarray}
M_{auto}(t,t_w) \sim L(t_w)^{d-a} f_2  \large( \frac{L(t_w)}{L(t)} \large)
\label{mpure}
\end{eqnarray}
with $a=1$ for Ising order parameter \cite{purex1,purex2,purex3} and  $a =d-2$ ($d>2$) for $O(N)$
model \cite{purex3}. Note the reduction of $M$ with respect to $C_1$ by a factor $1/L(t_w)$ in ferromagnets,
which immediately implies $X=0$ in the scaling regime $L(t_w) \sim L(t)$
($X=0$ as soon as $a>0$). As is usually argued, the origin of this factor can be seen
by considering the system at $t=t_w$ and focusing on the immediate response to
a small applied pulse field \cite{braytrieste} (here in $d=1$ for simplicity):
each interface responds by a factor $O(1)$ and since they occupy only a portion $1/L(t_w)$ of the system 
this yield a total response $1/L(t_w)$. By contrast,
note that {\it exactly at criticality}, $T=T_c$, there is a non trivial $X=X(t/t_w)$
which appears as a dimensionless amplitude ratio, by a different mechanism \cite{purex3}.

For the RFIM case, the {\it full} correlation function has the same scaling form as (\ref{cpure})
(as can be seen by generalizing the result for the autocorrelation computed in (\ref{resauto}))
but the {\it truncated} does not since we have obtained (\ref{c1ttw1}-\ref{c1ttw3}) 
\begin{eqnarray}
C_1(t,t_w) &=& \ln t_w \phi_1( \frac{\ln (t-t_w)}{\ln t_w} )
 \qquad \text{for} ~~  0< \frac{\ln (t-t_w)}{\ln t_w} <1 
\label{c0ttw1} \\
&=& \ln t_w \phi_2 ( \frac{t_w}{t} )
\qquad  \qquad  \qquad \text{for} ~~ \frac{\ln (t-t_w)}{\ln t_w} \geq 1 \label{c0ttw2}
\end{eqnarray}
whereas for the RFIM magnetization we have found (\ref{mattw1}-\ref{mattw3})
\begin{eqnarray}
&& m(\hat \Gamma, \Gamma_w) =
\ln t_w  \ \psi_1 ( \frac{\ln (t-t_w)}{\ln t_w} )
\qquad \qquad \hbox{for} \qquad \hat{\alpha}=\frac{\hat{\Gamma}}{\Gamma_w} <1 
 \\
&& m(\hat \Gamma, \Gamma_w) 
= \ln t_w  \ \psi_2 ( \frac{t_w}{t} )
\qquad \qquad \hbox{for} \qquad 
\hat{\alpha} =\frac{\hat{\Gamma}}{\Gamma_w} \simeq 1  \\
&&  m(\hat \Gamma, \Gamma_w)= 
\ln t_w  \ \psi_3 ( \frac{\ln t}{\ln t_w} )
\qquad \qquad \hbox{for} \qquad \hat{\alpha}=\frac{\hat{\Gamma}}{\Gamma_w} >1
\end{eqnarray}
These expressions are rather different from the ferromagnet (\ref{cpure}), first because
both $t$ and $\ln t$ appear in these scaling forms. The only domain growth-like 
non trivial scaling regime with $L(t) = T \ln^2 t$  occurs for $\alpha >1$. In this regime
the magnetisation has a form similar to (\ref{mpure}) with a reduction factor $1/\sqrt{L(t_w)}$
(rather than $1/L(t_w)$ in the pure system). Using the discussion of Section \ref{responselarge}
one understands that the origin is quite different from the pure case. In the immediate response to
an applied field (\ref{label}) only a small fraction of domains $1/\Gamma_w \sim 1/\sqrt{L(t_w)}$
respond but their response is {\it very large} since the full domain, of length
$\sim L(t_w)$ flips. The truncated correlation on the other hand is very small and does not
take the scaling form (\ref{cpure}). This yield a value $X=+\infty$ in this regime.

Similarly, the origin of the non trivial value of $X$ in the regime 
$t \sim t_w$ ( $\alpha=1$ ) is very different from the pure case.
Both the response and correlations originiate from rare events
and take the form
$\frac{1}{\ln t_w} L(t_w) f(t/t_w)$ where now the factor
$\frac{1}{\ln t_w}$ is the probability of the rare event and
$f(t/t_w)$ its contribution to activated dynamics. Since they
are of the same order this yield a non trivial $X$.

\section{Persistence properties of the RFIM}

\label{secpersist}

We now turn to a study of the {\it persistence} properties of the random
field Ising model which have received substantial attention for other systems
evolving towards equilibrium.  Two of the primary properties of interest in this context 
are the time decay of the probability that a {\it spin has  never
flipped} up to time $t$ and the time decay of the probability that a
{\it domain has survived} up to time $t$. The results for the single spin  persistence for the RFIM
(sections \ref{singleflip} and \ref{mflip} below)
are also valid for the spin glass. 
The large time limit of these quantities 
can be computed from the asymptotic full state (see fig. \ref{fig3})
on which we focus below. 

\subsection{Persistence of a single spin}

\label{singleflip}

In zero applied field the probability $\Pi(t)$ that a given {\it spin} at $x=0$
has never flipped up to time $t$ in a single run, 
is equal to the probability
that neither the nearest domain wall on one side, nor the nearest (opposite 
type) domain wall on the other side have crossed $x=0$.

In \cite{us_long}, we have found that the probability $\Pi_1(t)$ that a given 
Sinai particle does not cross its starting point up to time $t$ decays as
$\Pi_1(t) \sim
\overline{l}(t)^{-\theta_1}$ with $\theta_1=1/2$.
We thus obtain that $\Pi(t)$ in the zero field RFIM decays as
\begin{eqnarray} \label{theta}
\Pi(t)  \sim
\overline{l}(t)^{-\theta} \qquad \hbox{with} \qquad 
\theta= 2 \theta_1 = 1
\end{eqnarray}
This should be compared with the result \cite{derrida_hakim} for the
pure Ising model, where $\theta=\frac{3}{4}$ (corresponding there to the
characteristic length $l(t) \sim \sqrt{t}$).
This out of equilibrium behavior holds up to time $t=t_{eq}$ corresponding 
to renormalisation scale $\Gamma=\Gamma_J$ at which equilibrium is
reached.

In the presence of an applied field $H>0$, we can use our previous result 
\cite{us_long}
for the biased Sinai diffusion. We have found that the probabilities $\Pi_1^+(t)$ 
(resp. $\Pi_1^-(t)$) that a given 
Sinai particle remains on the right (resp. left) of its starting point up to time $t$,
in the case of a drift in the (+) direction, behave as
\begin{eqnarray}
&& \Pi^{+}_1(t) \approx \frac{2 \delta}{1 - e^{-2 \delta \Gamma}} \\
&& \Pi^{-}_1(t) \approx \frac{2 \delta}{e^{2 \delta \Gamma} - 1} 
\end{eqnarray}
This gives the the probabilities $\Pi^{\pm}(t)$ that a given spin
in the RFIM keeps the value $(\pm)$ up to time $t$, which
in the limit of large $\delta T \ln t \gg 1$ behave as
\begin{eqnarray}
&& \Pi^{+}(t) \approx \frac{H^2}{g^2} \\
&& \Pi^{-}_1(t) \approx  \frac{H^2}{g^2} t^{-2 H T/g}
\end{eqnarray}

\subsection{Persistence of the local time-averaged magnetization}

\label{mflip}

In addition to the persistence of a single spin,
one can also obtain the statistics of the flips
of the thermal average of the local magnetization, i.e $\langle S_x(t) \rangle$, at a given site
$x$. As explained in \cite{us_long} in the context of the Sinai model,
quantities averaged over many runs behave 
very differently than quantities for a single run; in particular while the spin of interest in one run may flip many times, if the same spin is examined in many runs with statistically similar initial conditions, the average over the runs at a given time into the runs will flip far less often.

The local magnetization will successively be equal to $1$ and
$-1$ with only very small probability, at large times, of being
be equal to an intermadiate value. The sequence of flips is given by 
the sequence of changes of orientation of a bond
during the renormalisation procedure
extensively studied in \cite{us_long}.
From that analysis, we obtain the following results for the RFIM.

In zero field $H=0$, we denote by $k$ the {\it number of changes of
sign} of $\langle S_x(t) \rangle$ 
at a given point $x$ between $0$ and $t$.
The distribution of the rescaled variable
\begin{eqnarray}
\kappa=\frac{k}{\ln \Gamma}=\frac{k}{\ln (T \ln t) }
\end{eqnarray}
is characterized by the asymptotic decay
\begin{eqnarray}
\mbox{Prob}(\kappa) \sim  \overline{l}(t)^{-\overline{\theta}(\kappa)}
\end{eqnarray}
where the generalized 
exponent $\overline{\theta}(\kappa)$ is
\begin{eqnarray}
 \label{multi} 
\overline{\theta}(\kappa) = \frac{\kappa}{2} \ln\left[2 \kappa\left(\kappa+
\sqrt{\kappa^2+\frac{5}{4}} \right)\right] + 
\frac{3}{4} - \frac{\kappa}{2}
- \frac{1}{2} \sqrt{\kappa^2+\frac{5}{4}}.
\end{eqnarray}
The exponent $\overline{\theta}(\kappa)$ is a positive convex function :
it decays from $\overline{\theta}(\kappa=0)=\frac{3-\sqrt{5}}{4}$
to $\overline{\theta}(\frac{1}{3})=0$, and then grows again for $\kappa>1/3$.
This implies in particular that
\begin{eqnarray}
\kappa = \frac{k}{\ln (T \ln t)} \rightarrow \frac{1}{3}  
\qquad \hbox{with probability 1 at large time}
\label{g13}
\end{eqnarray}
All of the moments of $\kappa$ will be dominated by the typical behavior;
i.e. $\langle \kappa^m\rangle\equiv 3^{-m}$ for all m. The full dependence on $\kappa$
of the $\overline{\theta}(\kappa)$ function describes the 
{\it tails} of the probability distribution of $\kappa$, 
i.e., the large deviations.
Note also that the probability that $\langle S_x(t) \rangle$
doesn't flip up to time $t$ decays as $\overline{l}(t)^{-\overline{\theta}}$
with exponent
\begin{eqnarray}
\overline{\theta}=\overline{\theta}(\kappa=0)=\frac{3-\sqrt{5}}{4}
\end{eqnarray}
which is significantly smaller than the corresponding decay
exponent $\theta=1$ in (\ref{theta}) for a single spin.

Since the renormalisation procedure has to be stopped at $\Gamma=\Gamma_J$
at which the equilibrium is reached, and that at later times
no more changes occur in the local magnetization we obtain that
the {\it total} number of flips is:
\begin{eqnarray}
k_{tot} \rightarrow \frac{1}{3}  \ln (4 J) 
\end{eqnarray}
the decay of the tails of the probability distribution of $\kappa = k_{tot}/\ln (4 J)$
being described by the same function $\overline{\theta}(\kappa)$ as above
in terms of the length $L_{IM}$.

Another result from \cite{us_long} is the characterisation
of the full sequence of the times
$\Gamma_1=T \ln t_1, ...\Gamma_k=T \ln t_k$
where the local magnetization $\langle S_x(t) \rangle$ flips.
The sequence of scales $\{\Gamma_k\}$ 
is {\it a multiplicative Markovian process} constructed
with the simple rule $\Gamma_{k+1}=\alpha_{k} \Gamma_k$
where $\{\alpha_k\}$ are independent identically 
distributed random variables
of probability distribution $\rho(\alpha)$
\begin{eqnarray}
\rho(\alpha)=  {1 \over \alpha} { 1 \over {\lambda_+ - \lambda_-}} 
\left( \alpha^{-\lambda_-} - \alpha^{-\lambda_+} \right)
\qquad \hbox{with} \qquad \lambda_{\pm}= \frac{3 \pm \sqrt 5}{2}
\end{eqnarray}
As a consequence, $\Gamma_k=\alpha_{k-1} \alpha_{k-2} 
\cdots \alpha_{2} \Gamma_1$ is simply the product of random variables,
so that we obtain using the central limit 
theorem that
\begin{eqnarray}
\lim_{k \to \infty} \left({ {\ln \Gamma_k} \over k} \right)
=\langle \ln \alpha\rangle =3
\end{eqnarray}
and we thus recover that the number $k$
of changes of $\langle S_x(t) \rangle$ grows as $\ln \Gamma = \ln \ln t$ 
and that the rescaled variable $\kappa={k \over {\ln (T \ln t)}}$
is equal to $1/3$ with probability $1$, as in
(\ref{g13}).

These results can be extended to the RFIM in the presence of an applied field $H>0$.
The total number of flips in this case saturates to a finite value given by
a scaling function of $H$ and $J$ identical to the one given in Section IV E
in \cite{us_long} for the total number of returns to the origin in the Sinai model.

\subsection{Domain persistence}

Persistence can also be defined for larger scale patterns. In the 
RFIM the simplest pattern (beyond a single spin) is
a domain. When a domain of e.g. consecutive $+$ spins disappears, the two nearest
domains of $-$ spins merge. Thus domains can either grow
by merging with other domains,
or die, and this leads to interesting survival properties,
that were studied by Krapivsky and Ben Naim 
\cite{krapivsky_benaim} for the pure Ising model. Here we slightly 
change their definitions of the exponents
to adapt to the logarithmic characteristic length scale 
$\overline{l}(t) = {\overline l}_{\Gamma} \sim (T \ln t)^2$ of the RFIM.

\begin{figure}[thb]

\centerline{\fig{8cm}{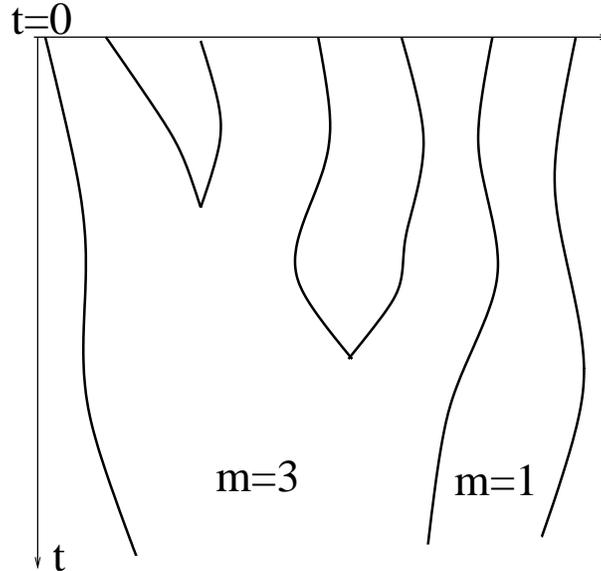}} 
\caption{Temporal evolution of domains in the RFIM. For each 
surviving domain at time $t$, $m$ denotes the number of
ancestor domains in the initial condition at $t=0$ \label{fig50} } 

\end{figure}


Of all the domains which still exist at time $t$,
we define $R_m(t)$ as the number of domains which were formed
out of $m$ of the initial domains. This is illustrated in figure \ref{fig50}.
Then $\sum_m R_m(t)= N(t)$, the
total number of domains at time $t$, and the
fraction of initial domains which have a descendent still ``alive" at $t$
is $S(t) = \sum_m m R_m(t) = <m> N(t)$. Following \cite{krapivsky_benaim}, one defines
the exponents:

\begin{eqnarray}
&& N(t) \sim \overline{l}(t)^{-1}  \\
&& S(t) \sim \overline{l}(t)^{- \psi} \\
&& R_1(t) \sim \overline{l}(t)^{- \delta}
\end{eqnarray}
in terms of the mean domain length $\overline{l}(t)$. Asymptotically,
$R_m(t)$ has the scaling form:

\begin{eqnarray}
R_m(t) \approx \overline{l}(t)^{\psi-2 } \tilde{R}\left(
\frac{m}{\overline{l}(t)^{1-\psi}} \right)
\end{eqnarray}

The scaling function behaves, in the pure case, as
$\tilde{R}(z) \sim z^{\sigma}$ for small $z$ and as an exponential
at large $z$. One finds  the exponent relation $\delta=\nu + (\nu-\psi) (1 + \sigma)$.

Let us now study the RFIM using decimation.
In the asymptotic state (full state), domains coincide with bonds,
and when a bond is decimated the
domain on this bond dies, and the two neighbors merge into a new domain
(the new bond). We thus associate to each bond an auxiliary variable
$m$ counting the number of initial domains which are ancestors
of the domain supported by the bond. Upon decimation of bond (2),
the rule for the variables $m$ is simply

\begin{eqnarray}
m' = m_1 + m_3
\end{eqnarray}

It turns out that this rule coincides with the magnetization rule of 
the RFTIC \cite{danfisher_rg2} and thus
one has the scaling $m \sim \Gamma^{\Phi}$ with $\phi=(1 + \sqrt{5})/2$
leading to 

\begin{eqnarray}
\psi = \frac{3 - \sqrt{5}}{4} = 0.190983..
\end{eqnarray}
This should be compared with the value for the pure Ising model,
which has only been determined numerically \cite{krapivsky_benaim}. 
as $\psi \approx 0.252$.

Here it is interesting to note that we have found that $\psi=\overline{\theta}$,
i.e. the general exact bound $\psi \leq \overline{\theta}$
is saturated \cite{us_rd}.
This bound for $\psi$ comes from the fact
that a point that has never been crossed by any domain wall
up to time $t$ for the effective dynamics, necessarily belongs
to a domain that has a descendant still living up to time $t$.
In the effective dynamics in the "full" renormalized landscape,
we have also the reciprocal property : a domain wall surviving
between $t'$ and $t$ necessarily contains points that have
 never been crossed by any domain wall between $t'$ and $t$.
Note that the equality $\psi=\theta$ also appears
in the coarsening of domains for the 1D $T=0$ Landau-Ginzburg $\phi^4$ soft-spin Ising model
\cite{derrida_coarsening_phi4} for the same reason.
The strict inequality $\psi < \overline{\theta}$
requires the possibility that a domain wall can survive between $t'$ and $t$
even if all points inside it at $t'$ get crossed by a domain wall
between $t'$ and $t$ (see \cite{us_rd} for examples).

We now study the probability $R_1(t)$ that a domain has survived
up to time $t$ without merging with any other domain.
This requires that the two domain walls A and B living at the boundaries
of this domain do not meet any other domain wall up to time $t$. 
In the asymptotic full state, these conditions imply that three
consecutive bonds cannot be decimated. Thus the domain in the middle
cannot grow and the probability that it survives upon decimation 
thus decays exponentially in $\Gamma$.

We thus obtain that $R_1(t)$ decays exponentially in $\Gamma$ 
(and thus with a nonuniversal power of time
determined by the initial condition which determines the convergence
towards the asymptotic state). It thus corresponds to:
\begin{eqnarray}
\delta = + \infty 
\end{eqnarray}
As a consequence, the scaling function $R(z)$, not computed here, 
has an essential singularity at the origin.

\medskip

\section{Finite size properties}

\label{secfinite}

The real-space renormalization procedure can also be used to study 
analytically large finite size systems \cite{danfisher_rg3,us_long}.
We will use extensively the analysis of the Sinai walker with
either absorbing or reflecting boundary counditions \cite{us_long}. 
For the random field Ising model, one may consider 
several boundary conditions :

(i) fixed spins at the ends: this corresponds
for the Glauber dynamics of the RFIM to domain walls which behave as Sinai's walkers 
with reflecting boundary conditions,

(ii) free boundary conditions: this corresponds to absorbing
boundary conditions for the domain walls.

We give here some results for (i), whose derivation is detailed
in Appendix \ref{appfinite} and discuss (ii) at the end.

\subsubsection{Fixed spins at both ends}

Assume for definitness that spins at both extremities are fixed
to the values $S_0=+1$ and $S_L=-1$, where $L$ is the system size.
There is thus an even number of domains
in the system, and one can describe its statistics at large time
using the following generating function. Let us define
$I_L(k; \Gamma)$ as probability that the system of size $L$ at time $t$ 
(i.e at scale $\Gamma=T \ln t$) contains exactly $n =2k+2$ domains,
with $k=0,1,2..$. One obtains (see Appendix \ref{appfinite}) the generating function 
in Laplace transform with respect to the length $L$ as:
\begin{eqnarray} 
\int_0^{\infty} d L e^{-q L}
\left( \sum_{k=0}^{\infty} z^k I_L(k; \Gamma) \right) = 
\frac{\sinh^2(\Gamma \sqrt{ p+\delta^2})}
{ p \cosh^2(\Gamma \sqrt{p+\delta^2})+\delta^2 -z(p+\delta^2)}
\end{eqnarray} 
where $p=\frac{q}{2 g}$.

In the case of zero applied field $H=0$ ($\delta=0$) one easily performs
the Laplace inversion and obtains:
\begin{eqnarray} 
 \sum_{k=0}^{\infty} z^k I_L(k; \Gamma) 
= &&  LT^{-1}_{p \to 2 g L} \left(
\frac{ \sinh^2(\Gamma \sqrt{p})}
{ p (\cosh^2(\Gamma \sqrt{p}) -z)} \right)  
= {\rm tan } (\alpha) \sum_{n=-\infty}^{+\infty} 
\frac{e^{-\frac{ 2 g L}{\Gamma^2}(\alpha+n \pi)^2} }{\alpha+n \pi}
\end{eqnarray}
where $\alpha =ArcCos \sqrt{z} \in (0,\frac{\pi}{2})$ for $z \in (0,1)$.
In particular we obtain the average and mean squared total
number of domains in the system at time $t$:
\begin{eqnarray} 
 <n> = 2 <k>_{L,\Gamma} + 2 && =  4 g \frac{L}{\Gamma^2} + \frac{4}{3} + o(L^0) \\
 <n^2>_{L,\Gamma} - (<n>_{L,\Gamma})^2
&& =  \frac{8 g L}{3 \Gamma^2}  + O(L^0)
\end{eqnarray}
with $\Gamma=T \ln t$.
Again, the same quantities at equilibrium are simply obtained
by setting $\Gamma=4 J$ in the above formulae.

In the presence of an applied field $H>0$ ($\delta>0$)
one obtains similarly:

\begin{eqnarray} 
&& <n> = 2 <k>_{L,\Gamma} + 2 \equiv 
4 g L \frac{\delta^2}{\sinh^2 \gamma} +
2 \frac{1-\gamma \coth \gamma}{\sinh^2 \gamma} + 2
+ o(L^0) \\
&& <n^2>_{L,\Gamma} - (<n>_{L,\Gamma})^2 
= 8 g L \frac{\delta^2}{\sinh^2 \gamma} 
\left(1+\frac{2(1-\gamma \coth \gamma)}{\sinh^2 \gamma} \right)
+ O(L^0) 
\end{eqnarray}
with $\gamma = \delta T \ln t$.

We have also obtained the probability distribution 
$F_{L}(M;\Gamma)$ of the total magnetization of the system:
\begin{eqnarray}
M_L = \sum_{j=1}^{L} <S_j> =  \sum_{i=1}^{i=2k+2} (-1)^{i+1} l_i
\end{eqnarray}
In Laplace transform with respect to $L$ and $M$ it is:
\begin{eqnarray}
\int_0^{+\infty} dL e^{-q L} 
\int_{-L}^{+L} dM e^{- r M}  F_{L}(M;\Gamma)
= \overline{l}_\Gamma  \frac{E_\Gamma^+(q+r) E_\Gamma^-(q-r)}
{1 - P_\Gamma^+(q+r) P_\Gamma^-(q-r)}
\end{eqnarray}
where:
\begin{eqnarray}
&& E_\Gamma^\pm(q) = \frac{\delta e^{\mp \delta \Gamma}}{
\sinh(\delta \Gamma) ( \sqrt{p + \delta^2} \coth(\Gamma \sqrt{p + \delta^2})
\mp \delta)} \\
&& P_\Gamma^\pm(q) = \frac{\sqrt{p + \delta^2} e^{\mp \delta \Gamma}}{
\sinh(\sqrt{p + \delta^2} \Gamma) ( \sqrt{p + \delta^2} \coth(\Gamma \sqrt{p + \delta^2})
\mp \delta)}
\end{eqnarray}
where $p=\frac{q}{2 g}$. Note that $q+r$ and $q-r$ are simply the Laplace variables associated
to the total rescaled positive length and negative length.

\subsubsection{Free boundary conditions}

Free boundary conditions in the RFIM correspond to 
absorbing boundary conditions for the diffusing domain walls.
However the study is slightly different from the one carried in \cite{us_long}
because now there are particles (A or B) both at bottom and tops.

\begin{figure}[thb]

\centerline{\fig{4cm}{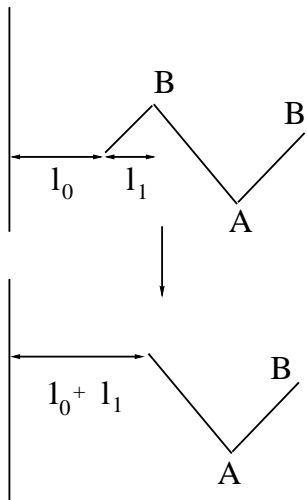}} 
\caption{RG rule near a free boundary as explained in the text.
The same rule holds when the first renormalized bond is descending by
exchanging $A$ and $B$.\label{fig6} } 

\end{figure}

The structure of the renormalized landscape at $\Gamma$ and the full state
near the boundary is now the following (see figure \ref{fig6}). There is an absorbing 
zone of length $l_0$. Then there is a first bond (barrier $F_1$ length $l_1$) 
of arbitrary sign (contrary to Sinai's case where the first bond was always descending).
The first particle at the common endpoint of the first bond and the second
bond and is either a A (descending first bond) or a B (ascending first bond).
The RG procedure is unchanged in the bulk. The new case is when bond $1$
is decimate. Then the absorbing zone becomes of length $l_0 + l_1$,
while the domain wall (A or B) gets out of the system (as the cluster formed
by the absorbing zone and the previously first bond flips). The bond number
$2$ keeps the same barrier and length (and becomes the new bond number $1$).

Let $R^\pm(l)$ be the probability distribution of the length
$l$ of the absorbing zone. It satisfies the RG equation 

\begin{eqnarray}
\partial_\Gamma R_\Gamma^\pm(l) = P_\Gamma^\pm(0,.)*_l R_\Gamma^\mp(.)
- R_\Gamma^\pm(l) \int dl' P_\Gamma^\mp(0,l')
\end{eqnarray}

In the symmetric case the solution in Laplace transform is:

\begin{eqnarray}
R_\Gamma(p) = \frac{2}{\Gamma \sqrt{p}} \tanh(\frac{\Gamma \sqrt{p}}{2})
\end{eqnarray}
Interestingly the shape of the size distribution is the same as
for the Sinai particle with absorbing boundaries but with
a global length rescaling by a factor $1/4$. A similat study can be made
in the biased case.

Let us close by noting that the approach to equilibrium will be 
modified near a free boundary, as compared with the bulk (discussed
in Section ). Indeed near the free boundary it is possible to create
a single domain wall with energy cost 2J. Thus, for times 
such that $2J < \Gamma=T \ln t < 4J$ one must consider different rules
for the first renormalized bond.

\section{Discussion and Future Prospects}

In this paper we have shown how a powerful real space dynamical renormalization group method
can be used to study the properties of the one dimensional random field Ising
model and 1D spin glasses in  a field. We have recovered many known results for the long time {\it equilibrium} behavior and obtained some new ones.  But the main advantages of this method is that it enables the computation and physical understanding of many non-equilibrium features of the coarsening process.  
Although the RSRG method is approximate, it retains all the
information needed to obtain  exact results for universal long time, large distance quantities. As exemplified
by the calculations of the response and truncated correlations, it has the great advantage over many more formal methods by virtue of providing
a clear physical interpretation of the  origins of the processes (e.g. particular types of rare events) that dominate many
properties of interest.

In this last section we address two issues: the potential observability of some of the one-dimensional physics that we have studied in detail; and the crucial issue of which features of these one-dimensional systems might apply in higher dimensional random systems.

\bigskip

\subsection{One-dimensional systems}

Although random field Ising ferromagnets are not directly realizable experimentally, other systems in the same universality class should be. For random antiferromagnets, an applied magnetic field couples to the antiferromagnetic order parameter as a random field, most simply if the local magnetic moments vary randomly; such systems have the obvious advantage of the strength of the dominant randomness being readily adjustable. Similarly, spin glasses, although unfrustrated in zero-field, do exist in quasi-one-dimensional systems.  In both of these types of systems, rapid quenches could be performed by decreasing the field at low temperatures down to a value of order $T\ll J$ at which coarsening can take place. In practice, other types of randomness in, for example, antiferromagnetic or frustrated two-leg spin-ladder systems in an applied field may have advantages.   

There is one physical phenomenon of which one must be wary: in many random systems --- especially spin-glasses \cite{fisher_huse} --- the equilibrium states toward which the system strives depend hypersensitively on macroscopic parameters such as the temperature and the magnetic field; this is often referred to --- somewhat confusingly ---  as ``chaos" \cite{chaos}.  If this were the case here, then changing the temperature or the magnetic field would not correspond simply to speeding up the dynamics, but would instead drive the system towards a new equilibrium which might differ on the scales being probed from the original one. Indeed, this effect is the origin of much of the most interesting aging phenomena seen in three-dimensional spin glasses \cite{fisher_huse,Vihaoc}.
Fortunately --- although less interestingly --- this effect does not occur in random field chains: although the effective local random fields could change in a non-uniform way with temperature or with applied field, this will lead only to smooth modifications of the large scale potential and thus, provided the changes are made sufficiently slowly, will not change the universal aspects of the coarsening provided all lengths are scaled appropriately. 

The most obvious difficulty for any experiment is one that is ubiquitous in random systems: how does one obtain a reasonable range of length scales if the correlations are growing only logarithmically in time?  The situation here is not quite as bad as it might seem as on macroscopic time scales $\xi \sim \ln^2 t$ is not all that short: with a microscopic time, $\tau_0$, of order picoseconds, a minute corresponds to coarsening by a factor of a thousand in length scale.   But for one-dimensional random-field systems, one can do much better: the correlation length is of order
\begin{equation}
\xi(t) \sim a_o \biggl(\frac{T \ln(t/\tau_0)}{h}\biggr)^2
\end{equation}
with $a_0$ a microscopic scale.  Thus the correlation length can be controlled by a combination of increasing time, increasing $T$ and decreasing $h$. To get a wide range of $\xi(t)$, it is beneficial to start with $h$ large enough that the shortest time scales which can be measured still correspond to microscopic lengths and then decrease the field gradually.   

While some of the non-equilibrium properties studied here should be amenable to conventional experimental probes in magnetic systems, some of the most interesting predictions  will probably not be: how the specific set of domain walls depends on equilibration time and how it varies --- or does not vary -- from run-to-run on the same sample.  For these one needs either microscopic imaging probes --- perhaps with magnetic atomic force or tunnelling microscopy --- or scattering data with enough resolution that speckle patterns from a finite spot size can be measured.  As mentioned earlier, the spatial Fourier transform of the two-point two-time correlation calculated here is, for spin glasses, related to the correlations between the speckle patterns at the two different times.  While measuring this with magnetic x-ray or neutron scattering may not be possible, it should be feasible via light scattering on systems in the same universality class in which the length   scales are sufficiently long.  

One  system on which light scattering measurements might be possible is a nematic liquid crystal in a long thin tubes with a square crossection and  heterogeneities on the surfaces which would couple randomly to the two possible symmetry related orientations of the nematic director. Whether this or other systems can be formed in a regime in which the dynamics are sufficiently fast to allow a wide range of length scales to equilibrate is a question whose answer must rely on a quantitative analysis of the physics on the scale of the domain wall structures that should occur.

\bigskip

\subsection{Higher-dimensional systems}

Many of the qualitative features of the coarsening process in RFIM chains are also expected to obtain in higher dimensional random systems.  These are particularly interesting in systems in which, in contrast to the one-dimensional models, true long-range order should exist in equilibrium.  Nevertheless, the characteristics of the coarsening process in the one-dimensional models with weak randomness should have much in common with such systems as long as  consideration is restricted, as we have primarily done, to time-scales such that the full equilibrium correlation length cannot be attained.

Features that are common to many higher dimensional random systems --- random exchange ferrommagnets, random field magnets and, although more controversially, spin glasses ---  include: growth of some kind of domain structure by thermal activation over barriers that grow with length scale and are broadly distributed; and the existence of aging and other history dependent phenomena. 
From a renormalization group point of view, these features are general characteristics of systems controlled by random zero-temperature fixed points  \cite{fisher_huse,dsf-act-dyn}. The notions of local equilibrium within constraints caused by large barriers and of domination of the dynamics at any given long time scale by rare regions of the sample in which the rate for going over barriers is of order the frequency being probed, are both  important. 
In general, this will, as in the one-dimensional models, result in three different regimes for two-time properties depending on whether the difference between the times is much smaller, of order of, or much larger than, the earlier time.

One of the most intriguing questions concerns the pseudo-determinism found in the one-dimensional random models.  Will this exist in some higher dimensional systems in which the domain walls are lines or surfaces whose evolution in time will involve topological changes, rather than points which can simply move or annihilate?  I.e., will domain walls tend to lie, at long times,  in similar sample-and-time-specific positions which are only weakly dependent on the initial conditions? Or will they evolve in different runs in very distinct ways-- perhaps by retaining much more memory of the initial conditions? 

Two examples of a {\it single} random walker diffusing in random potentials illustrate some of the difficulties of drawing any definitive conclusions.  Consider, first, a random potential which is independently random at each site with a long power law tail ot the distribution of the depth of the potential wells.  The deep wells will give rise to subdiffusive behavior, indeed, simple considerations of the time to find a deep trap and the time to escape it imply that the typical distance the walker will diffuse in time $t$ only grows as a power of $\ln t$.   But the statistics of the {\it set of sites} the walker visits by the time it has made a given number of jumps from one site to another will be identical to that of a normal free random walker; it is just the {\it time spent} on each site that causes the slow diffusion.  Since a pair of random walkers in  dimensions higher than four have a non-zero probability of never visiting any sites in common, it is clear that the long time behavior is {\it not} deterministic in high dimensions: two different runs starting from the same point will have a probablity that vanishes at long times of the two walkers being at the same site. In two dimensions, in contrast, the pair of walkers will be very likely to be trapped at the same site at long times.  [The three dimensional case is more subtle but will be more like the high dimensional that the low dimensional case.]  

A second type of random potential yields different behavior: a gaussian random potential with mean zero and $\overline{[V({\bf x})-V({\bf y})]^2} \sim |{\bf x - y}|^{2\zeta}$ with $\zeta>0$.  A pair of random walkers in such a potential will tend to get trapped in the {\it same} deep valley at long times.  This can be argued by considering the borders of valleys which are surrounded by barriers of at least a height $\Gamma$.  If these typically do not  have a lot of fine-scale structure, then they should act as effective traps for all walkers in the vicinity. All the walkers will then tend to eventualy leave  such a valley over the same barrier and hence end up in the same next-larger-scale valley as well. The behavior would then be asymptotically deterministic as in the one-dimensional Sinai model.  

The crucial feature that distinguishes these two cases, seems to be the smoothness of the potential: in the first example, the deep potential wells come from very short wavelength fluctuations and there is thus no useful notion of a coarse-grained potential.  But in the second example, the deep potential wells come, as in one-dimension, from long-wavelength slowly varying components; thus coarse-graining the potential to yield approximate dynamics which are asymptotically exact should be possible.  We should emphasize, however, that it is by no means established even for this simple model of a random walk in a smoothly varying gaussian random potential that the conclusion argued above --- that the dynamics will be asymptotically deterministic --- will be valid in high dimensions.

What happens in three dimensional random magnetic systems in which there are truly many degrees of freedom, we must leave as an important open question.

\section{Acknowledgements}

One of us, DSF, thanks the National Science Foundation for support via DMR-9976621 and Harvard's Materials Research Science and Engineering Center.

\appendix

\section{Convergence towards the full state}

\label{cvfull}

In this Appendix we analyze the rapid convergence towards the {\it full
state} that was  discussed in Section (\ref{secfull}). 

We consider, for simplicity, random initial configurations of the spins,
in which there is an independent probability $w$
that each intersite position is occupied by a domain wall,
and a probability $1-w$ that it is not.
Such random initial configurations describe, for example, initial equilibrium before a quench from
a temperature $T_0$ which is high enough that  
the random fiels are negligible  and the system behaves like a pure chain. This corresponds to $T_0=1/\beta_0 \sim J \gg \{h_n\}$
and  the choice $w=\frac{e^{-\beta_0 2J}}{1+e^{-\beta_0 2J}}$.
In particular, initial conditions where all spins are independent
and take value $\pm 1$ with probability $1/2$ corresponds to 
infinite temperature, $\beta_0=0$, and $w=1/2$.

For these type of initial configurations, the probability that there exist $n=0, \cdots l$
domain walls among $l$ consecutive lattice spacings is simply
given by the binomial distribution
\begin{eqnarray}
R(n \vert l)= \frac{l!}{n!(l-n)!} w^n (1-w)^{l-n} .
 \end{eqnarray}

In the renormalized landscape, the length $l$ of descending (resp. ascending)
bonds at scale $\Gamma$ is distributed with 
 $P_{\Gamma}^{+}(l)$ (resp. $P_{\Gamma}^{-}(l)$),
whose Laplace transforms are obtained after integration over $\zeta$
in (\ref{solu-biased}).
The probability that there had been initially
{\it no} domain walls in the interval occupied by a 
descending (resp. ascending) bond of the renormalized
landscape plays an important role; it is
\begin{eqnarray}
r^{\pm}_{zero}(\Gamma)= \sum_{l=1}^{\infty} P^{\pm}_{\Gamma}(l) R(0 \vert l) 
= \sum_{l=1}^{\infty} P^{\pm}_{\Gamma}(l) (1-w)^l
 \end{eqnarray}
The probability that there had been an odd number $n$ is
\begin{eqnarray}
r^{\pm}_{odd}(\Gamma)= \sum_{l=1}^{\infty} P^{\pm}_{\Gamma}(l) 
\sum_{n \ odd}^{l} R(n \vert l) 
= \sum_{l=1}^{\infty} P^{\pm}_{\Gamma}(l) \frac{1-(1-2 w)^l}{2}
 \end{eqnarray}
and the probability that there had been a non-vanishing even number is
\begin{eqnarray}
r^{\pm}_{even}(\Gamma)= 1- r^{\pm}_{odd}(\Gamma)- r^{\pm}_{zero}(\Gamma)
 \end{eqnarray}

To characterize the statistical properties of
the spin configurations on the renormalized landscape,
we focus on one renormalized bond at scale $\Gamma$ --- call it ``2" --- and its local environment which determines whether or not the maximum and minimum at the ends of bond $2$ are occupied by domain walls. We introduce the probabilities 
$V^{desc}_{\Gamma}(\epsilon_2 ; \{ \epsilon_1,\epsilon_3 \} )$
(resp. $V^{asc}_{\Gamma}(\epsilon_2 ; \{ \epsilon_1,\epsilon_3 \} )$),
where $\epsilon_i=\pm$ ,
that a descending (resp. ascending) bond of the renormalized landscape
at scale $\Gamma$ is in phase $\epsilon_2=\pm$,
with its left neighboring bond in  phase $\epsilon_1$
and its right neighboring bond in phase $\epsilon_3$.
The normalisation conditions are
\begin{eqnarray}
&& \sum_{\epsilon_1=\pm} \sum_{\epsilon_2=\pm} \sum_{\epsilon_3=\pm} V^{desc}_{\Gamma}(\epsilon_2 ; \{ \epsilon_1,\epsilon_3 \} ) =1 \\
&& \sum_{\epsilon_1=\pm} \sum_{\epsilon_2=\pm} \sum_{\epsilon_3=\pm} V^{asc}_{\Gamma}(\epsilon_2 ; \{ \epsilon_1,\epsilon_3 \} ) =1  .
\end{eqnarray}

If $\epsilon_2=-$ on a descending bond, then it is not possible to have $\epsilon_1=+$ since this would correspond to a domain wall of type A at
a maxima, and similarly it is not possible to have $\epsilon_3=+$
since this would correspond to a domain wall of type B at
a minima and thus we have immediately that 
\begin{eqnarray}
 V^{desc}_{\Gamma}(- ; \{+- \}) = V^{desc}_{\Gamma}(- ; \{-+ \}) 
=V^{desc}_{\Gamma}(- ; \{++ \})= 0 
 \end{eqnarray}
Similarly we have
\begin{eqnarray}
  V^{asc}_{\Gamma}(+ ; \{+- \}) = V^{asc}_{\Gamma}(- ; \{-+ \}) 
=V^{asc}_{\Gamma}(- ; \{-- \})= 0
 \end{eqnarray}

If, however, $\epsilon_2=+$ on a descending bond (resp. $\epsilon_2=-$ on a ascending bond), then all four possible neighborhoods of this bond can occur. 
Since it is a bit lengthy to give the full enumeration of all possible cases
with their corresponding probabilities, we give here only the final results

\begin{eqnarray}
&&  V^{desc}_{\Gamma}(- ; \{-- \}) = \frac{r^+_{zero}(\Gamma)}{2} \\
&&  V^{asc}_{\Gamma}(+ ; \{++ \}) = \frac{r^-_{zero}(\Gamma)}{2}
\end{eqnarray}
and
\begin{eqnarray}
&& V^{desc}_{\Gamma}(+ ; \{-- \}) = 
 (1-r^-_{zero}(\Gamma)) \left(1
- \frac{r^+_{zero}(\Gamma)}{2} 
- \frac{r^+_{zero}(\Gamma)r^-_{zero}(\Gamma)}{2} \right)
+\frac{(r^-_{zero}(\Gamma))^2}{2} r^+_{even}(\Gamma)  \\
&& V^{desc}_{\Gamma}(+ ; \{+- \}) = V^{desc}_{\Gamma}(+ ; \{-+ \}) 
= \frac{r^-_{zero}(\Gamma)}{2}
- \frac{(r^-_{zero}(\Gamma))^2}{2} ( r^+_{zero}(\Gamma)+r^+_{even}(\Gamma) )  \\
&& V^{desc}_{\Gamma}(+ ; \{++ \}) = 
\frac{(r^-_{zero}(\Gamma))^2}{2} (r^+_{zero}(\Gamma)+r^+_{even}(\Gamma)) \end{eqnarray}
and similarly for { \it ascending } bonds by exchanging $\pm \to \mp$.

The important property of these probabilities is that 
apart from $V^{desc}_{\Gamma}(+ ; \{-- \})$ and 
$V^{asc}_{\Gamma}(- ; \{++ \})$, which correspond to locally full configuration of the domain walls, all the allowed $V$ have
$r^{\pm}_{zero}(\Gamma)$ as a factor, i.e. to have a bond (2)
that does not have domain walls at both its extremities requires
that at least one of the three bonds (1,2,3) had exactly zero 
domain walls in the initial configuration.

Using the fixed point solution (\ref{solu-biased}), we find that 
\begin{eqnarray}
r^{\pm}_{zero}(\Gamma)
&& \simeq \int_0^{\infty} dl P_\Gamma^{\pm} (l) (1-w)^{l }
= \hat{P^{\pm}_{\Gamma}} \left( q=  \ln \left(\frac{1}{1-w} \right) \right)  
 \end{eqnarray}
and thus this and the probabilities of non-full bonds  converges exponentially in $\Gamma$ to $0$:
\begin{eqnarray}
r^{\pm}_{zero}(\Gamma)
\sim e^{-  \Gamma (\sqrt{ p+\delta^2} \pm \delta) }
 \end{eqnarray}
with $p= \frac{1}{2 g} \ln \left(\frac{1}{1-w} \right)$

Thus the system converges towards the "full" state
of the renormalized landscape exponentially in $\Gamma$,
with a non universal coefficient depending on the initial
concentration of domain walls through the parameter $w$,
and on the strength of the disorder through $g$.
For the symmetric case of no applied field ($\delta=0$), the fraction of the extrema of the renormalized landscape at time $t$ that are not occupied by a domain wall goes to zero as a {\it power} of time:
\begin{equation}
Prob[{\rm  missing \  domain\  wall}] \sim \frac{1}{t^{\eta_e}}
\end{equation}
with
\begin{equation}
\eta_e = T\sqrt{\frac{\ln2}{2\overline{h^2}}}.
\end{equation}

Note that in the initial state for the landscape,
and in the symmetric case $H=0$,
$h_n$ is positive or negative with probability $1/2$,
and thus after grouping the consecutive $h_n$ of the same sign to give the initial zig-zag landscape, we have for initial distribution of length
\begin{eqnarray}
P_0(l)=\frac{1}{2^l} \qquad \hbox{for} \qquad l = 1, 2, \cdots \infty
 \end{eqnarray}
This corresponds to 
\begin{eqnarray}
r_{zero}(\Gamma=0)= \frac{1-w}{1+w} 
 \end{eqnarray}
As an example, for $w=\frac{1}{2}$, we have $r_{zero}(\Gamma=0)=\frac{1}{3}$,
and $ V^{desc}_{\Gamma=0}(- ; \{-- \})=\frac{1}{6}$.

\section{autocorrelations in the RFIM with an applied field}
\label{autobias}

We have now to solve the RG equations (\ref{rg-auto}) given in the 
text with the initial condition
\begin{eqnarray}
&&  P^{++}_{\Gamma',\Gamma'} (\zeta)
=  \int_0^{\infty} dl { { l P^{+}_{\Gamma'}(\zeta,l) }
\over {\overline{l_{\Gamma'}} }} \\
&&  P^{-+}_{\Gamma',\Gamma'} (\zeta) =0
\end{eqnarray}

Since for large $\Gamma$, $P^{\pm}$ have reach their fixed point 
values (\ref{solu-biased}), we obtain ( using the simplified notations $u^{\pm}_{\Gamma}=u^{\pm}_{\Gamma}(p=0)=U^{\pm}_{\Gamma}(p=0)$ ) 
\begin{eqnarray}
\left(\partial_{\Gamma}-\partial_\zeta \right)
 P^{\pm +}_{\Gamma,\Gamma'} (\zeta)
=-2 u^{\mp}_{\Gamma} P^{\pm +}_{\Gamma,\Gamma'} (\zeta) 
+ 2 u^{\mp }_{\Gamma} u^{\pm}_{\Gamma} \int_0^{\infty} d\zeta' e^{- u^{\pm}_{\Gamma} 
(\zeta-\zeta')} P^{\pm+}_{\Gamma,\Gamma'} (\zeta')
+(u^{\pm}_{\Gamma})^2 \zeta  e^{- u^{\pm}_{\Gamma} \zeta } 
P^{\mp +}_{\Gamma,\Gamma'} (0)
\end{eqnarray}
together with the initial conditions
\begin{eqnarray}
&&  P^{++}_{\Gamma',\Gamma'} (\zeta)
= {1  \over {\overline{l_{\Gamma'}} }} e^{-\zeta u^{+}_{\Gamma'} }
\left(- \partial_p U^{+}_{\Gamma}(p)+\zeta U^{+}_{\Gamma}(p)
\partial_p u^{+}_{\Gamma}(p)  \right) \vert_{p=0}  \\
&&  P^{-+}_{\Gamma',\Gamma'} (\zeta) =0
\end{eqnarray}

The solutions are thus of the following form
\begin{eqnarray}
&& P^{\pm +}_{\Gamma,\Gamma'} (\zeta)
= {1  \over {\overline{l_{\Gamma}} }} e^{-\zeta u^{\pm }_{\Gamma} }
u^{\pm }_{\Gamma} \left(A^{\pm +}_{\Gamma,\Gamma'}+\zeta B^{\pm +}_{\Gamma,\Gamma}\right) 
\end{eqnarray}
where the coefficients satisfy
\begin{eqnarray}
&& B^{\pm +}_{\Gamma,\Gamma'} = \partial_{\Gamma} A^{\pm +}_{\Gamma,\Gamma'} \\
&&\partial_{\Gamma}^2 A^{\pm +}_{\Gamma,\Gamma'}
=u^{+}_{\Gamma} u^{-}_{\Gamma} \left(  A^{++}_{\Gamma,\Gamma'} 
+ A^{-+}_{\Gamma,\Gamma'} \right)
\end{eqnarray}
Introducing the sum $\Sigma^{+}_{\Gamma,\Gamma'}= A^{++}_{\Gamma,\Gamma'} 
+ A^{-+}_{\Gamma,\Gamma'} $ and the difference
$\Delta^{+}_{\Gamma,\Gamma'}= A^{++}_{\Gamma,\Gamma'} 
- A^{-+}_{\Gamma,\Gamma'} $, we obtain the decoupled equations
\begin{eqnarray}
&&\partial_{\Gamma}^2 \Delta^{+}_{\Gamma,\Gamma'} =0 \\
&& \partial_{\Gamma}^2 \Sigma^{+}_{\Gamma,\Gamma'}
= 2 u^{+}_{\Gamma} u^{-}_{\Gamma} \Sigma^{+}_{\Gamma,\Gamma'}
\end{eqnarray}
together with the boundary conditions
\begin{eqnarray}
&& \Delta^{+}_{\Gamma',\Gamma'}= \Sigma^{+}_{\Gamma',\Gamma'}
=A^{++}_{\Gamma',\Gamma'}= {1 \over {2 \delta^2}} \left(
\gamma' \coth \gamma'-1 \right) \\
&& \partial_{\Gamma} \Delta^{+}_{\Gamma,\Gamma'} \vert_{\Gamma=\Gamma'}= \partial_{\Gamma} \Sigma^{+}_{\Gamma,\Gamma'} \vert_{\Gamma=\Gamma'}
=B^{++}_{\Gamma',\Gamma'}= {1 \over {2 \delta}} \left(
 \coth \gamma'-{ {\gamma'} \over {\sinh^2\gamma'}} \right)
\end{eqnarray}
where $\gamma'=\delta \Gamma'$ as usual.
In terms of the function
\begin{eqnarray}
\rho(\gamma)=  \gamma \coth \gamma-1    
\end{eqnarray}
and its derivative with respect to $\gamma$:
\begin{eqnarray}
 {\cal M}(\gamma)=  \coth \gamma - { {\gamma} \over {\sinh^2\gamma}}    
\end{eqnarray}
we finally obtain
\begin{eqnarray}
&& A^{\pm +}_{\Gamma,\Gamma'}= {1 \over {4 \delta^2}}
\left(\rho(\gamma) \pm \gamma {\cal M}(\gamma')
\pm \rho(\gamma') \mp \gamma' {\cal M}(\gamma') \right) \\ 
&& B^{ \pm +}_{\Gamma,\Gamma'}= {1 \over {4 \delta}}
\left({\cal M}(\gamma) \pm {\cal M}(\gamma') \right)
\end{eqnarray}
In the same way we obtain the solutions for the minus initial condition
\begin{eqnarray}
&& P^{\pm -}_{\Gamma,\Gamma'} (\zeta)
= {1  \over {\overline{l_{\Gamma}} }} e^{-\zeta u^{\pm}_{\Gamma} }
u^{\pm}_{\Gamma} \left(A^{\pm -}_{\Gamma,\Gamma'}+\zeta B^{ \pm-}_{\Gamma,\Gamma}\right)  
\end{eqnarray}
with $ A^{\pm -}_{\Gamma,\Gamma'}= A^{\mp +}_{\Gamma,\Gamma'} $ and
$B^{\pm -}_{\Gamma,\Gamma'}= B^{\mp +}_{\Gamma,\Gamma'} $.
We may check the normalizations
\begin{eqnarray}
 \int_0^{\infty} d\zeta 
\left( P^{+\pm}_{\Gamma,\Gamma'} (\zeta)+   P^{-\pm}_{\Gamma,\Gamma'} (\zeta)\right) 
= {1 \over 2} \left(1 \pm {\cal M}(\gamma') \right)
\end{eqnarray}
The autocorrelation is
\begin{eqnarray}
&&  \overline{<S_i(t') S_i(t)>}=\int_0^{\infty} d\zeta 
\left( P^{++}_{\Gamma,\Gamma'} (\zeta)+   P^{--}_{\Gamma,\Gamma'} (\zeta)- 
P^{+-}_{\Gamma,\Gamma'} (\zeta)-P^{-+}_{\Gamma,\Gamma'} (\zeta) \right) \\
&&= {1  \over {\overline{l_{\Gamma}} }} \left( 2\left( A^{++}_{\Gamma,\Gamma'} -A^{-+}_{\Gamma,\Gamma'} \right)  + 
\left({1 \over { u^{+}_{\Gamma}}} +{1 \over { u^{-}_{\Gamma}}} \right)
\left( B^{++}_{\Gamma,\Gamma'} -B^{-+}_{\Gamma,\Gamma'} \right) \right) \\
&& = (\coth \gamma) {\cal M}(\gamma') +{ 1 \over {\sinh^2\gamma} }
\left(\gamma {\cal M}(\gamma')+ \rho(\gamma')
-\gamma' {\cal M}(\gamma') \right)
\end{eqnarray}
Note that $\overline{<S_i(t')>}= {\cal M}(\gamma')$
and $\overline{<S_i(t)>}= {\cal M}(\gamma)
=\coth \gamma-{ {\gamma} \over {\sinh^2\gamma}}$.

\section{Two-point two-time correlations 
$\overline{ \langle S_0(t) S_x(t) \rangle \langle S_0(t') S_x(t') \rangle  }$}

\label{appssss}

\subsection{Definitions}

Two points at $0$ and $x>0$ are given.  To compute the two spin two time correlation
we introduce the following quantities. Let 
 $\Omega^{2n,+}_{\Gamma, \Gamma'}(\zeta_0,l_0;\zeta_1,l_1,...;\zeta_{2 n}, l_{2 n};x)$,
$n=0,1,2..$,
(and respectively 
$\Omega^{2n,-}_{\Gamma, \Gamma'}(\zeta_0,l_0;\zeta_1,l_1,...;\zeta_{2 n}, l_{2 n} ;x)$)
be the probability that at $\Gamma'$, $x$ and $0$ are on bonds of same
orientation (respectively different orientation) 
AND that at $\Gamma$, $x$ and $0$ are on bonds of same
orientation, with a configuration $(\zeta_0,l_0;\zeta_1,l_1,...;\zeta_{2 n}, l_{2 n})$
(see figure).
One defines similarly 
$\Omega^{2n+1,+}_{\Gamma, \Gamma'}(\zeta_0,l_0;\zeta_1,l_1,...;\zeta_{2 n+1}, l_{2 n+1};x)$,
$n=0,1,2..$,(respectively 
$\Omega^{2n+1,-}_{\Gamma, \Gamma'}(\zeta_0,l_0;\zeta_1,l_1,...;\zeta_{2 n+1}, l_{2 n+1};x)$)
be the probability that at $\Gamma'$, $x$ and $0$ are on bonds of same
orientation (respectively different orientation) 
AND that at $\Gamma$, $x$ and $0$ are on bonds of different
orientation, with a configuration $(\zeta_0,l_0;\zeta_1,l_1,...;\zeta_{2 n+1}, l_{2 n+1})$.

Initial condition at $\Gamma=\Gamma'$ (for all $n \ge 0$):

\begin{eqnarray}
&& \Omega^{2n,+}_{\Gamma', \Gamma'}(\zeta_0,l_0;\zeta_1,l_1,...;\zeta_{2 n}, l_{2 n};x) \\
&&  =
P_{\Gamma'}(\zeta_0,l_0) P_{\Gamma'}(\zeta_1,l_1) \ldots P_{\Gamma'}(\zeta_{2 n}, l_{2 n})
 W_{\Gamma'}( l_0, l_{2n} , \sum_{i=0}^{2n} l_i -x ) \\
&& \Omega^{2n+1,-}_{\Gamma', \Gamma'}(\zeta_0,l_0;\zeta_1,l_1,...;\zeta_{2 n+1}, l_{2 n+1};x) \\ &&= 
P_{\Gamma'}(\zeta_0,l_0) P_{\Gamma'}(\zeta_1,l_1) \ldots P_{\Gamma'}(\zeta_{2 n+1}, l_{2 n+1}) 
W_{\Gamma'}( l_0, l_{2n+1}, \sum_{i=0}^{2n+1} l_i -x )
\end{eqnarray}
with the notation
\begin{eqnarray} \label{defW}
&& W_{\Gamma}( l_1, l_2 , L)= 
 \frac{2}{\Gamma^2} \int_0^{l_1} dy_1 \int_0^{l_2} dy_2 
\delta(L- y_1 - y_2) \\
&& = \frac{2}{\Gamma^2} \left[ min(l_1,L)-max(0,L-l_2) \right]
\theta \left[ min(l_1,L)-max(0,L-l_2) \right]
\end{eqnarray}

We have of course $\Omega^{2n,-}_{\Gamma', \Gamma'}=\Omega^{2n+1,+}_{\Gamma', \Gamma'}=0$.

In the end we are interested in 
the probabilities $P^{\epsilon,\epsilon'}_{\Gamma, \Gamma'}(x)$
that $\langle S_0(t') S_x(t') \rangle =\epsilon'$ and
$\langle S_0(t) S_x(t) \rangle =\epsilon$.
We have the normalisation
\begin{eqnarray} 
P^{+,+}_{\Gamma, \Gamma'}(x)+P^{-,-}_{\Gamma, \Gamma'}(x)
+P^{-,+}_{\Gamma, \Gamma'}(x) +P^{+,-}_{\Gamma, \Gamma'}(x)=1
\end{eqnarray}
The correlation function is
\begin{eqnarray} 
\overline{ \langle S_0(t) S_x(t) \rangle \langle S_0(t') S_x(t') \rangle  } = P^{+,+}_{\Gamma, \Gamma'}(x)+P^{-,-}_{\Gamma, \Gamma'}(x)
-P^{-,+}_{\Gamma, \Gamma'}(x) -P^{+,-}_{\Gamma, \Gamma'}(x)
\end{eqnarray}
In terms of the functions $\Omega$ we have
\begin{eqnarray} 
&& P^{+,+}_{\Gamma, \Gamma'}(x)
= \sum_{n=0}^{\infty} \int_{\zeta_0,l_0, \ldots \zeta_{2n}, l_{2n}}
\Omega^{2n,+}_{\Gamma, \Gamma'}(\zeta_0,l_0;\zeta_1,l_1,...;\zeta_{2 n}, l_{2 n};x) \\
&& P^{+,-}_{\Gamma, \Gamma'}(x)
= \sum_{n=0}^{\infty} \int_{\zeta_0,l_0, \ldots \zeta_{2n}, l_{2n}}
\Omega^{2n,-}_{\Gamma, \Gamma'}(\zeta_0,l_0;\zeta_1,l_1,...;\zeta_{2 n}, l_{2 n};x) \\
&& P^{-,+}_{\Gamma, \Gamma'}(x)
= \sum_{n=0}^{\infty} \int_{\zeta_0,l_0, \ldots \zeta_{2n+1}, l_{2n+1}}
\Omega^{2n+1,+}_{\Gamma, \Gamma'}(\zeta_0,l_0;\zeta_1,l_1,...;\zeta_{2 n+1}, l_{2 n+1};x) \\
&& P^{-,-}_{\Gamma, \Gamma'}(x)
= \sum_{n=0}^{\infty} \int_{\zeta_0,l_0, \ldots \zeta_{2n+1}, l_{2n+1}}
\Omega^{2n+1,-}_{\Gamma, \Gamma'}(\zeta_0,l_0;\zeta_1,l_1,...;\zeta_{2 n+1}, l_{2 n+1};x)
\end{eqnarray}

\subsection{RG equations}

The RG equations for $\Omega^{m,\epsilon'}_{\Gamma,\Gamma'}$ 
for $\epsilon'=\pm$ and $m =0,1,2, \ldots$ read :

\begin{eqnarray} 
&& \left(\partial_\Gamma - \sum_{k=0}^{m} \partial_{\zeta_k} \right) 
\Omega^{m,\epsilon'}_{\Gamma,\Gamma'}(\zeta_0,l_0;\zeta_1,l_1;
\ldots\zeta_{m},l_{m};x) = 
\nonumber \\
&& - 2 P_\Gamma(\zeta=0) 
\Omega^{m,\epsilon'}_{\Gamma,\Gamma'}(\zeta_0,l_0;\zeta_1,l_1;
\ldots\zeta_{m},l_{m};x)  \\
&& +\sum_{k=0}^{m} 
\int_{z,l+l'+l''=l_k}
\Omega^{m+2,\epsilon'}_{\Gamma,\Gamma'}(\zeta_0,l_0;\ldots;\zeta_{k-1},l_{k-1};
z,l;0,l';\zeta_k-z,l'';\ldots;\zeta_{m},l_{m};x) \\
&& + \int_{z,l+l_0'+l_1'=l_0}
P_\Gamma(\zeta_0-z,l) 
\Omega^{m+1,\epsilon'}_{\Gamma,\Gamma'}(0,l_0';z,l_1';\zeta_1,l_1;
\ldots;\zeta_{m},l_{m};x) \\
&& 
+ \int_{z,l+l_{m}'+l_{m+1}'=l_{m}}
\Omega^{m+1,\epsilon'}_{\Gamma,\Gamma'}(\zeta_0,l_0;\ldots;\zeta_{m-1}, l_{m-1};
z,l_{m}';0,l_{m+1}';x) P_\Gamma(\zeta_{m}-z,l) \\
&& + \int_{z,l+l'+l_0'=l_0}
P_\Gamma(z,l)
P_\Gamma(0,l')
\Omega^{m,\epsilon'}_{\Gamma,\Gamma'}(\zeta_0-z,l_0';\zeta_1,l_1;\ldots;
\zeta_{m},l_{m};x) \\
&&
+ \int_{z,l_{m}'+l'+l=l_{m}}
\Omega^{m,\epsilon'}_{\Gamma,\Gamma'}(\zeta_0,l_0;\ldots;
\zeta_{m-1},l_{m-1};z,l_{m}';x)
P_\Gamma(0,l')
P_\Gamma(\zeta_{p}-z,l)
\end{eqnarray}

In particular, for $m=0$ we have

\begin{eqnarray} 
&& \left(\partial_\Gamma - \partial_{\zeta_0} \right) 
\Omega^{0,\epsilon'}_{\Gamma,\Gamma'}(\zeta_0,l_0;x) = 
\nonumber \\
&& - 2 P_\Gamma(\zeta=0) 
\Omega^{0,\epsilon'}_{\Gamma,\Gamma'}(\zeta_0,l_0;x)  \\
&& + \int_{z,l+l'+l''=l_0}
\Omega^{2,\epsilon'}_{\Gamma,\Gamma'}(z,l;0,l';\zeta_0-z,l'';x) \\
&& + \int_{z,l+l_0'+l_1'=l_0}
P_\Gamma(\zeta_0-z,l) 
\Omega^{1,\epsilon'}_{\Gamma,\Gamma'}(0,l_0';z,l_1';x) \\
&& 
+ \int_{z,l+l''+l'=l_0}
\Omega^{1,\epsilon'}_{\Gamma,\Gamma'}(z,l'';0,l';x) P_\Gamma(\zeta_{0}-z,l) \\
&& + \int_{z,l+l'+l_0'=l_0}
P_\Gamma(z,l)
P_\Gamma(0,l')
\Omega^{0,\epsilon'}_{\Gamma,\Gamma'}(\zeta_0-z,l_0';x) \\
&&
+ \int_{z,l_{0}'+l'+l=l_{0}}
\Omega^{0,\epsilon'}_{\Gamma,\Gamma'}(z,l_0';x)
P_\Gamma(0,l')
P_\Gamma(\zeta_{0}-z,l) \\
 &&
+ \int_{z,l_{0}'+l'+l=l_{0}}
\Omega^{0,\epsilon'}_{\Gamma,\Gamma'}(0,l_0';x)
P_\Gamma(z,l')
P_\Gamma(\zeta_{0}-z,l)
\end{eqnarray}

\subsection{Form of solutions}

For $n \ge 1$ and $\epsilon'=\pm$, we set
\begin{eqnarray} 
&& 
\Omega^{2n,\epsilon'}_{\Gamma,\Gamma'}(\zeta_0,l_0;\ldots ;\zeta_{2 n},l_{2 n};x) = E^{+,\epsilon'}_{\Gamma,\Gamma'}(\zeta_0,l_0;\zeta_{2 n},l_{2 n}; 
\sum_{i=0}^{2n} l_i -x)
P_{\Gamma}(\zeta_1,l_1) \ldots P_{\Gamma}(\zeta_{2n-1},l_{2n-1})
 \\
&& 
\Omega^{2n+1,\epsilon'}_{\Gamma,\Gamma'}(\zeta_0,l_0;\ldots ;\zeta_{2 n+1},l_{2 n+1};x) = E^{-,\epsilon'}_{\Gamma,\Gamma'}(\zeta_0,l_0;\zeta_{2 n+1},l_{2 n+1}
; \sum_{i=0}^{2n+1} l_i -x)
P_{\Gamma}(\zeta_1,l_1) \ldots P_{\Gamma}(\zeta_{2n},l_{2n})
 \end{eqnarray}
and also
\begin{eqnarray} 
\Omega^{1,\epsilon'}_{\Gamma,\Gamma'}(\zeta_0,l_0;\zeta_{1},l_{1};x) = E^{-,\epsilon'}_{\Gamma,\Gamma'}(\zeta_0,l_0;\zeta_{1},l_{1}
; l_0+l_1 -x)
 \end{eqnarray}
where the $P_{\Gamma}(\zeta,l)$ satisfy the bond equation,
and where the E satisfy the RG equations 

\begin{eqnarray} 
&& (\partial_{\Gamma} -\partial_{\zeta_1} -\partial_{\zeta_2} )
E^{\pm,\epsilon'}_{\Gamma,\Gamma'}(\zeta_1,l_1;\zeta_2,l_2; L)
= -2 P_{\Gamma}(\zeta=0) E^{\pm,\epsilon'}_{\Gamma,\Gamma'}(\zeta_1,l_1;\zeta_2,l_2; L) \\
&& +
 P_{\Gamma}(0,.)*_{l_1}  P_{\Gamma}(.,.) *_{\zeta_1,l_1}
E^{\pm,\epsilon'}_{\Gamma,\Gamma'}(.,.;\zeta_2,l_2; L) \\
&& +   P_{\Gamma}(0,.)*_{l_2}  P_{\Gamma}(.,.) *_{\zeta_2,l_2}
E^{\pm,\epsilon'}_{\Gamma,\Gamma'}(\zeta_1,l_1;.,.; L) \\
&& +\int_{l+l'+l_1'=l_1} E^{\mp,\epsilon'}_{\Gamma,\Gamma'}
(0,l_1';\zeta_2,l_2; L-l) 
P_{\Gamma}(.,l)*_{\zeta_1}  P_{\Gamma}(.,l') \\
&& +\int_{l+l'+l_2'=l_2} E^{\mp,\epsilon'}_{\Gamma,\Gamma'}
(\zeta_1,l_1;0,l_2'; L-l) 
P_{\Gamma}(.,l)*_{\zeta_2}  P_{\Gamma}(.,l') \\
&& +\int_{l+l'+l_1'=l_1} E^{\pm,\epsilon'}_{\Gamma,\Gamma'}
(.,l_1';\zeta_2,l_2; L-l-l') *_{\zeta_1} 
P_{\Gamma}(.,l)  P_{\Gamma}(0,l') \\
&& +\int_{l+l'+l_2'=l_2} E^{\pm,\epsilon'}_{\Gamma,\Gamma'}
(\zeta_1,l_1;.,l_2'; L-l-l') *_{\zeta_2} 
P_{\Gamma}(.,l)  P_{\Gamma}(0,l')
 \end{eqnarray}

with the initial conditions

\begin{eqnarray} 
E^{\epsilon,\epsilon'}_{\Gamma',\Gamma'}(\zeta_1,l_1;\zeta_2,l_2; L)
= \delta_{\epsilon,\epsilon'}
P_{\Gamma'}(\zeta_1,l_1 )  P_{\Gamma'}(\zeta_2,l_2 )
W_{\Gamma'}( l_1, l_2 , L )
\end{eqnarray}

The RG equation for $\Omega^{0,\epsilon'}_{\Gamma,\Gamma'}$ 
becomes

\begin{eqnarray} 
&& \left(\partial_\Gamma - \partial_{\zeta_0} \right) 
\Omega^{0,\epsilon'}_{\Gamma,\Gamma'}(\zeta_0,l_0;x) = 
- 2 P_\Gamma(\zeta=0) 
\Omega^{0,\epsilon'}_{\Gamma,\Gamma'}(\zeta_0,l_0;x) \\
&& + \int_{z,l+l'+l''=l_0}
E^{+,\epsilon'}_{\Gamma,\Gamma'}(z,l;\zeta_0-z,l''; l_0-x) P_{\Gamma}(0,l')\\
&& + \int_{l+l_0'+l_1'=l_0}
P_\Gamma(.,l) *_{\zeta_0}
\Omega^{1,\epsilon'}_{\Gamma,\Gamma'}(0,l_0';.,l_1';x) \\
&& 
+ \int_{l+l_0'+l_1'=l_0}
\Omega^{1,\epsilon'}_{\Gamma,\Gamma'}(.,l_0';0,l_1';x)*_{\zeta_0} P_\Gamma(.,l) \\
&& + 2 
P_\Gamma(0,. ) *_{l_0} P_\Gamma(.,.) *_{\zeta_0,l_0}
\Omega^{0,\epsilon'}_{\Gamma,\Gamma'}(.,.;x) \\
&& +  
P_\Gamma(.,. ) *_{\zeta_0,l_0} P_\Gamma(.,.) *_{l_0}
\Omega^{0,\epsilon'}_{\Gamma,\Gamma'}(0,.;x)
\end{eqnarray}

with the initial condition

\begin{eqnarray}
&& \Omega^{0,\epsilon'}_{\Gamma', \Gamma'}(\zeta_0,l_0;x)
=  \delta_{\epsilon',+1}
P_{\Gamma'}(\zeta_0,l_0) W_{\Gamma'}( l_0, l_{0} , l_0 -x )
\end{eqnarray}

Using the form of the solutions, we thus obtain

\begin{eqnarray} 
&& P^{+,+}_{\Gamma, \Gamma'}(x)
=  \int_{\zeta_0,l_0} \Omega^{0,+}_{\Gamma, \Gamma'}(\zeta_0,l_0;x) \\
&& +\sum_{n=1}^{\infty} \int_{\zeta_0, \zeta_{2n}}
\int_{L,l_0,l_1 \ldots l_{2n}}
E^{++}_{\Gamma, \Gamma'}(\zeta_0,l_0;\zeta_{2n},l_{2n}; L)
\delta( L-(\sum_{i=0}^{2n} l_i-x)) P_{\Gamma}(l_1) \ldots P_{\Gamma} (l_{2n-1}) \\
&& P^{+,-}_{\Gamma, \Gamma'}(x)
= \int_{\zeta_0,l_0} \Omega^{0,-}_{\Gamma, \Gamma'}(\zeta_0,l_0;x) \\
&& +\sum_{n=1}^{\infty} \int_{\zeta_0, \zeta_{2n}}
\int_{L,l_0,l_1 \ldots l_{2n}}
E^{+-}_{\Gamma, \Gamma'}(\zeta_0,l_0;\zeta_{2n},l_{2n}; L)
\delta( L-(\sum_{i=0}^{2n} l_i-x)) P_{\Gamma}(l_1) \ldots P_{\Gamma} (l_{2n-1})
 \\
&& P^{-,+}_{\Gamma, \Gamma'}(x)
=  \\
&& \sum_{n=0}^{\infty} \int_{\zeta_0, \zeta_{2n+1}}
\int_{L,l_0, l_1 , l_{2n+1}}
E^{-,+}_{\Gamma, \Gamma'}(\zeta_0,l_0;\zeta_{2 n+1}, l_{2 n+1};L)
\delta( L-(\sum_{i=0}^{2n+1} l_i-x)) P_{\Gamma}(l_1) \ldots P_{\Gamma} (l_{2n})
\\
&& P^{-,-}_{\Gamma, \Gamma'}(x)
= \\
&& \sum_{n=0}^{\infty} \int_{\zeta_0, \zeta_{2n+1}}
\int_{L,l_0, l_1 , l_{2n+1}}
E^{-,-}_{\Gamma, \Gamma'}(\zeta_0,l_0;\zeta_{2 n+1}, l_{2 n+1};L)
\delta( L-(\sum_{i=0}^{2n+1} l_i-x)) P_{\Gamma}(l_1) \ldots P_{\Gamma} (l_{2n})
\label{relations}
\end{eqnarray}

\subsection{Laplace transforms}

It is convenient to introduce the Laplace transforms
\begin{eqnarray} 
 E^{\epsilon,\epsilon'}_{\Gamma,\Gamma'}(\zeta_1,p_1;\zeta_2,p_2;p)
=\int_0^{\infty} dl_1 \int_0^{\infty} dl_2 \int_0^{\infty} dL
e^{-p_1 l_1 -p_2 l_2-p L} 
E_{\Gamma,\Gamma'}(\zeta_1,l_1;\zeta_2,l_2; L)
\end{eqnarray}
Using the fixed point solution
\begin{eqnarray} 
P_{\Gamma}(\zeta,p)=U_{\Gamma}(p) e^{-\zeta u_{\Gamma}(p)} 
\end{eqnarray}
the RG equations for $E$ become

\begin{eqnarray} 
&& (\partial_{\Gamma} -\partial_{\zeta_1} -\partial_{\zeta_2} )
E^{\pm,\epsilon'}_{\Gamma,\Gamma'}(\zeta_1,p_1;\zeta_2,p_2;p)
= -2 U_{\Gamma}(0) E^{\pm,\epsilon'}_{\Gamma,\Gamma'}(\zeta_1,p_1;\zeta_2,p_2;p) \\
&& + U_{\Gamma}^2(p_1) \int_0^{\zeta_1} dz  e^{-(\zeta_1-z) u_{\Gamma}(p_1)} 
E^{\pm,\epsilon'}_{\Gamma,\Gamma'}(z,p_1;\zeta_2,p_2;p) \\
&& +  U_{\Gamma}^2(p_2) \int_0^{\zeta_2} dz  e^{-(\zeta_2-z) u_{\Gamma}(p_2)} 
E^{\pm,\epsilon'}_{\Gamma,\Gamma'}(\zeta_1,p_1;z,p_2; p) \\
&& + U_{\Gamma}(p_1) U_{\Gamma}(p_1+p)
\frac{ e^{-\zeta_1 u_{\Gamma}(p_1+p)} - e^{-\zeta_1 u_{\Gamma}(p_1)} }{u_{\Gamma}(p_1) - u_{\Gamma}(p_1+p)}
E^{\mp,\epsilon'}_{\Gamma,\Gamma'}(0,p_1;\zeta_2,p_2; p) \\
&& + U_{\Gamma}(p_2) U_{\Gamma}(p_2+p)
\frac{ e^{-\zeta_2 u_{\Gamma}(p_2+p)} - e^{-\zeta_2 u_{\Gamma}(p_2)} }{u_{\Gamma}(p_2) - u_{\Gamma}(p_2+p)}
E^{\mp,\epsilon'}_{\Gamma,\Gamma'}(\zeta_1,p_1;0,p_2; p) \\
&& + U_{\Gamma}^2(p_1+p) \int_0^{\zeta_1} dz  e^{-(\zeta_1-z) u_{\Gamma}(p_1+p)} E^{\pm,\epsilon'}_{\Gamma,\Gamma'}(z,p_1;\zeta_2,p_2;p) \\
&& +  U_{\Gamma}^2(p_2+p) \int_0^{\zeta_2} dz  e^{-(\zeta_2-z) u_{\Gamma}(p_2+p)} 
E^{\pm,\epsilon'}_{\Gamma,\Gamma'}(\zeta_1,p_1;z,p_2; p) 
\end{eqnarray}

with the initial conditions

\begin{eqnarray} 
 E^{\epsilon,\epsilon'}_{\Gamma',\Gamma'}(\zeta_1,p_1;\zeta_2,p_2; p)
= \delta_{\epsilon,\epsilon'}
\frac{2}{\Gamma'^2 p^2}
&& \left[ U_{\Gamma'}(p_1)  e^{-\zeta_1 u_{\Gamma'}(p_1) }
- U_{\Gamma'}(p_1+p)  e^{-\zeta_1 u_{\Gamma'}(p_1+p) }\right] \\
&& \left[ U_{\Gamma'}(p_2)  e^{-\zeta_2 u_{\Gamma'}(p_2) }
- U_{\Gamma'}(p_2+p)  e^{-\zeta_2 u_{\Gamma'}(p_2+p) }\right]
\end{eqnarray}

Also defining
\begin{eqnarray}
&& \Omega^{2n=0,\epsilon'}_{\Gamma', \Gamma'}(\zeta_0,p_0;p)
= \int_0^{\infty} dl_0 e^{-p_0 l_0}
 \int_0^{\infty} dx e^{-p x}
 \Omega^{2n=0,\epsilon'}_{\Gamma', \Gamma'}(\zeta_0,l_0;x)  
\end{eqnarray}
the RG equation becomes

\begin{eqnarray} 
&& \left(\partial_\Gamma - \partial_{\zeta_0} \right) 
\Omega^{0,\epsilon'}_{\Gamma,\Gamma'}(\zeta_0,p_0;p) = 
- 2 u_\Gamma 
\Omega^{0,\epsilon'}_{\Gamma,\Gamma'}(\zeta_0,p_0;p) \\
&& + \int_{z}
E^{+,\epsilon'}_{\Gamma,\Gamma'}(z,p_0+p;\zeta_0-z,p_0+p;-p) P_{\Gamma}(0,p_0+p)\\
&& + 
U_\Gamma(p_0) \int_0^{\zeta_0} dz e^{-(\zeta_0-z) u_\Gamma(p_0) }
\Omega^{1,\epsilon'}_{\Gamma,\Gamma'}(0,p_0;z,p_0;p) \\
&& 
+ U_\Gamma(p_0) \int_0^{\zeta_0} dz e^{-(\zeta_0-z) u_\Gamma(p_0) }
\Omega^{1,\epsilon'}_{\Gamma,\Gamma'}(z,p_0;0,p_0;p) \\
&& + 2 
U^2_\Gamma(p_0 ) \int_0^{\zeta_0} dz e^{-(\zeta_0-z) u_\Gamma(p_0) } 
\Omega^{0,\epsilon'}_{\Gamma,\Gamma'}(z,p_0;p) \\
&& +  U^2_\Gamma(p_0 ) \zeta_0 e^{-\zeta_0 u_\Gamma(p_0) }
\Omega^{0,\epsilon'}_{\Gamma,\Gamma'}(0,p_0;p)
\end{eqnarray}

with the initial condition

\begin{eqnarray}
 \Omega^{2n=0,\epsilon'}_{\Gamma', \Gamma'}(\zeta_0,p_0;p)
= \delta_{\epsilon',+1} \frac{2}{\Gamma'^2 p^2} && [
U_{\Gamma'}(p_0+p)  e^{-\zeta_0 u_{\Gamma'}(p_0+p) }
- U_{\Gamma'}(p_0)  e^{-\zeta_0 u_{\Gamma'}(p_0) }
\\
&&
- p \partial_{p_0} 
\left( U_{\Gamma'}(p_0)  e^{-\zeta_0 u_{\Gamma'}(p_0) } \right)  ]  
\end{eqnarray}

In Laplace

\begin{eqnarray} 
P^{\epsilon,\epsilon'}_{\Gamma, \Gamma'}(q) =
 \int_0^{\infty} dx e^{-q x}
P^{\epsilon,\epsilon'}_{\Gamma, \Gamma'}(x)
\end{eqnarray}

the relations (\ref{relations}) become
\begin{eqnarray} 
 P^{+,+}_{\Gamma, \Gamma'}(q)
= &&  
\int_{\zeta_0} \Omega^{0,+}_{\Gamma, \Gamma'}(\zeta_0,0;q)
+\frac{P_{\Gamma}(q)}{1-P^2_{\Gamma}(q)} 
\int_{\zeta_1,\zeta_2}
E^{++}_{\Gamma, \Gamma'}(\zeta_1,q;\zeta_2,q; -q) \\
 P^{+,-}_{\Gamma, \Gamma'}(q)
= &&  
\int_{\zeta_0} \Omega^{0,-}_{\Gamma, \Gamma'}(\zeta_0,0;q)
+\frac{P_{\Gamma}(q)}{1-P^2_{\Gamma}(q)} 
\int_{\zeta_1,\zeta_2}
E^{+-}_{\Gamma, \Gamma'}(\zeta_1,q;\zeta_2,q; -q) \\
 P^{-,+}_{\Gamma, \Gamma'}(q)
= && 
\frac{1}{1-P^2_{\Gamma}(q)} 
\int_{\zeta_1,\zeta_2}
E^{-+}_{\Gamma, \Gamma'}(\zeta_1,q;\zeta_2,q; -q) \\
 P^{-,-}_{\Gamma, \Gamma'}(q)
= && 
\frac{1}{1-P^2_{\Gamma}(q)} 
\int_{\zeta_1,\zeta_2}
E^{--}_{\Gamma, \Gamma'}(\zeta_1,q;\zeta_2,q; -q)
\end{eqnarray}

Thus for the functions $E$, we need to solve only the case $p_1=p_2=q=-p$, i. e.
with the notation $U_{\Gamma}(p=0)=u_{\Gamma}(p=0)=u_{\Gamma}$

\begin{eqnarray} 
&& (\partial_{\Gamma} -\partial_{\zeta_1} -\partial_{\zeta_2} )
E^{\pm,\epsilon'}_{\Gamma,\Gamma'}(\zeta_1,q;\zeta_2,q;-q)
= -2 u_{\Gamma} E^{\pm,\epsilon'}_{\Gamma,\Gamma'}(\zeta_1,q;\zeta_2,q;-q) \\
&& + U_{\Gamma}^2(q) \int_0^{\zeta_1} dz  e^{-(\zeta_1-z) u_{\Gamma}(q)} 
E^{\pm,\epsilon'}_{\Gamma,\Gamma'}(z,q;\zeta_2,q;-q) \\
&& +  U_{\Gamma}^2(q) \int_0^{\zeta_2} dz  e^{-(\zeta_2-z) u_{\Gamma}(q)} 
E^{\pm,\epsilon'}_{\Gamma,\Gamma'}(\zeta_1,q;z,q; -q) \\
&& + U_{\Gamma}(q) u_{\Gamma}
\frac{ e^{-\zeta_1 u_{\Gamma}} - e^{-\zeta_1 u_{\Gamma}(q)} }{u_{\Gamma}(q) - u_{\Gamma}}
E^{\mp,\epsilon'}_{\Gamma,\Gamma'}(0,q;\zeta_2,q; -q) \\
&& + U_{\Gamma}(q) u_{\Gamma}
\frac{ e^{-\zeta_2 u_{\Gamma}} - e^{-\zeta_2 u_{\Gamma}(q)} }{u_{\Gamma}(q) - u_{\Gamma}}
E^{\mp,\epsilon'}_{\Gamma,\Gamma'}(\zeta_1,q;0,q;-q) \\
&& + u_{\Gamma}^2 \int_0^{\zeta_1} dz  e^{-(\zeta_1-z) u_{\Gamma}} E^{\pm,\epsilon'}_{\Gamma,\Gamma'}(z,q;\zeta_2,q;-q) \\
&& +  u_{\Gamma}^2 \int_0^{\zeta_2} dz  e^{-(\zeta_2-z) u_{\Gamma}} 
E^{\pm,\epsilon'}_{\Gamma,\Gamma'}(\zeta_1,q;z,q; -q) 
\end{eqnarray}

with the initial conditions

\begin{eqnarray} 
 E^{\epsilon,\epsilon'}_{\Gamma',\Gamma'}(\zeta_1,q;\zeta_2,q; -q)
= \delta_{\epsilon,\epsilon'}
\frac{2}{\Gamma'^2 q^2}
&& \left[ U_{\Gamma'}(q)  e^{-\zeta_1 u_{\Gamma'}(q) }
- u_{\Gamma'}  e^{-\zeta_1 u_{\Gamma'} }\right] \\
&& \left[ U_{\Gamma'}(q)  e^{-\zeta_2 u_{\Gamma'}(q) }
- u_{\Gamma'}  e^{-\zeta_2 u_{\Gamma'} }\right]
\end{eqnarray}

For $n=0$ we need to solve only the cases $p_0=p_1=0 $ and $p=q$,
i.e. the equation

\begin{eqnarray} 
&& \left(\partial_\Gamma - \partial_{\zeta_0} \right) 
\Omega^{0,\epsilon'}_{\Gamma,\Gamma'}(\zeta_0,0;q) = 
- 2 u_\Gamma 
\Omega^{0,\epsilon'}_{\Gamma,\Gamma'}(\zeta_0,0;q) \\
&& + U_{\Gamma}(q) \int_{z}
E^{+,\epsilon'}_{\Gamma,\Gamma'}(z,q;\zeta_0-z,q;-q) \\
&& + 
u_\Gamma \int_0^{\zeta_0} dz e^{-(\zeta_0-z) u_\Gamma }
\Omega^{1,\epsilon'}_{\Gamma,\Gamma'}(0,0;z,0;q) \\
&& 
+ u_\Gamma \int_0^{\zeta_0} dz e^{-(\zeta_0-z) u_\Gamma }
\Omega^{1,\epsilon'}_{\Gamma,\Gamma'}(z,0;0,0;q) \\
&& + 2 
u^2_\Gamma \int_0^{\zeta_0} dz e^{-(\zeta_0-z) u_\Gamma } 
\Omega^{0,\epsilon'}_{\Gamma,\Gamma'}(z,0;q) \\
&& +  u^2_\Gamma \zeta_0 e^{-\zeta_0 u_\Gamma }
\Omega^{0,\epsilon'}_{\Gamma,\Gamma'}(0,0;q)
\end{eqnarray}

with the initial conditions 

\begin{eqnarray}
 \Omega^{0,\epsilon'}_{\Gamma', \Gamma'}(\zeta_0,0;q)
= \delta_{\epsilon',+1} \frac{2}{\Gamma'^2 q^2} && [
U_{\Gamma'}(q)  e^{-\zeta_0 u_{\Gamma'}(q) }
 - U_{\Gamma'}  e^{-\zeta_0 u_{\Gamma'} } \\
&& 
- p \partial_{p_0} 
\left( U_{\Gamma'}(p_0)  e^{-\zeta_0 u_{\Gamma'}(p_0) } \right)\vert_{p_0=0}
 ]  
\end{eqnarray}

\subsection{Solution for the functions $E$}

Using the symmetry in $(\zeta_1,\zeta_2)$, we set

\begin{eqnarray} 
&& E^{\epsilon,\epsilon'}_{\Gamma,\Gamma'}(\zeta_1,q;\zeta_2,q; -q)
= \frac{2}{\Gamma^2} (A^{\epsilon,\epsilon'}_{\Gamma,\Gamma'}(q)
e^{-\zeta_1 u_{\Gamma}(q) } e^{-\zeta_2 u_{\Gamma}(q) }
+  B^{\epsilon,\epsilon'}_{\Gamma,\Gamma'}(q)
e^{-\zeta_1 u_{\Gamma} } e^{-\zeta_2 u_{\Gamma}(q) } \\
&& + B^{\epsilon,\epsilon'}_{\Gamma,\Gamma'}(q)
e^{-\zeta_1 u_{\Gamma}(q) } e^{-\zeta_2 u_{\Gamma} }
+  D^{\epsilon,\epsilon'}_{\Gamma,\Gamma'}(q)
e^{-\zeta_1 u_{\Gamma} } e^{-\zeta_2 u_{\Gamma} } )
\end{eqnarray}

It is in fact more convenient to use the following combinations
\begin{eqnarray} 
 I^{\epsilon,\epsilon'}_{\Gamma,\Gamma'}(q)
= && \frac{\Gamma^2}{2}
E^{\epsilon,\epsilon'}_{\Gamma',\Gamma'}(\zeta_1=0,q;\zeta_2=0,q; -q) \\
= && A^{\epsilon,\epsilon'}_{\Gamma,\Gamma'}(q)
+2  B^{\epsilon,\epsilon'}_{\Gamma,\Gamma'}(q)
+ D^{\epsilon,\epsilon'}_{\Gamma,\Gamma'}(q) \\
 J^{\epsilon,\epsilon'}_{\Gamma,\Gamma'}(q)
= && \frac{\Gamma^2}{2} \int_0^{\infty} d \zeta_2 E^{\epsilon,\epsilon'}_{\Gamma',\Gamma'}(\zeta_1=0,q;\zeta_2,q; -q) \\
= && \frac{A^{\epsilon,\epsilon'}_{\Gamma,\Gamma'}(q)}{u_{\Gamma}(q)}
+ B^{\epsilon,\epsilon'}_{\Gamma,\Gamma'}(q) \left(\frac{1}{u_{\Gamma}(q)} + \frac{1}{u_{\Gamma}}\right)
+\frac{D^{\epsilon,\epsilon'}_{\Gamma,\Gamma'}(q)}{u_{\Gamma}}  \\
K^{\epsilon,\epsilon'}_{\Gamma,\Gamma'}(q)
= && \frac{\Gamma^2}{2} \int_0^{\infty} d \zeta_1 \int_0^{\infty} d \zeta_2 E^{\epsilon,\epsilon'}_{\Gamma',\Gamma'}(\zeta_1,q;\zeta_2,q; -q) \\
= && \frac{A^{\epsilon,\epsilon'}_{\Gamma,\Gamma'}(q)}{u_{\Gamma}^2(q)}
+ 2 \frac{B^{\epsilon,\epsilon'}_{\Gamma,\Gamma'}(q) }
{u_{\Gamma}(q) u_{\Gamma}}
+\frac{D^{\epsilon,\epsilon'}_{\Gamma,\Gamma'}(q)}{u_{\Gamma}^2}
\end{eqnarray}

They satisfy the system
\begin{eqnarray} 
&& \partial_{\Gamma} I^{\pm,\epsilon'}_{\Gamma,\Gamma'}(q)
=-2 ( u_{\Gamma}+u_{\Gamma}(q)) I^{\pm,\epsilon'}_{\Gamma,\Gamma'}(q)
+ 2  u_{\Gamma}u_{\Gamma}(q) J^{\pm,\epsilon'}_{\Gamma,\Gamma'}(q)  \\
&& \partial_{\Gamma} J^{\pm,\epsilon'}_{\Gamma,\Gamma'}(q)
= \left( \frac{U_{\Gamma}^2(q)}{u_{\Gamma}(q)} -u_{\Gamma}(q) \right) J^{\pm,\epsilon'}_{\Gamma,\Gamma'}(q)
-I^{\pm,\epsilon'}_{\Gamma,\Gamma'}(q)
+   u_{\Gamma} u_{\Gamma}(q) K^{\pm,\epsilon'}_{\Gamma,\Gamma'}(q)
+\frac{U_{\Gamma}(q)}{u_{\Gamma}(q)} I^{\mp,\epsilon'}_{\Gamma,\Gamma'}(q)  \\
&& \partial_{\Gamma} K^{\pm,\epsilon'}_{\Gamma,\Gamma'}(q)
=\left(2 \frac{U_{\Gamma}^2(q)}{u_{\Gamma}(q)} + 2  u_{\Gamma} \right) K^{\pm,\epsilon'}_{\Gamma,\Gamma'}(q)
- 2 J^{\pm,\epsilon'}_{\Gamma,\Gamma'}(q)
+ 2 \frac{U_{\Gamma}(q)}{u_{\Gamma}(q)} J^{\mp,\epsilon'}_{\Gamma,\Gamma'}(q)
\end{eqnarray}
and the initial conditions
\begin{eqnarray} 
&& I^{\epsilon,\epsilon'}_{\Gamma',\Gamma'}(q)
= \delta_{\epsilon,\epsilon'}
\frac{1}{ q^2} \left[ U_{\Gamma'}(q)   - u_{\Gamma'} \right]^2 \\
&& J^{\epsilon,\epsilon'}_{\Gamma',\Gamma'}(q)
= \delta_{\epsilon,\epsilon'}
\frac{1}{ q^2} \left[ U_{\Gamma'}(q)   - u_{\Gamma'} \right]
\left[ \frac{ U_{\Gamma'}(q)   }{ u_{\Gamma'}(q) } -1 \right] \\
&& K^{\epsilon,\epsilon'}_{\Gamma',\Gamma'}(q)
= \delta_{\epsilon,\epsilon'}
\frac{1}{ q^2}\left[ \frac{ U_{\Gamma'}(q)   }{ u_{\Gamma'(q)} } -1 \right]^2
\end{eqnarray}

It is convenient to introduce the following sums and differences
\begin{eqnarray} 
&& I^{S,\epsilon'}_{\Gamma',\Gamma'}(q)
= I^{+,\epsilon'}_{\Gamma',\Gamma'}(q)+I^{-,\epsilon'}_{\Gamma',\Gamma'}(q) \\
&& I^{D,\epsilon'}_{\Gamma',\Gamma'}(q)
= I^{+,\epsilon'}_{\Gamma',\Gamma'}(q)-I^{-,\epsilon'}_{\Gamma',\Gamma'}(q) \\
&& J^{S,\epsilon'}_{\Gamma',\Gamma'}(q)
= J^{+,\epsilon'}_{\Gamma',\Gamma'}(q)+J^{-,\epsilon'}_{\Gamma',\Gamma'}(q) \\
&& J^{D,\epsilon'}_{\Gamma',\Gamma'}(q)
= J^{+,\epsilon'}_{\Gamma',\Gamma'}(q)-J^{-,\epsilon'}_{\Gamma',\Gamma'}(q) \\
&& K^{S,\epsilon'}_{\Gamma',\Gamma'}(q)
= K^{+,\epsilon'}_{\Gamma',\Gamma'}(q)+K^{-,\epsilon'}_{\Gamma',\Gamma'}(q) \\
&& K^{D,\epsilon'}_{\Gamma',\Gamma'}(q)
= K^{+,\epsilon'}_{\Gamma',\Gamma'}(q)-K^{-,\epsilon'}_{\Gamma',\Gamma'}(q) \\
\end{eqnarray}

Then for $\epsilon'$ fixed, the three functions $I^{S,\epsilon'}$,
$J^{S,\epsilon'}$ and $K^{S,\epsilon'}$ satisfy the system
\begin{eqnarray} 
&& \partial_{\Gamma} I^{S,\epsilon'}_{\Gamma,\Gamma'}(q)
=-2 ( u_{\Gamma}+u_{\Gamma}(q)) I^{S,\epsilon'}_{\Gamma,\Gamma'}(q)
+ 2  u_{\Gamma}u_{\Gamma}(q) J^{S,\epsilon'}_{\Gamma,\Gamma'}(q)  \\
&& \partial_{\Gamma} J^{S,\epsilon'}_{\Gamma,\Gamma'}(q)
= \left( \frac{U_{\Gamma}^2(q)}{u_{\Gamma}(q)} -u_{\Gamma}(q) \right) J^{S,\epsilon'}_{\Gamma,\Gamma'}(q)
-I^{S,\epsilon'}_{\Gamma,\Gamma'}(q)
+   u_{\Gamma} u_{\Gamma}(q) K^{S,\epsilon'}_{\Gamma,\Gamma'}(q)
+\frac{U_{\Gamma}(q)}{u_{\Gamma}(q)} I^{S,\epsilon'}_{\Gamma,\Gamma'}(q)  \\
&& \partial_{\Gamma} K^{S,\epsilon'}_{\Gamma,\Gamma'}(q)
=\left(2 \frac{U_{\Gamma}^2(q)}{u_{\Gamma}(q)} + 2  u_{\Gamma} \right) K^{S,\epsilon'}_{\Gamma,\Gamma'}(q)
- 2 J^{S,\epsilon'}_{\Gamma,\Gamma'}(q)
+ 2 \frac{U_{\Gamma}(q)}{u_{\Gamma}(q)} J^{S,\epsilon'}_{\Gamma,\Gamma'}(q)
\end{eqnarray}
with the initial conditions
\begin{eqnarray} 
&& I^{S,\epsilon'}_{\Gamma',\Gamma'}(q)
= \frac{1}{ q^2} \left[ U_{\Gamma'}(q)   - u_{\Gamma'} \right]^2 \\
&& J^{S,\epsilon'}_{\Gamma',\Gamma'}(q)
= \frac{1}{ q^2} \left[ U_{\Gamma'}(q)   - u_{\Gamma'} \right]
\left[ \frac{ U_{\Gamma'}(q)   }{ u_{\Gamma'}(q) } -1 \right] \\
&& K^{S,\epsilon'}_{\Gamma',\Gamma'}(q)
= \frac{1}{ q^2}\left[ \frac{ U_{\Gamma'}(q)   }{ u_{\Gamma'(q)} } -1 \right]^2
\end{eqnarray}
and thus the solution is simply
\begin{eqnarray} 
&& I^{S,\epsilon'}_{\Gamma,\Gamma'}(q)
= \frac{1}{ q^2} \left[ U_{\Gamma}(q)   - u_{\Gamma} \right]^2 \\
&& J^{S,\epsilon'}_{\Gamma,\Gamma'}(q)
= \frac{1}{ q^2} \left[ U_{\Gamma}(q)   - u_{\Gamma} \right]
\left[ \frac{ U_{\Gamma}(q)   }{ u_{\Gamma}(q) } -1 \right] \\
&& K^{S,\epsilon'}_{\Gamma,\Gamma'}(q)
= \frac{1}{ q^2}\left[ \frac{ U_{\Gamma}(q)   }{ u_{\Gamma}(q) } -1 \right]^2
\end{eqnarray}

For $\epsilon'$ fixed,
the three functions $I^{D,\epsilon'}$,
$J^{D,\epsilon'}$ and $K^{D,\epsilon'}$ satisfy the system
\begin{eqnarray} 
&& \partial_{\Gamma} I^{D,\epsilon'}_{\Gamma,\Gamma'}(q)
=-2 ( u_{\Gamma}+u_{\Gamma}(q)) I^{D,\epsilon'}_{\Gamma,\Gamma'}(q)
+ 2  u_{\Gamma}u_{\Gamma}(q) J^{D,\epsilon'}_{\Gamma,\Gamma'}(q)  \\
&& \partial_{\Gamma} J^{D,\epsilon'}_{\Gamma,\Gamma'}(q)
= \left( \frac{U_{\Gamma}^2(q)}{u_{\Gamma}(q)} -u_{\Gamma}(q) \right) J^{D,\epsilon'}_{\Gamma,\Gamma'}(q)
-I^{D,\epsilon'}_{\Gamma,\Gamma'}(q)
+   u_{\Gamma} u_{\Gamma}(q) K^{D,\epsilon'}_{\Gamma,\Gamma'}(q)
-\frac{U_{\Gamma}(q)}{u_{\Gamma}(q)} I^{D,\epsilon'}_{\Gamma,\Gamma'}(q)  \\
&& \partial_{\Gamma} K^{D,\epsilon'}_{\Gamma,\Gamma'}(q)
=\left(2 \frac{U_{\Gamma}^2(q)}{u_{\Gamma}(q)} + 2  u_{\Gamma} \right) K^{D,\epsilon'}_{\Gamma,\Gamma'}(q)
- 2 J^{D,\epsilon'}_{\Gamma,\Gamma'}(q)
- 2 \frac{U_{\Gamma}(q)}{u_{\Gamma}(q)} J^{D,\epsilon'}_{\Gamma,\Gamma'}(q)
\end{eqnarray}
with the initial conditions
\begin{eqnarray} 
&& I^{D,\epsilon'}_{\Gamma',\Gamma'}(q)
= (-1)^{\epsilon'}
\frac{1}{ q^2} \left[ U_{\Gamma'}(q)   - u_{\Gamma'} \right]^2 \\
&& J^{D,\epsilon'}_{\Gamma',\Gamma'}(q)
= (-1)^{\epsilon'}
\frac{1}{ q^2} \left[ U_{\Gamma'}(q)   - u_{\Gamma'} \right]
\left[ \frac{ U_{\Gamma'}(q)   }{ u_{\Gamma'}(q) } -1 \right] \\
&& K^{D,\epsilon'}_{\Gamma',\Gamma'}(q)
= (-1)^{\epsilon'}
\frac{1}{ q^2}\left[ \frac{ U_{\Gamma'}(q)   }{ u_{\Gamma'(q)} } -1 \right]^2
\end{eqnarray}

Three linearly independent solutions
of the system read 
\begin{eqnarray} 
&& I^{D1}_{\Gamma}(q)
=  \left[ U_{\Gamma}(q)   + u_{\Gamma} \right]^2 \\
&& J^{D1}_{\Gamma}(q)
=  \left[ U_{\Gamma}(q)   + u_{\Gamma} \right]
\left[ \frac{ U_{\Gamma}(q)   }{ u_{\Gamma}(q) } +1 \right] \\
&& K^{D1}_{\Gamma}(q)
= \left[ \frac{ U_{\Gamma}(q)   }{ u_{\Gamma(q)} } +1 \right]^2
\end{eqnarray}

\begin{eqnarray} 
&& I^{D2}_{\Gamma}(q)
=  2 u_{\Gamma} \frac{U_{\Gamma}(q)   + u_{\Gamma}}
{U_{\Gamma}(q)   + u_{\Gamma}(q)}  \\
&& J^{D2}_{\Gamma}(q)
=\frac{ (2 u_{\Gamma}+U_{\Gamma}(q)) \left[ \frac{ U_{\Gamma}(q)   }{ u_{\Gamma}(q) } +1 \right] } {U_{\Gamma}(q)   + u_{\Gamma}(q)}
 \\
&& K^{D2}_{\Gamma}(q)
= 2 \frac{\left[ \frac{ U_{\Gamma}(q)   }{ u_{\Gamma(q)} } +1 \right]} {u_{\Gamma}(q)}
\end{eqnarray}

\begin{eqnarray} 
&& I^{D3}_{\Gamma}(q)
= \frac{ U_{\Gamma}(q) [ 2 u_{\Gamma}^2+(2 u_{\Gamma}+U_{\Gamma}(q))(u_{\Gamma}(q)+U_{\Gamma}(q))]}{(u_{\Gamma}(q)+U_{\Gamma}(q)) (u_{\Gamma}(q)^2-U_{\Gamma}(q)^2)}  \\
&& J^{D3}_{\Gamma}(q)
=  \frac{ U_{\Gamma}(q) [2 u_{\Gamma}+u_{\Gamma}(q)+U_{\Gamma}(q)]}
{u_{\Gamma}(q) (u_{\Gamma}(q)^2-U_{\Gamma}(q)^2))} \\
&& K^{D3}_{\Gamma}(q)
= \frac{2 U_{\Gamma}(q)}{ u_{\Gamma}(q)^2 (u_{\Gamma}(q)-U_{\Gamma}(q)) }
\end{eqnarray}

It is useful to consider the matrix formed by these solutions

\begin{eqnarray*}
N_\Gamma =  \pmatrix{ 
&  I^{D1}_{\Gamma}(q) &  I^{D2}_{\Gamma}(q) &  I^{D3}_{\Gamma}(q) \\ 
& J^{D1}_{\Gamma}(q) &  J^{D2}_{\Gamma}(q) &  J^{D3}_{\Gamma}(q) \\
& K^{D1}_{\Gamma}(q) &  K^{D2}_{\Gamma}(q) &  K^{D3}_{\Gamma}(q)
} 
\end{eqnarray*}
whose determinant gives the
Wronskian $W_\Gamma$ of the three linear independent solutions:
\begin{eqnarray}
W_\Gamma= \det[N_\Gamma]= - \frac{U^3_{\Gamma}(q)}{u^3_{\Gamma}(q)}
\end{eqnarray}

The solution satisfying the initial conditions
we are interested in will be obtained as a linear combination
\begin{eqnarray}
\pmatrix{ 
& I^{D,\epsilon'}_{\Gamma,\Gamma'}(q) \nonumber \\
& J^{D,\epsilon'}_{\Gamma,\Gamma'}(q) \nonumber\\
& K^{D,\epsilon'}_{\Gamma,\Gamma'}(q)  } 
= \sum_{i=1}^3 \lambda^{i \epsilon'}_{\Gamma'}(q)
\pmatrix{ 
& I^{D i}_{\Gamma}(q) \nonumber \\
& J^{D i}_{\Gamma}(q) \nonumber \\
& K^{D i}_{\Gamma}(q) \nonumber }  \equiv
N_\Gamma \cdot \pmatrix{ 
& \lambda^{1 \epsilon'}_{\Gamma'}(q) \nonumber \\
& \lambda^{2 \epsilon'}_{\Gamma'}(q) \nonumber \\
& \lambda^{3 \epsilon'}_{\Gamma'}(q) \nonumber }
\end{eqnarray}
where the coefficients $\lambda^{1 \epsilon'}_{\Gamma'}(q)$
are determined by the initial conditions
\begin{eqnarray}
\pmatrix{ 
& \lambda^{1 \epsilon'}_{\Gamma'}(q) \nonumber \\
& \lambda^{2 \epsilon'}_{\Gamma'}(q) \nonumber \\
& \lambda^{3 \epsilon'}_{\Gamma'}(q) \nonumber } 
= N_{\Gamma'}^{-1} \cdot 
\pmatrix{ 
& I^{D,\epsilon'}_{\Gamma',\Gamma'}(q) \nonumber \\
& J^{D,\epsilon'}_{\Gamma',\Gamma'}(q) \nonumber\\
& K^{D,\epsilon'}_{\Gamma',\Gamma'}(q)  }
\end{eqnarray}

The inverse of the matrix $N_{\Gamma}$ is

\begin{eqnarray}
N^{-1}_\Gamma =  
\pmatrix{ 
& \frac{ 2 u_{\Gamma}(q)}{U_{\Gamma}(q)^2 [u_{\Gamma}(q) - U_{\Gamma}(q)] }
& - \frac{ 2 u_{\Gamma}(q)(2 u_{\Gamma} u_{\Gamma}(q) + 
U_{\Gamma}(q) [u_{\Gamma}(q) + U_{\Gamma}(q) ])}
     {U_{\Gamma}(q)^2 (u_{\Gamma}(q)^2 -U_{\Gamma}(q)^2)}
& \frac{ u_{\Gamma}(q)^2 (2 u_{\Gamma}^2 u_{\Gamma}(q) 
+ U_{\Gamma}(q) [2 u_{\Gamma} + U_{\Gamma}(q)] [u_{\Gamma}(q)+ U_{\Gamma}(q) ])}
     {U_{\Gamma}(q)^2 [u_{\Gamma}(q) + U_{\Gamma}(q)]  [u_{\Gamma}(q)^2 - U_{\Gamma}(q)^2 ]} 
\nonumber \\ 
&  - \frac{ (u_{\Gamma}(q) + U_{\Gamma}(q))}{U_{\Gamma}(q)^2} 
& \frac{ u_{\Gamma}(q)(2 u_{\Gamma} + U_{\Gamma}(q)) }{U_{\Gamma}(q)^2} 
& - \frac{ u_{\Gamma} u_{\Gamma}(q)^2 [u_{\Gamma} + U_{\Gamma}(q)] }{
       U_{\Gamma}(q)^2 [u_{\Gamma}(q) + U_{\Gamma}(q)]}
\nonumber \\
& - \frac{[u_{\Gamma}(q) + U_{\Gamma}(q)]^2}{U_{\Gamma}(q)^2}
& \frac{ 2 u_{\Gamma}(q)[u_{\Gamma} + U_{\Gamma}(q)] [u_{\Gamma}(q) + U_{\Gamma}(q)]}{U_{\Gamma}(q)}
& - \frac{ u_{\Gamma}(q)^2 [u_{\Gamma} + U_{\Gamma}(q)]^2}{U_{\Gamma}(q)^2} 
} 
\end{eqnarray}

and finally we get the coefficients

\begin{eqnarray}
&& \lambda^{1 \epsilon'}_{\Gamma'}(q) =
- (-1)^{\epsilon'} \frac{(-8 u_{\Gamma'}^2 u_{\Gamma'}(q)
+[u_{\Gamma'}(q) + U_{\Gamma'}(q) ] [ - 5 u_{\Gamma'}(q)^2 + U_{\Gamma'}^2 (q)+ 
        4 u_{\Gamma'} (3 u_{\Gamma'}(q)  - U_{\Gamma'}(q))])}
      {q^2 ( u_{\Gamma'}(q) - U_{\Gamma'}(q)) ( u_{\Gamma'}(q) + U_{\Gamma'}(q))^2)}
 \\
&& \lambda^{2 \epsilon'}_{\Gamma'}(q) =
- (-1)^{\epsilon'}\frac{ 2 (u_{\Gamma'} - u_{\Gamma'}(q)) (2 u_{\Gamma'} - u_{\Gamma'}(q) - U_{\Gamma'}(q))}
    {q^2 (u_{\Gamma'}(q) + U_{\Gamma'}(q))} \\
&& \lambda^{3 \epsilon'}_{\Gamma'}(q) =
- (-1)^{\epsilon'} \frac{4 (u_{\Gamma'} -u_{\Gamma'}(q))^2}{q^2}
\end{eqnarray}

In the end we are interested in 
\begin{eqnarray} 
 P^{+,+}_{\Gamma, \Gamma'}(q)
= &&  
\int_{\zeta_0} \Omega^{0,+}_{\Gamma, \Gamma'}(\zeta_0,0;q)
+  \frac{2}{\Gamma^2 q^2} \frac{P_{\Gamma}(q)}{1-P^2_{\Gamma}(q)} 
K^{+,+}_{\Gamma,\Gamma'}(q) \\
 P^{+,-}_{\Gamma, \Gamma'}(q)
= &&  
\int_{\zeta_0} \Omega^{0,-}_{\Gamma, \Gamma'}(\zeta_0,0;q)
+ \frac{2}{\Gamma^2 q^2} \frac{P_{\Gamma}(q)}{1-P^2_{\Gamma}(q)} 
K^{+,-}_{\Gamma,\Gamma'}(q) \\
 P^{-,+}_{\Gamma, \Gamma'}(q)
= && 
 \frac{2}{\Gamma^2 q^2} \ \frac{1}{1-P^2_{\Gamma}(q)}
K^{-,+}_{\Gamma,\Gamma'}(q)
 \\
 P^{-,-}_{\Gamma, \Gamma'}(q)
= && \frac{2}{\Gamma^2 q^2} \ \frac{1}{1-P^2_{\Gamma}(q)}
K^{-,-}_{\Gamma,\Gamma'}(q)
\end{eqnarray}

Since we have the relations
\begin{eqnarray} 
&& K^{S,+}_{\Gamma,\Gamma'}(q) = K^{S,-}_{\Gamma,\Gamma'}(q) \\
&& K^{D,+}_{\Gamma,\Gamma'}(q) = - K^{D,-}_{\Gamma,\Gamma'}(q) \\
\end{eqnarray}
we can express the four functions $K^{\epsilon,\epsilon'}_{\Gamma,\Gamma'}(q)$
in terms of $K^{S,+}_{\Gamma,\Gamma'}(q)$ and $K^{D,+}_{\Gamma,\Gamma'}(q)$
only as
\begin{eqnarray} 
K^{\epsilon,\epsilon'}_{\Gamma,\Gamma'}(q)
= \frac{1}{2} \left( 
K^{S,+}_{\Gamma,\Gamma'}(q) +
(-1)^{\epsilon} (-1)^{\epsilon'} K^{D,+}_{\Gamma,\Gamma'}(q)\right)  \end{eqnarray}

Since we have the following constraints
\begin{eqnarray} 
&& P^{+,+}_{\Gamma, \Gamma'}(q) +P^{+,-}_{\Gamma, \Gamma'}(q) = P^{+}_{\Gamma}(q)
  \\
&&  P^{-,+}_{\Gamma, \Gamma'}(q) +P^{-,-}_{\Gamma, \Gamma'}(q) = P^{-}_{\Gamma}(q)  
 \\
&& P^{+,+}_{\Gamma, \Gamma'}(q) +P^{-,+}_{\Gamma, \Gamma'}(q) = P^{+}_{\Gamma'}(q) 
 \\
&& P^{+,-}_{\Gamma, \Gamma'}(q) +P^{-,-}_{\Gamma, \Gamma'}(q) = P^{-}_{\Gamma'}(q) 
  \end{eqnarray}
where
\begin{eqnarray} 
&&  P^{+}_{\Gamma}(q) = \frac{1}{q} - \frac{2}{\Gamma^2 q^2}
 \ \frac{1-P_{\Gamma}(q) }{1+P_{\Gamma}(q)}
  \\
&& P^{-}_{\Gamma}(q) = \frac{2}{\Gamma^2 q^2}
 \ \frac{1-P_{\Gamma}(q) }{1+P_{\Gamma}(q)} 
 \end{eqnarray}
we do not have to compute separately the functions $\Omega^{0,-}_{\Gamma, \Gamma'}(\zeta_0,0;q)$.

Indeed, we get
\begin{eqnarray} 
&&
P^{+}_{\Gamma}(q)
=   
\int_{\zeta_0} \Omega^{0,+}_{\Gamma, \Gamma'}(\zeta_0,0;q)
+\int_{\zeta_0} \Omega^{0,-}_{\Gamma, \Gamma'}(\zeta_0,0;q)
+  \frac{2}{\Gamma^2 } \frac{P_{\Gamma}(q)}{1-P^2_{\Gamma}(q)} 
K^{S,+}_{\Gamma,\Gamma'}(q) \\
&&
P^{-}_{\Gamma}(q)
=   \frac{2}{\Gamma^2 } \ \frac{1}{1-P^2_{\Gamma}(q)} 
K^{S,+}_{\Gamma,\Gamma'}(q) \\
&& 
P^{+}_{\Gamma'}(q) =
\int_{\zeta_0} \Omega^{0,+}_{\Gamma, \Gamma'}(\zeta_0,0;q)
+ \frac{1}{\Gamma^2 } \ \frac{1}{1-P^2_{\Gamma}(q)}
\left( (1+P_{\Gamma}(q)) K^{S,+}_{\Gamma,\Gamma'}(q)
- (1-P_{\Gamma}(q)) K^{D,+}_{\Gamma,\Gamma'}(q) \right) \\
&& 
P^{-}_{\Gamma'}(q) =
\int_{\zeta_0} \Omega^{0,-}_{\Gamma, \Gamma'}(\zeta_0,0;q)
+ \frac{1}{\Gamma^2 } \ \frac{1}{1-P^2_{\Gamma}(q)}
\left( (1+P_{\Gamma}(q)) K^{S,+}_{\Gamma,\Gamma'}(q)
+ (1-P_{\Gamma}(q)) K^{D,+}_{\Gamma,\Gamma'}(q) \right)
\end{eqnarray}

The second equation is satisfied since $K^{S,+}_{\Gamma,\Gamma'}(q)
= \frac{1}{ q^2}\left[ 1- P_{\Gamma}(q)  \right]^2 $.
The three other are compatible and give
\begin{eqnarray} 
&& \int_{\zeta_0} \Omega^{0,+}_{\Gamma, \Gamma'}(\zeta_0,0;q)
=\frac{1}{q} - \frac{2}{\Gamma'^2 q^2} \ \frac{1-P_{\Gamma'}(q)}{1+P_{\Gamma'}(q)}
-\frac{1}{\Gamma^2 q^2} (1-P_{\Gamma}(q))
+ \frac{1}{\Gamma^2 } \ \frac{1}{1+P_{\Gamma}(q)}
K^{D,+}_{\Gamma,\Gamma'}(q)\\
&& 
\int_{\zeta_0} \Omega^{0,-}_{\Gamma, \Gamma'}(\zeta_0,0;q)
=\frac{2}{\Gamma'^2 q^2} \ \frac{1-P_{\Gamma'}(q)}{1+P_{\Gamma'}(q)}
-\frac{1}{\Gamma^2 q^2} (1-P_{\Gamma}(q))
- \frac{1}{\Gamma^2 } \ \frac{1}{1+P_{\Gamma}(q)}
K^{D,+}_{\Gamma,\Gamma'}(q)
\end{eqnarray}

\subsection{ Final result}

The Laplace transform of the correlation function can now be obtained as
\begin{eqnarray} 
&& \int_0^{\infty} dx e^{-qx} 
\overline{ \langle S_0(t) S_x(t) \rangle \langle S_0(t') S_x(t') \rangle  } = P^{+,+}_{\Gamma, \Gamma'}(q)+P^{-,-}_{\Gamma, \Gamma'}(q)
-P^{-,+}_{\Gamma, \Gamma'}(q) -P^{+,-}_{\Gamma, \Gamma'}(q)  \\
&& \qquad \qquad \qquad \qquad =
\frac{1}{q} 
-\frac{4}{\Gamma'^2 q^2} \ \frac{1-P_{\Gamma'}(q)}{1+P_{\Gamma'}(q)} 
+\frac{4}{\Gamma^2} \ \frac{1}{1-P^2_{\Gamma}(q)} K^{D,+}_{\Gamma,\Gamma'}(q) 
\end{eqnarray}
and thus the final explicit expression is
\begin{eqnarray} 
&& \int_0^{\infty} dx e^{-qx} \overline{ \langle S_0(t) S_x(t) \rangle \langle S_0(t') S_x(t') \rangle  } 
 = \frac{1}{q} 
-\frac{4}{\Gamma'^2 q^2} {\rm tanh}^2 \left( \frac{\Gamma' \sqrt q }{2} \right) \\
&& + \frac{4}{\Gamma^2 } {\rm cotanh}^2 \left( \frac{\Gamma \sqrt q }{2} \right)
\lambda^{1 +}_{\Gamma'}(q) +\frac{8}{\Gamma^2 } 
\frac{ {\rm cotanh} \left( \frac{\Gamma \sqrt q }{2} \right)} { \sqrt q}
\lambda^{2 +}_{\Gamma'}(q)
+\frac{4}{\Gamma^2 } \frac{1}
{ q \sinh^2\left( \frac{\Gamma \sqrt q }{2} \right) }
\lambda^{3 +}_{\Gamma'}(q)
\end{eqnarray}

where

\begin{eqnarray}
&& \lambda^{1 +}_{\Gamma'}(q) =
\frac{1}{2 \Gamma'^2 q^3 \sinh^2(\Gamma' {\sqrt q} )} 
\left[ 8+3 \Gamma'^2 q-16 \cosh(\Gamma' {\sqrt q})+
8 \Gamma' {\sqrt q} \sinh(\Gamma' {\sqrt q}) \right. \\
&& \left. \qquad  \qquad  \qquad  \qquad  \qquad
+(8+5 \Gamma'^2 q) \cosh(2 \Gamma' {\sqrt q})
-12 \Gamma' {\sqrt q} \sinh ( 2 \Gamma' {\sqrt q}) \right] \\
&& \lambda^{2 +}_{\Gamma'}(q) 
=- \frac{2}{q^2} \left( \frac{1}{\Gamma'} - 
{\sqrt q} {\rm coth} (\Gamma' {\sqrt q})\right)
\left( \frac{2}{\Gamma' \sqrt q}  {\rm tanh} \left( \frac{\Gamma' {\sqrt q}}{2} \right)-1\right)
 \\
&& \lambda^{3 \epsilon'}_{\Gamma'}(q) =
- \frac{4 }{q^2} \left( \frac{1}{\Gamma'} - 
{\sqrt q} {\rm coth} (\Gamma' {\sqrt q})\right)^2
\end{eqnarray}
This leads to the scaling form given in the text in equations (\ref{scalssss}) and
(\ref{resssss}). 

\section{Finite size properties}

\label{appfinite}

In this Appendix we sketch the derivation of the finite size 
results from the finite size measure for the RFIM.

If spins at both ends are fixed to have the same (resp. opposite) value,
there are an odd (resp. even) number of bonds in the finite size measure.

Assuming for definitness that spins at both extremities are fixed
to the values $S_0=+1$ and $S_L=-1$.
Then domains closest to the boundaries are domain walls
of type $A$ which as Sinai walkers see reflecting
boundary conditions. In this case, as explained in \cite{us_long},
the probability $N_{\Gamma,L}^{2k+2}(l_1,l_2,\ldots,l_{2k+2})$
that the system at scale $\Gamma$ has $(2k+2)$ bonds
($k=0,1, \ldots$), with respective lengths $(l_1, \ldots, l_{2k+2})$
is  
\begin{eqnarray} \label{decoupledrr}
 N_{\Gamma,L}^{2k+2}(l_1,l_2,\ldots,l_{2k+2}) 
= E^{+}_{\Gamma}(l_1) P^{-}_{\Gamma}(l_2) P^{+}_{\Gamma}(l_3)\ldots
 P^{+}_{\Gamma}( l_{2k+1}) 
 E^{-}_{\Gamma}(l_{2k+2}) 
\overline{l}_\Gamma \delta\left(L-\sum_{i=1}^{2k+2} l_i\right) 
\end{eqnarray}
where 
$P_{\Gamma}^{\pm}(l)$ are the bulk length distributions 
 (\ref{lengthdis}), $\overline{l}_\Gamma=\overline{l}_\Gamma^+ + \overline{l}_\Gamma^-$ ,
and the $E_{\Gamma}^{\pm}(l)$ the length distribution
of boundary bonds (see \cite{us_long}).
The normalisation is
\begin{eqnarray} \label{decoupledrr2}
\sum_{k=0}^{\infty} \int_{l_1, \ldots l_{2k+2} }
N_{\Gamma,L}^{2k+2}(l_1,l_2,\ldots,l_{2k+2})  = 1 
\end{eqnarray}

The probability that the system has $(2k+2)$ domains
at scale $\Gamma$ is
\begin{eqnarray} 
I_L(k; \Gamma) = \int_{l_1,... l_{2k+2}}
E_{\Gamma}^+(l_1) P_{\Gamma}^-(l_2) ... P_{\Gamma}^+( l_{2k+1} ) E_{\Gamma}^-(l_{2k+2})
\overline{l_{\Gamma}} \delta(L-\sum_{i=1}^{2k+2} l_i)
\end{eqnarray}
so that the Laplace transform with respect to the length
of the generating function is
\begin{eqnarray} 
\int_0^{\infty} d L e^{-q L}
\left( \sum_{k=0}^{\infty} z^k I_L(k; \Gamma) \right)
= && \overline{l_{\Gamma}} \frac{E_{\Gamma}^+(q/2g) E_{\Gamma}^-(q/2g)}
{1-z P_{\Gamma}^+(q/2g) P_{\Gamma}^-(q/2g)} =
\frac{2 g}{q} \frac{1-P_{\Gamma}^+(q/2g) P_{\Gamma}^-(q/2g)}
{1-z P_{\Gamma}^+(q/2g) P_{\Gamma}^-(q/2g)} 
\end{eqnarray}
where $p=\frac{q}{2 g} $. This leads to the results given in the text.

The magnetization can also be obtained.
In the "full" renormalized landscape, the magnetisation 
is given by $ M_L = \sum_{i=1}^{i=2k+2} (-1)^{i+1} l_i$.
The probability that it has value $M$ simply is
\begin{eqnarray}
 F_{L}(M;\Gamma) 
&& = \sum_{k=0}^{k=+\infty} 
\int_{l_1, \ldots , l_{2k+2}} 
E^{+}(l_1) P^{-}(l_2) P^{+}(l_3)
.. P^{+}(l_{2 k-1}) E^{-}(l_{2 k+2}) \\
&& \overline{l}_\Gamma \delta(L-\sum_{i=1}^{2 k+2} l_i) 
\delta(M - \sum_{i=1}^{i=2k+2} (-1)^{i+1} l_i)
\end{eqnarray}
This leads to the result given in the text.


\begin{thebibliography}{10}
\bibitem{rareb}
The type (b) rare events of \cite{us_long} are here equivalent to (a) type events.
\bibitem{misprint}
Note the following misprint in the published version of \cite{us_long}. In the
  formula below Eq. (187) for $c_k(T)$ the lower bound of the integral should
  be $0$ rather than $- \infty$ (since $w$ for a bond is a positive quantity),
  and thus one gets $c_k(T) = T/k$. The only consequence is to divide by $2$
  the first term in the end result Eq. (193) and Eq. (194), which should now
  read $\overline{\langle |x(t) - \langle x(t) \rangle | \rangle} =
  \frac{2}{45} (34 + 41 \ln 2) T (T \ln t)$ and $\overline{\langle x(t)^2
  \rangle - \langle x(t) \rangle^2} = \frac{4}{315} (\frac{95}{2} + 68 \ln 2) T
  (T \ln t)^3$.
\bibitem{footnote_drift}
In the presence of an applied field $H>0$, corresponding to the biased case, one finds that
  the correlation length for the truncated correlations is $1/\xi^{eq}_{th} =
  (\pi^2 g/8 J^2) + (H^2/2 g)$ in agreement with \cite{farhi_gutman}. It
  becomes different from the non-truncated correlation length. The change due
  to the bias in the motion of the domain walls, $1/\xi \to 1/\xi(\delta=0) + \delta^2$, is similar to the one
  obtained in \cite{us_long} Section F.
\bibitem{footnotesg} 
In mean field models the non truncated correlation is computed, but this should not
make a difference for spin glasses which have vanishing magnetization.
\end{thebibliography}

\end{document}